\documentclass[letterpaper,twocolumn]{aastex62}
\shorttitle{K2 Cool Dwarfs IV}
\shortauthors{Dressing et al.}
\usepackage{epsfig}
\usepackage{amsmath}
\usepackage{rotating}
\usepackage{natbib} 
\usepackage{enumerate}
\usepackage{hyperref}
\usepackage{url}
\usepackage{longtable}
\usepackage{color}
\accepted{2019 May 24}
\submitjournal{\emph{The Astronomical Journal}}

\begin{document}

\def\mearth{{\rm\,M_\oplus}}                                                    
\def\msun{{\rm\,M_\odot}}                                                       
\def\rsun{{\rm\,R_\odot}}                                                       
\def\rearth{{\rm\,R_\oplus}} ннн
\def\fearth{{\rm\,F_\oplus}}
\def\lsun{{\rm\,L_\odot}}                                                          
\def\kepler {{\emph{Kepler}\,}}                                              
\newcommand{\teff}{\ensuremath{T_{\mathrm{eff}}}}                               
\newcommand{\logg}{\ensuremath{\log g}}

%\newcommand{\rev}[2][purple]{\textcolor{#1}{\textbf{#2}}}

%%%%%%%%%%%%%%%%%%%%%%%%%%%%%%%%%%%%%%%%%%%%%%%%%
% THE DOCUMENT BEGINS HERE                      %
%%%%%%%%%%%%%%%%%%%%%%%%%%%%%%%%%%%%%%%%%%%%%%%%%

\title{Characterizing \emph{K2} Candidate Planetary Systems Orbiting Low-Mass Stars IV:\\ Updated Properties for 86 Cool Dwarfs Observed During Campaigns 1--17}
\author{Courtney D. Dressing}
\affiliation{Astronomy Department, University of California, Berkeley, CA 94720, USA}

\correspondingauthor{Courtney D. Dressing}
\email{dressing@berkeley.edu}

\author{Kevin Hardegree-Ullman}
\affiliation{Department of Physics and Astronomy, University of Toledo, Toledo, OH, 43606, USA}
\affiliation{IPAC-NExScI, Mail Code 100-22, Caltech, 1200 E. California Blvd., Pasadena, CA 91125, USA}

\author{Joshua E. Schlieder}
\affiliation{INASA Goddard Space Flight Center, Greenbelt, MD 20771, USA}

\author{Elisabeth Newton}
\affiliation{Department of Physics, and Kavli Institute for Astrophysics and  Space  Research,  Massachusetts Institute of Technology, Cambridge, MA 02139, USA}
\affiliation{Department of Physics and Astronomy, Dartmouth College, Hanover, NH 03755 USA}

\author{Andrew Vanderburg}
\altaffiliation{NASA Sagan Fellow}
\affiliation{Department of Astronomy, The University of Texas at Austin, Austin, TX 78712, USA}

\author{Adina D. Feinstein}
\affiliation{Department of Astronomy \& Astrophysics, The University of Chicago, Chicago, IL 60637 USA}

\author{Girish M. Duvvuri}
\affiliation{Department of Astrophysical \& Planetary Sciences, University of Colorado, Boulder, CO 80309 USA}

\author{Lauren Arnold}
\affiliation{Marine Biology Graduate Program, University of Hawai`i at M\=anoa, 2525 Correa Rd., Honolulu, HI 96822}

\author{Makennah Bristow}
\affiliation{Department of Physics, University of North Carolina at Asheville, Asheville, NC USA}

\author{Beverly Thackeray}
\affiliation{Department of Astronomy, University of Maryland College Park, College Park, MD USA}

\author{Ellianna Schwab Abrahams}
\altaffiliation{National Science Foundation Graduate Research Fellow}
\affiliation{Astronomy Department, University of California, Berkeley, CA 94720, USA}

\author{David Ciardi}
\affiliation{IPAC-NExScI, Mail Code 100-22, Caltech, 1200 E. California Blvd., Pasadena, CA 91125, USA}

\author{Ian Crossfield}
\affiliation{Department of Physics, and Kavli Institute for Astrophysics and  Space  Research,  Massachusetts Institute of Technology, Cambridge, MA 02139, USA}

\author{Liang Yu}
\affiliation{Department of Physics, and Kavli Institute for Astrophysics and  Space  Research,  Massachusetts Institute of Technology, Cambridge, MA 02139, USA}

\author{Arturo O. Martinez}
\affiliation{Department of Physics and Astronomy, Georgia State University, 25 Park Pl NE \#605, Atlanta, GA, 30303, USA}

\author{Jessie L. Christiansen}
\affiliation{IPAC-NExScI, Mail Code 100-22, Caltech, 1200 E. California Blvd., Pasadena, CA 91125, USA}

\author{Justin R. Crepp}
\affiliation{Department of Physics, University of Notre Dame, Notre Dame, IN 46556, USA}

\author{Howard Isaacson}
\affiliation{Astronomy Department, University of California, Berkeley, CA 94720, USA}

%\date{\today}
%\slugcomment{Accepted to the Astronomical Journal}

\begin{abstract}
We present revised stellar properties for 172 \emph{K2} target stars that were identified as possible hosts of transiting planets during Campaigns~1--17. Using medium-resolution near-infrared spectra acquired with the NASA Infrared Telescope Facility/SpeX and Palomar/TripleSpec, we found that 86 of our targets were bona fide cool dwarfs, 74 were hotter dwarfs, and 12 were giants. Combining our spectroscopic metallicities with Gaia parallaxes and archival photometry, we derived photometric stellar parameters and compared them to our spectroscopic estimates. Although our spectroscopic and photometric radius and temperature estimates are consistent, our photometric mass estimates are systematically $\Delta M_\star = 0.11\msun$ (34\%) higher than our spectroscopic mass estimates for the least massive stars ($M_{\star, phot} < 0.4\msun$). Adopting the photometric parameters and comparing our results to parameters reported in the Ecliptic Plane Input Catalog, our revised stellar radii are $\Delta R_\star = 0.15\rsun$ ($40\%$) larger and our revised stellar effective temperatures are roughly $\Delta T_{\rm eff} = 65$K cooler. Correctly determining the properties of K2 target stars is essential for characterizing any associated planet candidates, estimating the planet search sensitivity, and calculating planet occurrence rates. Even though Gaia parallaxes have increased the power of photometric surveys, spectroscopic characterization remains essential for determining stellar metallicities and investigating correlations between stellar metallicity and planetary properties. 
 \end{abstract}


%\maketitle

\section{Introduction}
In 2013, the \emph{Kepler} spacecraft was repurposed for the \emph{K2} mission: a survey for transiting planets in a series of observation fields along the ecliptic plane. For each ``Campaign,'' the \emph{K2} Guest Observer office solicited proposals for target stars from the community. The selected targets are therefore a conglomeration of stars chosen for a variety of science programs. Recognizing that the short duration of each \emph{K2} campaign (65--90~days) precluded the detection of most small, cool planets transiting Sunlike stars, many \emph{K2} proposers interested in those planets nominated stars believed to be cool dwarfs. Despite the short \emph{K2} observation periods, the smaller radii and lower temperatures of cool dwarfs permit the detection of multiple transits of potentially habitable planets. As noted by \citet{huber_et_al2016}, 41\% of stars observed by \emph{K2} during Campaigns 1--8 were initially classified as cool dwarfs and \emph{K2} has already observed many more cool dwarfs than the primary \emph{Kepler} mission: during Campaigns 0--14, \emph{K2} observed 50,159~potential cool dwarfs.\footnote{\url{https://keplerscience.arc.nasa.gov/keplers-k2-mission-reaches-300000-standard-targets-and-200-confirmed-planets.html}}

Due to the fast-paced nature of the \emph{K2} mission many of the proposed targets were not well-characterized prior to observation by \emph{K2}. Accordingly, a variety of teams pursue ground-based characterization of \emph{K2} target stars after possible planets are detected. In this paper, we present revised characterizations for 172~\emph{K2} Objects of Interest (K2OIs). As in the first paper in this series \citep{dressing_et_al2017a}, we use empirically-based relations \citep{newton_et_al2014, newton_et_al2015, mann_et_al2013a, mann_et_al2014, mann_et_al2015} to determine the properties of potential cool dwarfs. 

All of our targets have initial characterizations from the \emph{K2} Ecliptic Plane Input Catalog \citep[EPIC,][]{huber_et_al2016}, a  database containing photometry, kinematics, and stellar properties for stars in and near the fields targeted by \emph{K2}. For the vast majority of stars, the parameters were estimated from photometry and proper motions by using the \emph{Galaxia} galactic model \citep{sharma_et_al2011} and Padova isochrones \citep{girardi_et_al2000, marigo+girardi2007, marigo_et_al2008} to generate mock realizations of each \emph{K2} field. \emph{Galaxia} produces synthetic stellar catalogs based on the adopted galactic conditions and survey parameters. The baseline version uses the \emph{Besan\c{c}on} analytical model \citet{robin_et_al2003} for the disk of the Milky Way and N-body models by \citet{bullock+johnston2005} for the stellar halo. The Padova ischrones are an established set of models for stars with initial masses of $0.15 \msun < M_\star < 7 \msun$ and a range of metallicities ($0.0004 \leq Z \leq 0.03$). For Sun-like stars, the Padova models agree well with other models such as Yonsei-Yale models \citep{yi_et_al2003, yi_et_al2004} and Dartmouth models \citep{dotter_et_al2008}, but the Padova models predict lower luminosities for cooler stars.

In the EPIC, a subset of stars have slightly more accurate properties based on \emph{Hipparcos} parallaxes \citep{vanleeuwen2007} and spectroscopy from RAVE~DR4 \citep{kordopatis_et_al2013}, LAMOST DR1 \citep{luo_et_al2015}, and APOGEE DR12 \citep{alam_et_al2015}. Due to the reliance on Padova isochrones, \citet{huber_et_al2016} cautioned that radius estimates for cool dwarfs may be up to 20\% too small because the Padova isochrones are known to systematically underestimate the radii of cool dwarfs \citep{boyajian_et_al2012}. 

In \citet{dressing_et_al2017a}, we acquired and analyzed NIR spectra of 144~candidate cool dwarfs observed by \emph{K2} during Campaigns~1--7. We found that half of the candidate cool dwarfs were giant stars or hotter dwarfs. For the 72~stars classified as cool dwarfs, we determined their radii to be roughly $0.13\rsun$ (39\%) larger than the estimates provided in the EPIC. Our cool dwarf sample included stars with spectral types of K3 to M4, stellar effective temperatures of 3276--4753K, and stellar radii of $0.19-0.78 \rsun$.

Similarly, in \citet{martinez_et_al2017} we refined the properties of 34~cool dwarfs using NIR spectra acquired using the SOFI spectrograph \citep{moorwood_et_al1998} at the New Technology Telescope and found a median radius difference of $0.15\rsun$ compared to the values in the EPIC. We saw no systematic difference between our revised temperatures and those estimated in the EPIC, which suggests that the problem is primarily due to the overly petite model radii. This could result from the underlying model assumptions of the Padova isochrones such as treatment of opacities, convection, magnetic fields, star spots, and other phenomena intrinsic to low-mass stars \citep[e.g.,][]{feiden+chaboyer2012, feiden+chaboyer2013}.

Our work is one of many complementary efforts to improve the characterization of planetary systems and target stars observed by \emph{K2}. For instance, \citet{wittenmyer_et_al2018} presented revised properties for 46~\emph{K2} target stars by obtaining high-resolution spectra with the HERMES multi-object spectrograph on the Anglo-Australian Telescope and \citet{hirano_et_al2018} acquired AO imaging and optical spectra to characterize 16 planets orbiting 12 low-mass \emph{K2} target stars. 

The overall goals of our multi-semester project are to characterize the set of cool dwarf planetary systems detected by the \emph{K2} mission and investigate the overall prevalence and properties of cool dwarf planetary systems. In Paper I \citep{dressing_et_al2017a}, we established the project and characterized the first set of candidate cool dwarfs observed by our program. We then revised the properties of the associated planet candidates by combining our revised stellar characterizations with new fits of the \emph{K2} transit photometry \citep[][Paper II]{dressing_et_al2017b}. In Paper III \citep{dressing_et_al2018a}, we focused on K2-55b, a surprisingly massive Neptune-sized planet for which we had refined the orbital ephemerides by observing an additional transit with \emph{Spitzer} and measured the mass with Keck/HIRES. The next paper in this series (Dressing, Vanderburg et al., \emph{in prep}) will present updated transit fits, false positive probabilities, and bulk properties for the planet candidates associated with the stars classified in this paper. 

This paper is focused on the characterization of the second set of stars observed by our program: 172  candidate cool dwarfs identified as possible planet host stars based on data acquired during \emph{K2}~campaigns \mbox{1--17}. We characterize these stars using a combination of archival photometry, new spectroscopic observations obtained by our team, and recently released astrometric data from Gaia~DR2 \citep{gaia_et_al2018, babusiaux_et_al2018, cropper_et_al2018, evans_et_al2018, hambly_et_al2018, luri_et_al2018, mignard_et_al2018, riello_et_al2018, sartoretti_et_al2018, soubiran_et_al2018}. In Section~\ref{sec:targets}, we describe our target sample. We then present our spectroscopic observations in Section~\ref{sec:observations} and discuss our stellar classification procedure in Section~\ref{sec:analysis}. In Section~\ref{sec:discussion} we review the revised properties of the target sample and compare our new parameter estimates to the results of previous studies. We conclude in Section~\ref{sec:conclusions}. 

\section{Target Sample}
\label{sec:targets}
\begin{figure}[tbp]
\centering
\includegraphics[width=0.5\textwidth]{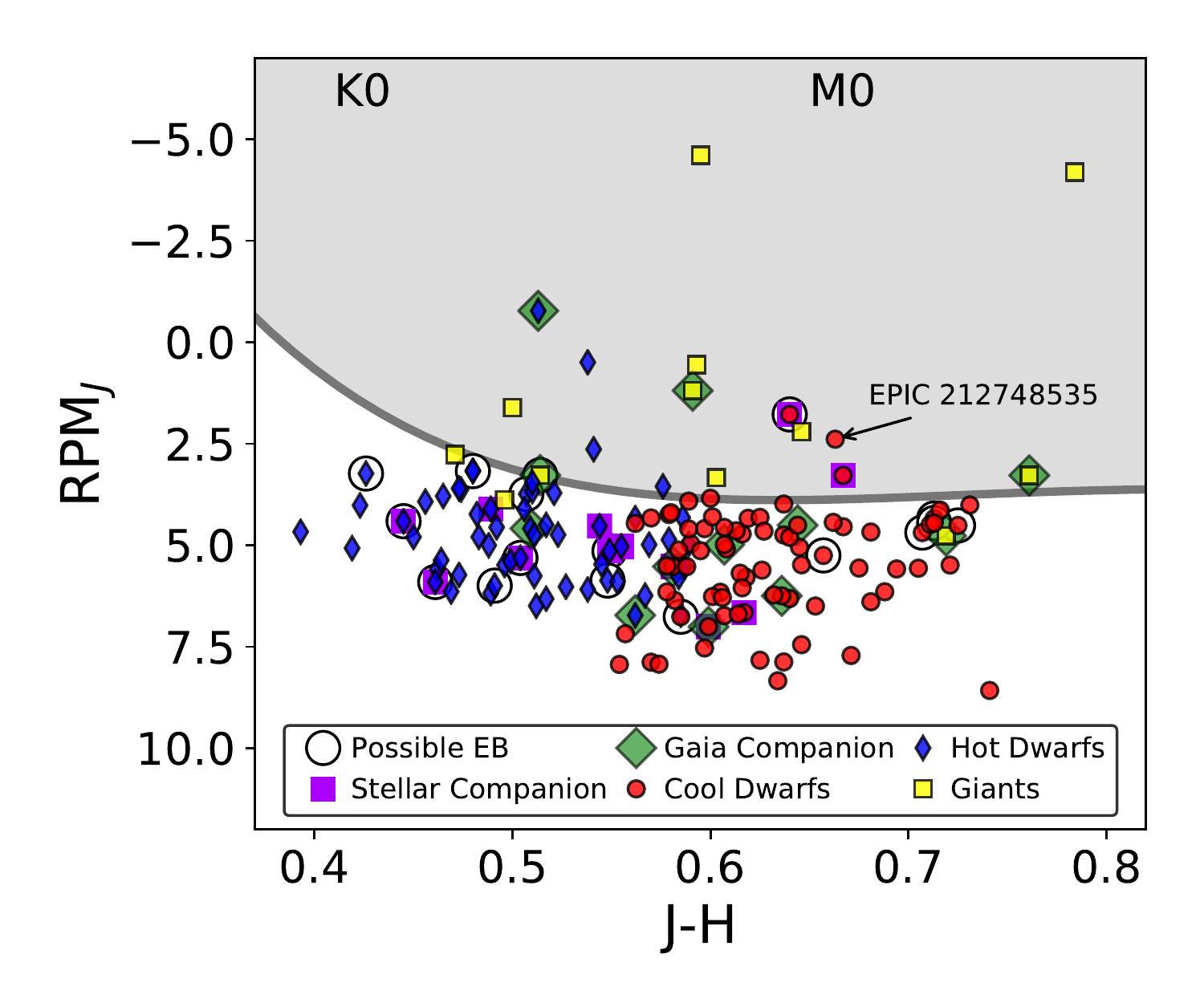}
\caption{Reduced proper motion in $J$ band vs. $J - H$ color for all of the stars we observed and later classified as giants (yellow squares), hotter dwarfs (blue diamonds), or cool dwarfs (red circles). The gray line marks the dwarf/giant cut suggested by \citet{collier-cameron_et_al2007}; stars lying above this line (in the gray shaded region) are more likely to be giants while targets below the line are more likely to be dwarfs. For reference, we note the approximate $J-H$ colors of K0 and M0 stars. Note that some stars do not appear on this plot because they did not have proper motions reported in the EPIC. \label{fig:rpm_color}} 
\end{figure}

The overarching goal of our project is to characterize planetary systems orbiting cool dwarfs observed by \emph{K2}. Accordingly, we selected our targets from the set of planet candidate host stars. The majority of our targets were the putative hosts of candidate planets discovered by A.~Vanderburg and the \emph{K2} California Consortium (K2C2), but we also consulted the public repository of \emph{K2} candidates provided by the NASA Exoplanet Archive\footnote{\url{https://exoplanetarchive.ipac.caltech.edu/cgi-bin/TblView/nph-tblView?app=ExoTbls&config=k2candidates}}  \citep{akeson_et_al2013}. We aimed to characterize all stars with proper motions and colors consistent with those of cool dwarfs (see Figure~\ref{fig:rpm_color}) as well as those for which the planet candidate discoverers estimated host star properties of $T_{\rm eff} \leq 5000$K and $\log g \geq 4.0$.

Over the 37~nights listed in Table~\ref{tab:observing}, we observed 172~possible cool dwarfs that were identified as likely planet host stars. Many of these targets were selected from unpublished lists provided by A. Vanderburg (75~stars, 45\%) or from the K2C2 candidate lists (93~stars, 56\%) later published by \citet{petigura_et_al2018}, \citet{livingston_et_al2018}, \citet{yu_et_al2018}, and \citet{crossfield_et_al2018}. The majority of the targets appear on the \emph{K2} Candidates Table on the NASA Exoplanet Archive (109~stars, 65\%). Those candidates were previously published by \citet[][7~stars]{montet_et_al2015},  \citet[][5~stars]{adams_et_al2016}, \citet[][24~stars]{barros_et_al2016},  \citet[][37~stars]{crossfield_et_al2016},  \citet[][2~stars]{libralato_et_al2016}, \citet[][13~stars]{pope_et_al2016}, \citet[][56~stars]{vanderburg_et_al2016}, \citet[][2~stars]{schmitt_et_al2016},  \citet[][1~star]{mann_et_al2017b}, \citet[][3~stars]{rizzuto_et_al2017}, \citet[][22~stars]{mayo_et_al2018}, and \citet[][19~stars]{petigura_et_al2018}. Note that there is substantial overlap across \emph{K2} candidate lists and that many stars appear on multiple lists.

One of the goals of our overall program is to determine the bulk and atmospheric composition of small planets. Accordingly, we tended to prioritize follow-up observations of  candidate planets orbiting bright stars because brighter stars are more amenable to radial velocity mass measurement and subsequent atmospheric characterization. We also investigated candidate reliability by inspecting the \emph{K2} photometry of possible targets and consulting the ExoFOP-K2 follow-up website\footnote{\url{https://exofop.ipac.caltech.edu/k2/}} for notes from other observers. We avoided observing candidates already classified as eclipsing binaries and favored targets without nearby stellar companions. See the companion paper for a detailed discussion of the \emph{K2} Objects of Interest associated with each target star (Dressing, Vanderburg et al., \emph{in prep}).

\section{Observations}
\label{sec:observations}
As in \citet{dressing_et_al2017a}, we conducted our observations using two medium resolution spectrographs: SpeX on the NASA Infrared Telescope Facility (IRTF) and TripleSpec on the Palomar 200". Our SpeX observations were acquired during the 2016B - 2018A semesters as part of programs 2016B057, 2017A019, 2017B064, and 2018A073 (PI: Dressing). Our TripleSpec observations were obtained during 2016A - 2017B through programs P08, P03, P11, and P08 (PI: Dressing). 

We provide additional details about the weather and targets observed during each run in Table~\ref{tab:observing}. We reserved our faintest targets for the most pristine conditions and observed our brighter targets during poor weather. In all cases, we removed the telluric features from our science spectra by observing nearby A0V stars within one hour of our science observations \citep{vacca_et_al2003}. We strove to find A0V stars at similar airmasses (difference $< 0.1$ airmasses) and within $15^{\circ}$ of our target stars. 

\subsection{IRTF/SpeX}
We conducted our SpeX observations in SXD mode using the $0\farcs3 \times 15\arcsec$ slit to acquire moderate resolution ($R \approx 2000$) spectra \citep{rayner_et_al2003, rayner_et_al2004}. All of these observations were obtained after the SpeX upgrade in 2014 and therefore cover a broad wavelength range of 0.7 to 2.55~$\mu$m.

For each set of observations, we used an ABBA nod pattern with the default configuration of $7\farcs5$ distance between the A and B positions. Each position was $3\farcs75$ from the respective end of the slit. Unless a target was accompanied by a nearby companion, we aligned the slit to the parallactic angle to reduce systematics. For close binaries, we instead rotated the slit so that both stars could be observed simultaneously or so that the light from the nearby star would not contaminate the spectrum of the target star. We set integration times for each target based on the observing conditions and the stellar magnitude. We repeated the ABBA nod pattern as many times as needed so that the reduced spectra would have S/N $> 100$ per resolution element. In order to minimize systematic effects due to hot pixels and alpha-particle decays from the ThF$_4$ anti-reflective coatings, we repeated the ABBA pattern at least three times regardless of the brightness of the star.\footnote{This procedure is recommended in the SpeX manual, which is available online at \url{http://irtfweb.ifa.hawaii.edu/~spex/spex_manual/SpeX_manual_06Oct17.pdf}.} We also limited individual exposure times to $< 200$~s. The total exposure times varied from a few minutes to an hour depending on target brightness and observing conditions.

Throughout the night, we ran the standardized IRTF calibration sequence to acquire flats and wavelength calibration spectra. Both types of calibration data were taken using lamps; the flats were illuminated by an internal quartz lamp while the wavelength calibration spectra feature lines from an internal thorium-argon lamp. We usually acquired two sets of calibration spectra at the start and end of the night as well as at least one set per region of the sky. On nights when we observed the same part of the sky for many hours, we ran the calibration sequence multiple times per region so that each science spectrum could be reduced using calibration frames acquired within a few hours of the science spectrum. 

\subsection{Palomar/TripleSpec}
We acquired our TripleSpec (TSPEC) observations using the fixed $1\arcsec \times 30\arcsec$ slit and therefore obtained simultaneous coverage between 1.0 and 2.4~$\mu$m at a spectral resolution of 2500-2700 \citep{herter_et_al2008}. We mitigated contamination from bad pixels by conducting our observations using a four-position ABCD nod pattern rather than a two-position ABBA pattern more similar to our SpeX pattern. We adopted the same ABCD nod pattern as \citet{muirhead_et_al2014} and \citet{dressing_et_al2017a}. As explained in \citet{dressing_et_al2017a}, we left the slit in the east-west orientation unless we were attempting to capture light from two stars simultaneously or avoid contamination from nearby stars. In order to calibrate our TripleSpec data, we obtained dome darks and dome flats at the start and end of each night. 

\section{Data Analysis \&\\ Stellar Characterization}
\label{sec:analysis}
We reduced the NIR spectra of IRTF/SpeX targets using the publicly available {\tt Spextool} pipeline \citep{cushing_et_al2004}. For Palomar/TSPEC targets, we used a specialized version of {\tt Spextool} adapted for using with TripleSpec data (available upon request from M. Cushing). We corrected all of our spectra for telluric contamination using the {\tt xtellcor} package \citep{vacca_et_al2003}, which is included in both versions of the {\tt Spextool} pipeline. As in \citet{dressing_et_al2017a}, we used the Paschen $\delta$ line at 1.005~$\mu$m to create the convolution kernel needed to correct the Vega model spectrum for the instrumental profile and rotational broadening.

\subsection{Initial Classification}
\begin{figure*}[tbp]
\centering
\includegraphics[width=1\textwidth]{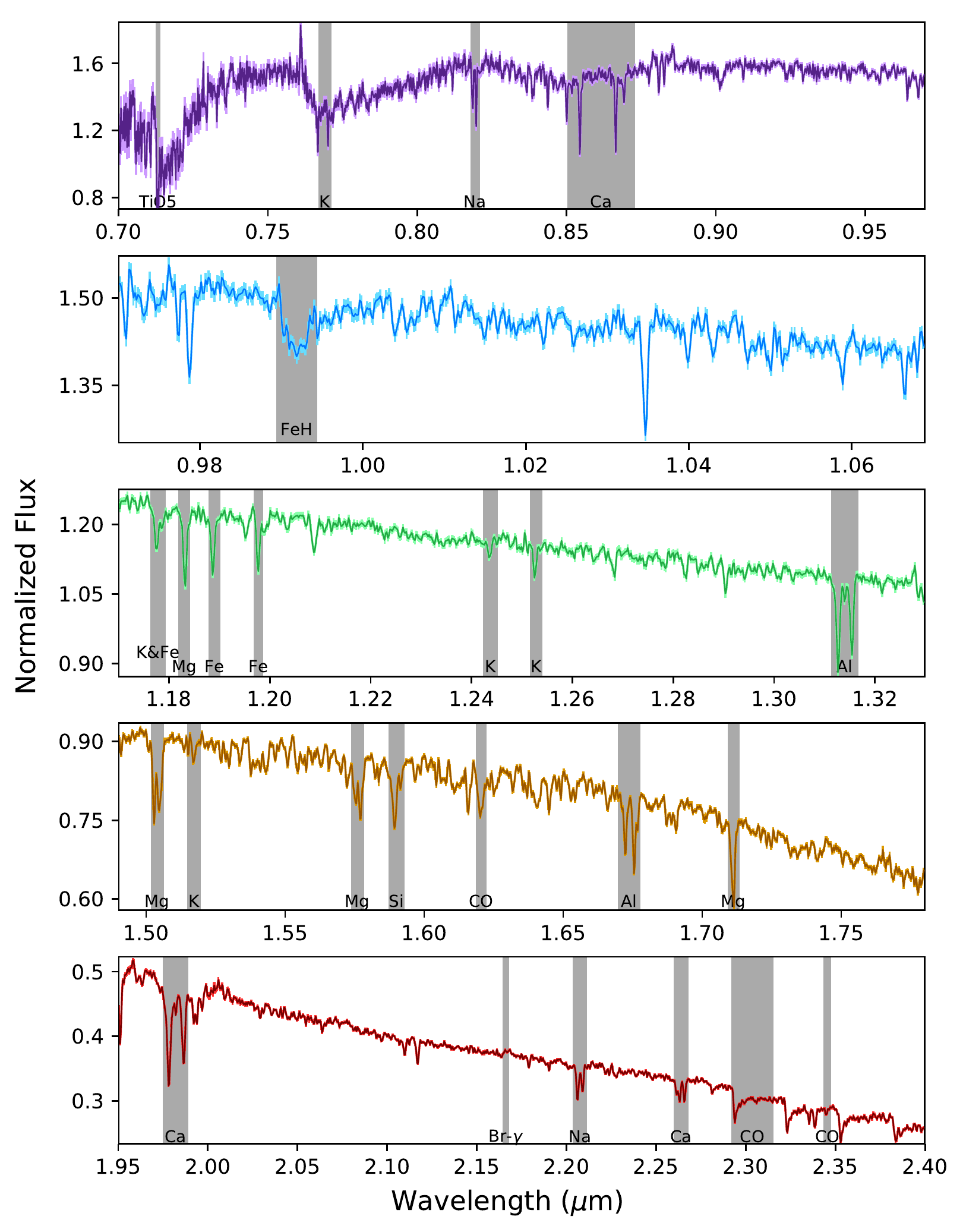}
\caption{Reduced IRTF/SpeX spectrum (dark lines) of EPIC~206298289, which we classified as a cool dwarf with a spectral type of M1. The errors are depicted by the light shading around the spectrum. \label{fig:spex}} 
\end{figure*}

\begin{figure*}[tbp]
\centering
\includegraphics[width=1\textwidth]{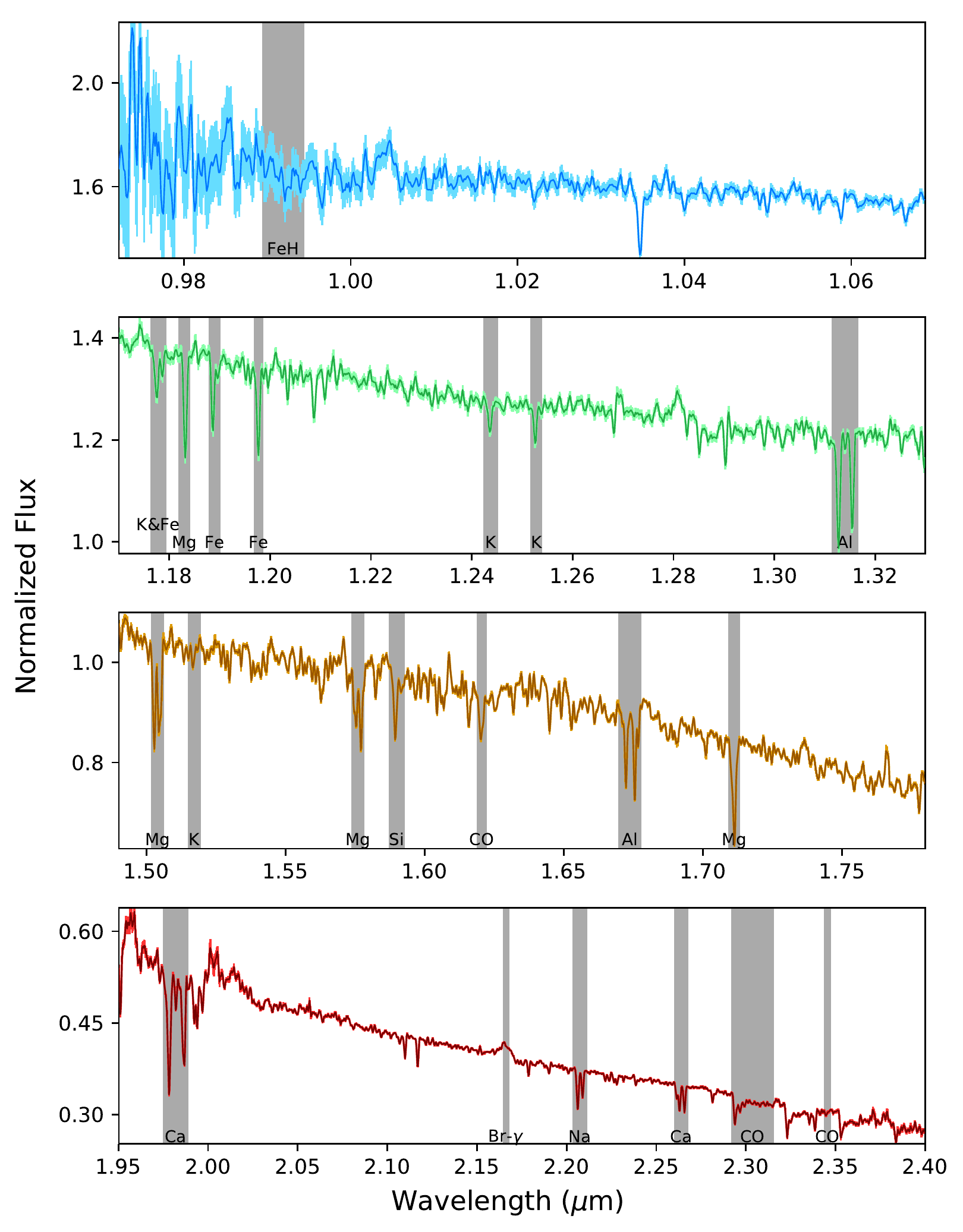}
\caption{Reduced Palomar/TSPEC spectrum (dark lines) of EPIC~248433930, which we classified as a cool dwarf with a spectral type of M1. The errors are depicted by the light shading around the spectrum.  \label{fig:tspec}} 
\end{figure*}

After reducing the spectra, we determined the spectral types and luminosity classes of our target stars by comparing the reduced spectra to spectra of stars with known spectral types obtained from the IRTF Spectral Library \citep{rayner_et_al2009}. We display representative SpeX and TSPEC spectra in Figures~\ref{fig:spex} and \ref{fig:tspec}, respectively. We performed the comparison using the same interactive Python-based plotting interface described in \citet{dressing_et_al2017a}. Correcting for differences in stellar radial velocities and treating the $J$, $H$, and $K$ bandpasses individually, we computed the $\chi^2$ of a fit of each library spectrum to our data and recorded the best match for each bandpass. We then visually compared our spectra to the library spectra producing the lowest $\chi^2$ and selected the best match. We verified these final by-eye assignments by using parallaxes from the second Gaia data release \citep[hereafter Gaia DR2;][]{gaia_et_al2018} to place our full sample on the Hertzsprung-Russell diagram (see Section~\ref{ssec:gaia}).

As a further cross-check, we also measured the strength of three gravity-sensitive indices: K~, Na~I, and Ca~II. All three indices were used by \citet{mann_et_al2012} to investigate the luminosity classes of \kepler targets that were originally classified as M dwarfs. For our equivalent calculations, we adopted the band and continuum wavelength ranges shown in the final three rows of Table~2 of \citet{mann_et_al2012}.

The K~I regions were defined in \citet{mann_et_al2012}, but the Na~I and Ca~II regions were chosen by \citet{schiavon_et_al1997} and \citet{cenarro_et_al2001}, respectively. \ As shown by \citet{torres-dodgen+weaver1993} and \citet{schiavon_et_al1997}, the Na~I doublet (8172 - 8197\AA) and K~I (7669 - 7705\AA) lines are significantly deeper in the spectra of dwarf stars than in the spectra of giant stars. In contrast, the Ca~II triplet (8484-8662\AA) is more pronounced in giant spectra than in dwarf spectra \citep[e.g.,][]{jones_et_al1984, carter_et_al1986, cenarro_et_al2001}. All three of these indices are too blue to be measured in TSPEC data (wavelength range = 1.0 - 2.4~$\mu$m), but the agreement between the indices computed for our SpeX targets and our initial luminosity classifications suggests that our TSPEC targets are also correctly classified. Moreover, both our TSPEC targets and our SpeX targets have positions on the H-R diagram that are consistent with their assigned luminosity classes (see Section~\ref{ssec:gaia}).

Although all of our targets were initially selected because they were believed to be likely cool dwarfs, we found that the sample was contaminated by giant stars and hotter dwarf stars. Overall, 86 (50\%) of our targets were classified as cool dwarfs,  74 (43\%) as hotter dwarfs (i.e., spectral types earlier than K5, effective temperatures above 4800K, or radii larger than $0.8\rsun$), and 12 (7\%) as giant stars. We list the classifications in Table~\ref{tab:classifications}. We exclude the contaminating giants and hotter dwarfs from the detailed analyses in the rest of the paper, but the reduced spectra for all targets are posted on the ExoFOP-K2 website. For reference, Table~\ref{tab:giants} includes the relevant spectral indices for the seven SpeX targets classified as giant stars. The remaining five giants were observed with TSPEC and therefore lack coverage blueward of 1~$\mu$m. When available, Table~\ref{tab:giants} also includes proper motions and parallaxes from Gaia~DR2.

Compared to the initial stellar sample classified in \citet{dressing_et_al2017a}, this sample was slightly less contaminated by giant stars and slightly more contaminated by hotter stars. Of the 146~targets analyzed in \citet{dressing_et_al2017a}, 74 (51\%) were classified as cool dwarfs, 49 (34\%) as hotter dwarfs and 23 (16\%) as giant stars. We attribute the reduced giant contamination in this paper to our stricter use of reduced proper motion cuts when selecting targets. The increase in the fraction of hotter dwarfs is likely due to our bias in prioritizing bright targets on nights with relatively poor weather conditions. 

\begin{figure}[tbp]
\centering
\includegraphics[width=0.48\textwidth]{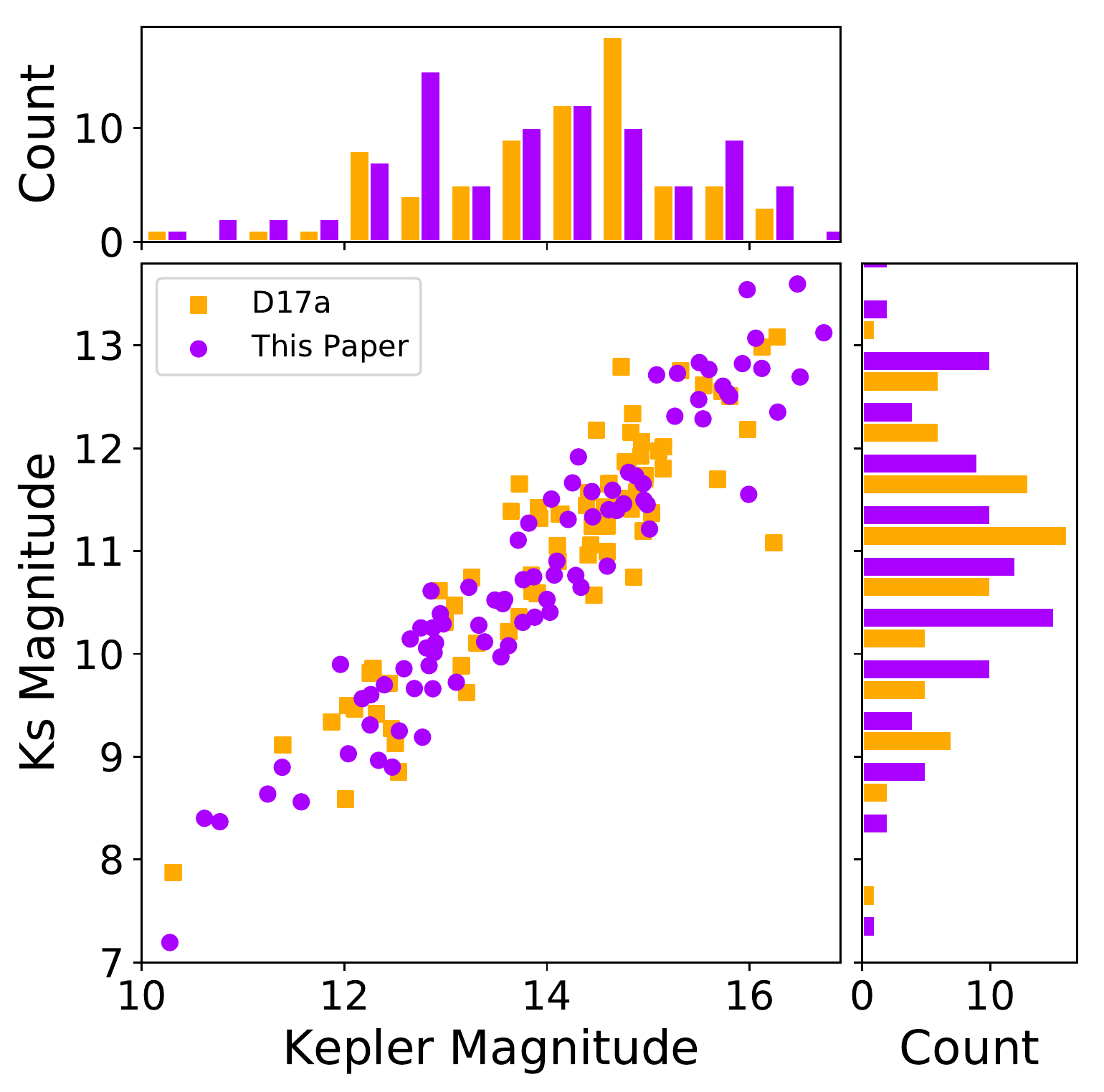}
\includegraphics[width=0.48\textwidth]{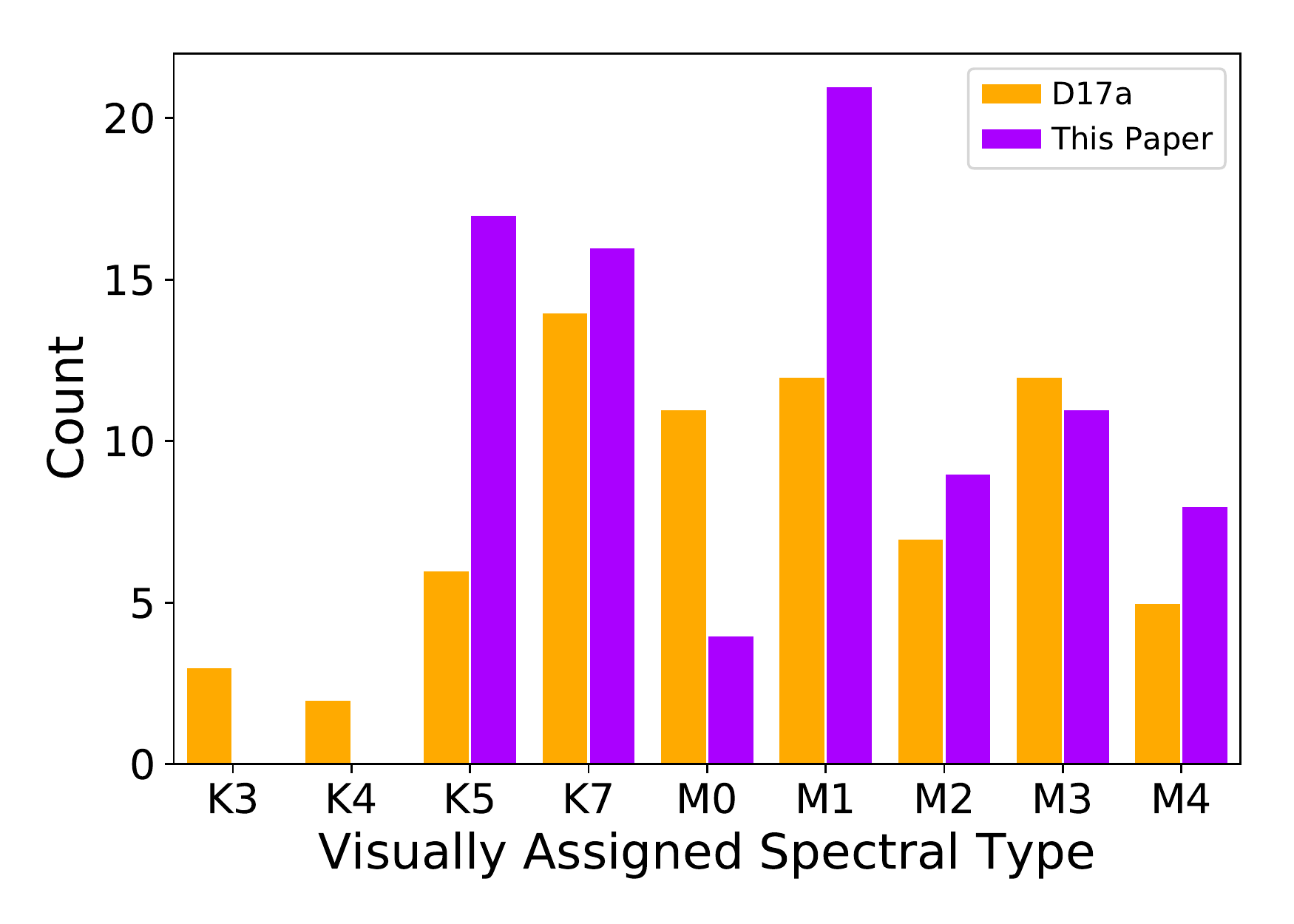}
\caption{Distribution of magnitudes (top) and spectral types (bottom) for stars classified as cool dwarfs in this paper (purple) and in \citet[][orange]{dressing_et_al2017a}. \label{fig:mag_sptype_dist}} 
\end{figure}

We display the magnitude and spectral type distribution for the 86~stars classified as cool dwarfs in Figure~\ref{fig:mag_sptype_dist}. Compared to the sample of 146~stars studied in \citet{dressing_et_al2017a}, this sample covers a slightly broader magnitude range (\mbox{$7.2 < Ks < 13.6$} versus \mbox{$7.9 < Ks < 13.1$}) and has a brighter median magnitude (\mbox{$Ks = 10.7$} versus \mbox{$Ks = 11.3$}). Although the \citet{dressing_et_al2017a} cool dwarf sample included K3 and K4 dwarfs, this sample intentionally excludes stars earlier than K5. Relative to our earlier sample, this sample includes more K5 dwarfs, fewer M0 dwarfs, and more M1 dwarfs. However, our spectral type assignments are only accurate to $\pm1$ spectral type and some of the stars classified as K7 or M1 may actually be M0 dwarfs.   

\subsection{Detailed Spectroscopic Classification}
\label{ssec:details}
\begin{figure*}[tbp]
\centering
\includegraphics[width=0.49\textwidth]{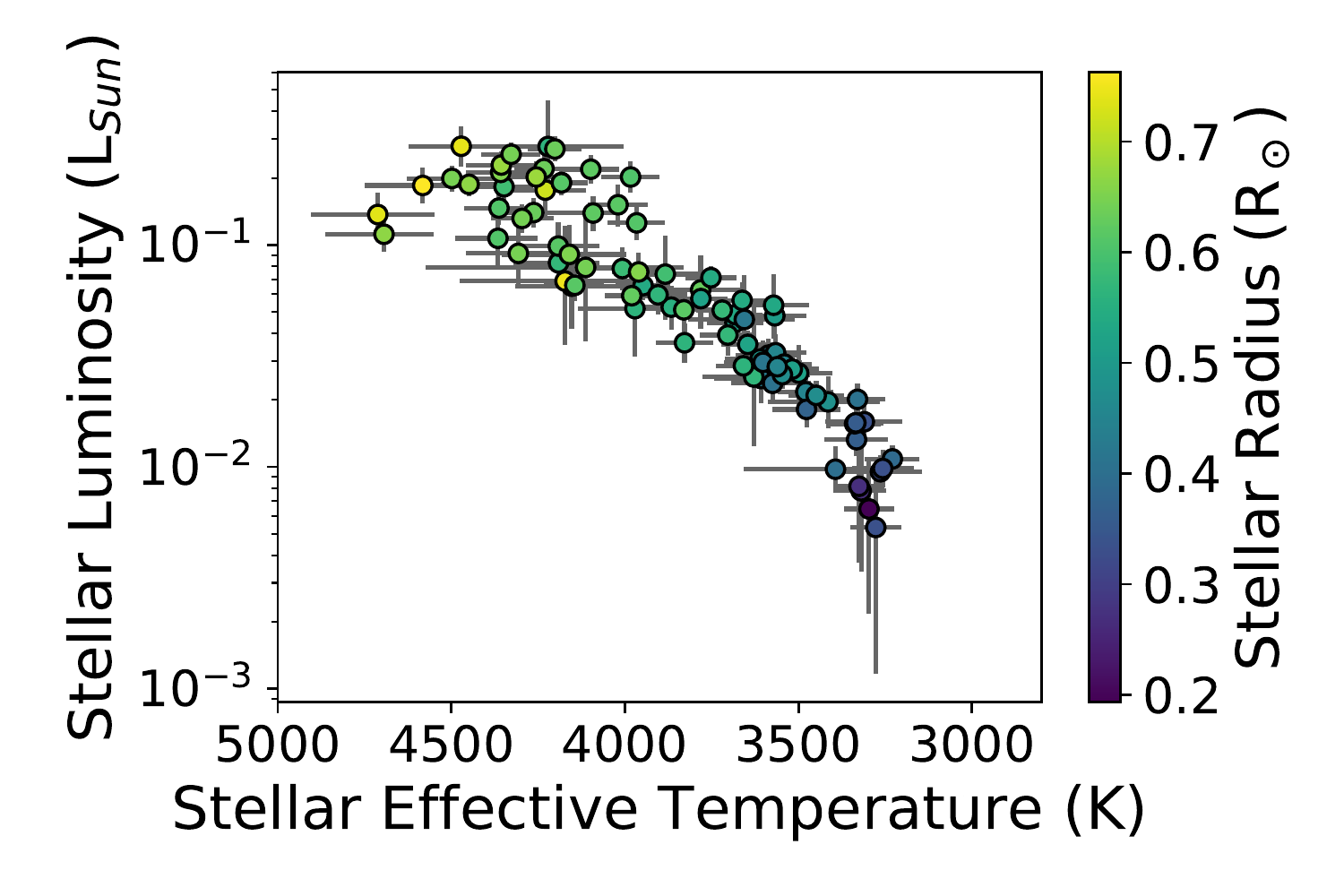}
\includegraphics[width=0.49\textwidth]{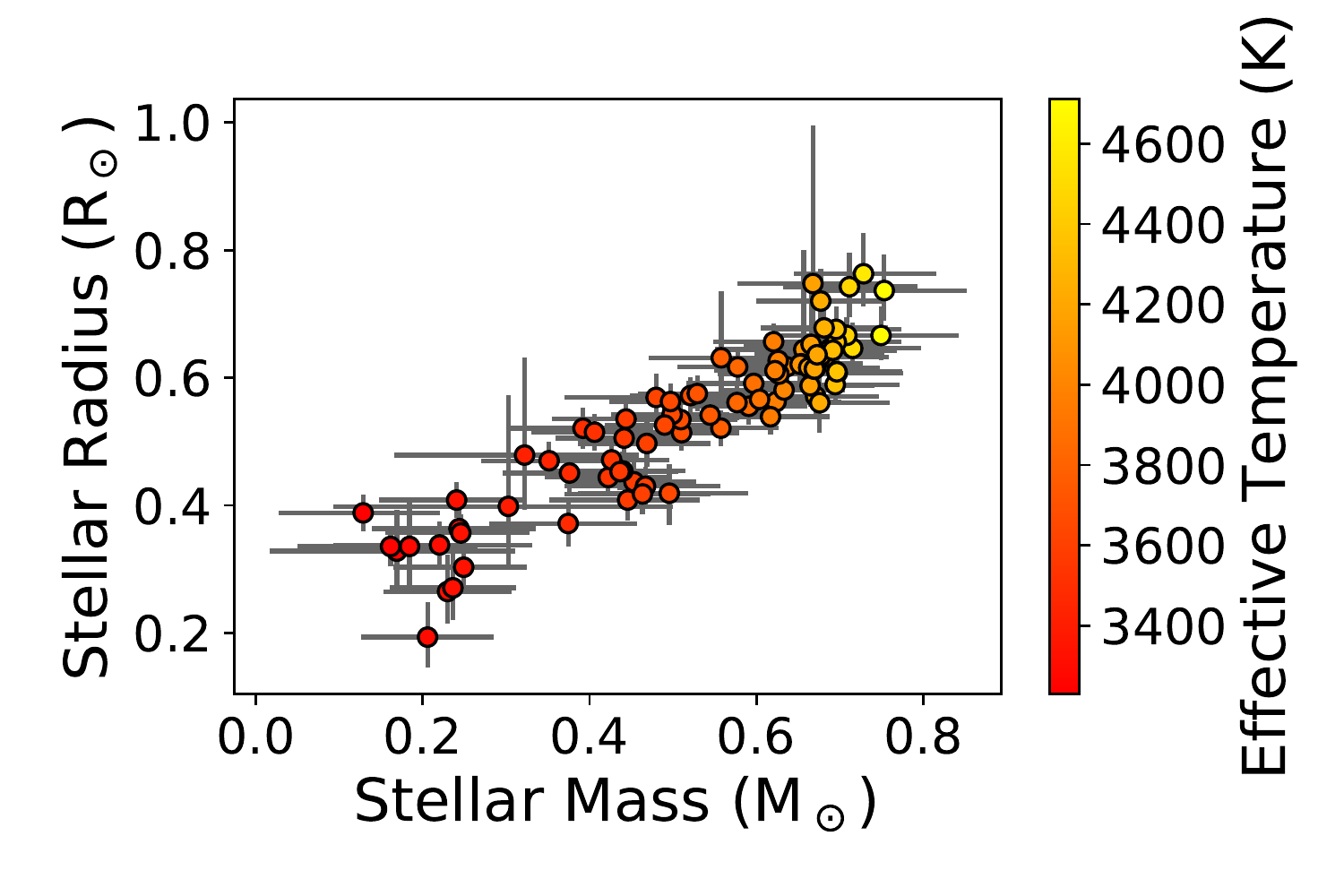}\\
\caption{Parameters for the cool dwarf sample inferred from NIR spectroscopy. \emph{Left: } Stellar luminosity vs. stellar effective temperature with points shaded according to revised stellar radii. \emph{Right: } Radii and masses with points shaded according to revised stellar effective temperatures.  \label{fig:specfits}} 
\end{figure*}

We initially estimated the physical properties of the cool dwarfs using the same procedures as in \citet{dressing_et_al2017a} and display the results in Figure~\ref{fig:specfits} and Table~\ref{tab:spectroscopic}. Specifically, we used the publicly available, IDL packages {\tt tellrv} and {\tt nirew} \citep{newton_et_al2014,newton_et_al2015} to implement the empirical relations established by \citet{newton_et_al2015}. These relations predict the stellar effective temperatures, radii, and luminosities of cool dwarfs from the equivalent widths of Al and Mg features measured in medium-resolution $H$-band spectra. The relations are valid for cool dwarfs with 3200~K$ < T_{\rm eff} < $4800~K, $0.18\rsun < R_\star < 0.8\rsun$, and $-2.5 < \log L_\star/\lsun < -0.5$. As in \citet{dressing_et_al2017a}, we estimated the masses of the cool dwarfs by using the stellar effective temperature-mass relation from \citet{mann_et_al2013c} to convert our temperature estimates into masses. We then calculated surface gravities from the estimated masses and radii.

Our Palomar/TSPEC spectra were obtained at higher resolution than the IRTF/SpeX spectra used to calibrate the \citet{newton_et_al2015} relations, so we downgraded the resolution of the Palomar/TSPEC spectra to that of the IRTF/SpeX data before measuring the equivalent widths. Ignoring the change in resolution could introduce a systematic $0.1{\emph\AA}$ difference in the measured equivalent widths \citep{newton_et_al2015}; analyzing either downsampled Palomar/TSPEC data or unaltered IRTF/SpeX data yields consistent results \citep{dressing_et_al2017a}. 

\begin{figure}[tbp]
\centering
\includegraphics[width=0.48\textwidth]{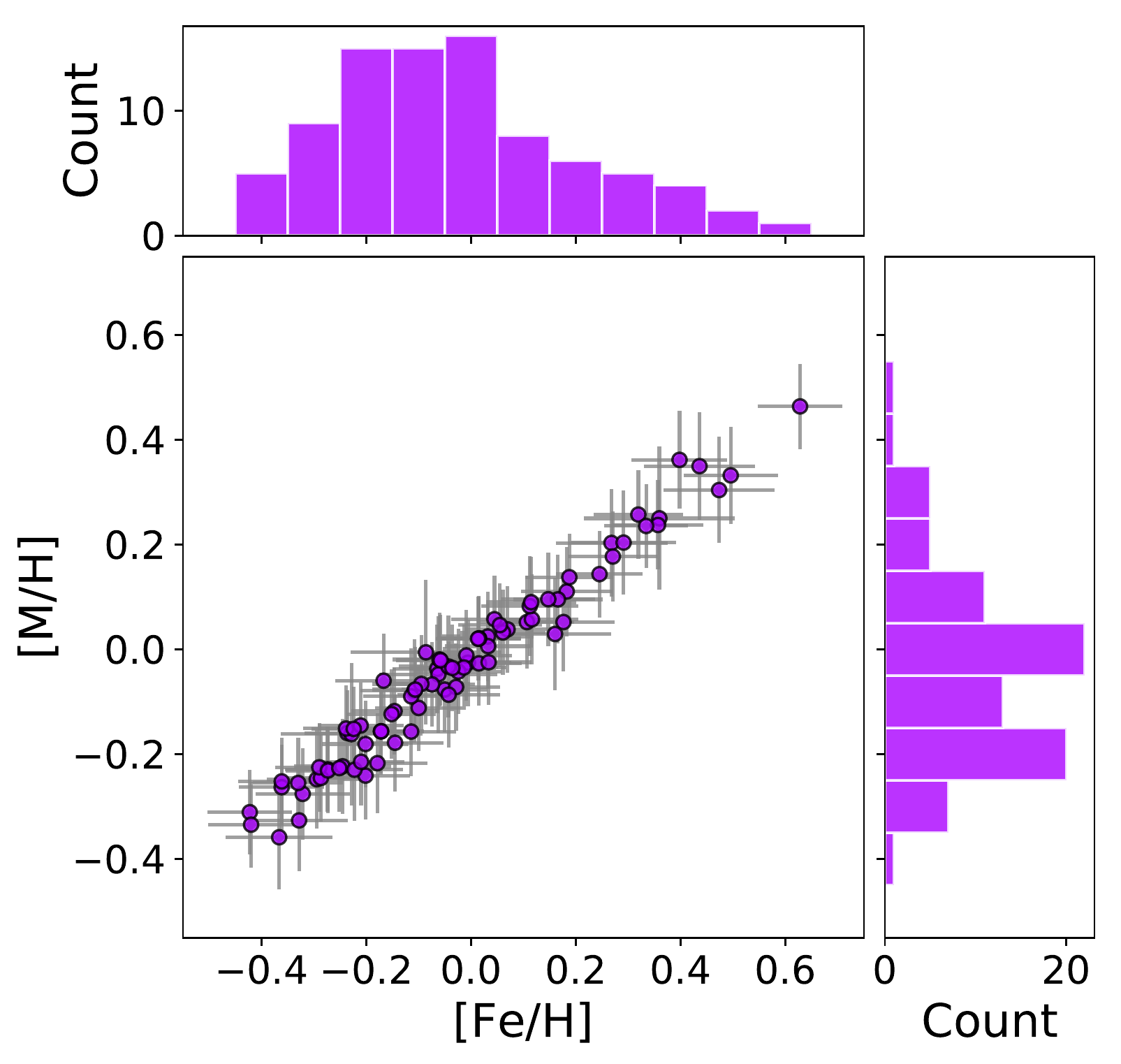}
\caption{Estimated [Fe/H] and [M/H] for the cool dwarfs with spectral types of K7 or later.  \label{fig:metallicity}} 
\end{figure}

In order to determine stellar metallicities, we implemented the relations defined in \citet{mann_et_al2013a} for cool dwarfs with spectral types between K7 and M5. We first calculated metallicities using the $H$-band and $K$-band spectra separately and compared the results. Although the [Fe/H] and [M/H] estimates calculated from the $Ks$-band spectra were well-correlated, we found that the $H$-band [Fe/H] estimates displayed significant scatter relative to the $Ks$-band estimates. The \mbox{$H$-band} [M/H] estimates were consistent with the $Ks$-band [M/H] estimates, suggesting that the $H$-band [Fe/H] estimates are less reliable than the $Ks$-band estimates and more affected by telluric contamination. As shown in Table~\ref{tab:observing}, many of our observations were obtained in partially cloudy conditions. To reduce weather-dependent systematics, we adopt the [Fe/H] and [M/H] estimates calculated from the $Ks$-band spectra instead of averaging the results from both bands. We display the resulting metallicities in Figure~\ref{fig:metallicity}. 

\begin{figure*}[tbp]
\centering
\includegraphics[width=1\textwidth]{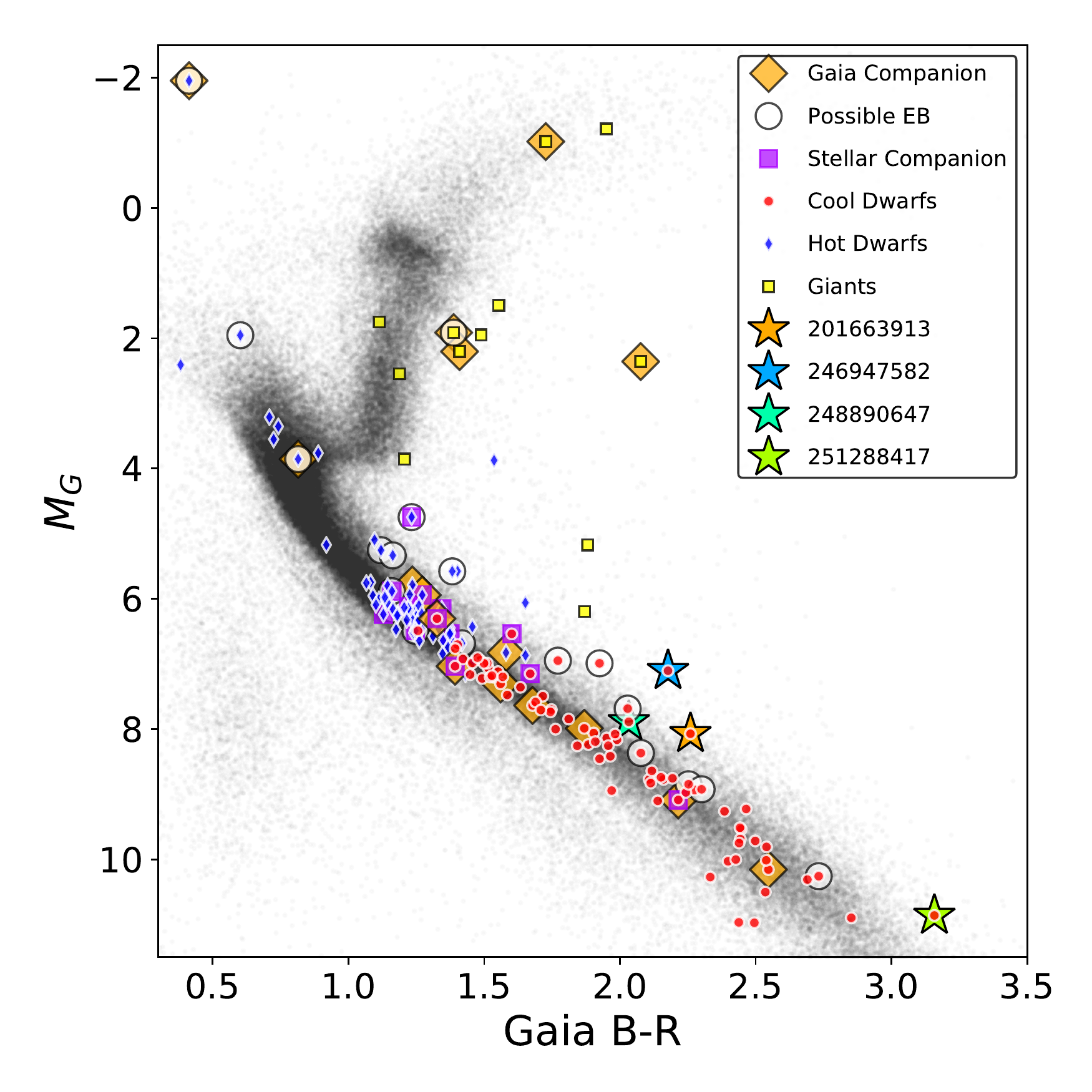}
\caption{Color-magnitude diagram in $M_{G}$ vs. Gaia $B-R$ for all \emph{K2} targets with Gaia parallaxes (translucent black dots). The larger symbols mark K2OIs with Gaia parallaxes that we classify in this paper as giants (yellow squares), hotter dwarfs (blue diamonds), or cool dwarfs (red circles). Stars with nearby stellar companions are marked by purple squares and those suspected to be eclipsing binaries are surrounded by white circles.}
\label{fig:cmd}
\end{figure*}

\subsection{Incorporating Gaia Distance Estimates}
\label{ssec:gaia}
Our targets are moderately bright stars and might therefore be expected to have parallaxes reported in the second Gaia data release \citep{gaia_et_al2018}. We checked for Gaia DR2 matches by using the Advanced ADQL tab of the Gaia Archive\footnote{\url{https://gea.esac.esa.int/archive/}} to create a list of all stars within $20"$ of the positions reported in the EPIC and compute their positions at the same reference epoch. We then selected all stars within $5"$ of our target stars and verified that the matches were genuine by comparing the proper motions and visual magnitudes ($G$ and $Kp$) of the target stars. 

Of our 172 targets, 171 (99\%) matched with stars in Gaia~DR2 and 168 of those stars have reported parallaxes. We also identified 24~possible companion stars within $5"$ of 17 of our target stars. The star without a match in DR2 is the cool dwarf EPIC~204888276 and the stars with Gaia crossmatches but no reported parallaxes are EPIC~211783206 (hotter dwarf), EPIC~220187552 (cool dwarf), and EPIC~249483541 (cool dwarf). In total, 12 giant stars (100\%), 73 hotter dwarfs (99\%), and 83 cool dwarfs (97\%) have parallaxes reported in Gaia~DR2. 

\subsubsection{Possible Stellar Binaries}
One cool dwarf (EPIC~210693462) appears to be a close binary because it has two matches within $1"$ of the stellar position reported by \citet{huber_et_al2016} in the EPIC: Gaia~DR2~44838019758175488 ($0\farcs4$ away) and Gaia~DR2~44838019756570112 ($0\farcs3$ away). The two stars have similar parallaxes and proper motions. Based on the multiple Gaia matches and the presence of two stars in follow-up AO images obtained by D.~Ciardi with Keck/NIRC2\footnote{\url{https://exofop.ipac.caltech.edu/k2/edit_target.php?id=210693462}}, we classify EPIC~210693462 as a binary and exclude it from the rest of the stellar characterization process in this paper. We performed a detailed characterization of the system in a separate paper \citep{feinstein_et_al2019}.

Seventeen additional cool dwarfs have more distant candidate companions at angular separations of $1\farcs5-5\farcs0$. In order to assess whether any of these stars are physically associated with our target stars, we compared the relative proper motions and angular separations of each possible pair to the \citet{lepine+bongiorno2007} criterion for likely co-movers. As shown in Table~\ref{tab:companions},  the candidate stellar companions to  EPIC~201650711 and EPIC~210693462 have parallaxes and proper motions similar to those of the primary star and are likely to be physically associated. In addition, EPIC~202071401 has one candidate companion that is likely to be physically associated (Gaia DR2 3378104379464943104) and one that is a likely interloper (Gaia DR2 3378104379464943232). The remaining candidate companions either have proper motions that are inconsistent with physical association (13~stars) or unknown proper motions (8~stars).
As part of our check for possible stellar binaries, we consulted the ExoFOP-K2 website. Of the 172~stars in the full sample, 15~stars (9\%; 7~cool dwarfs, 7~hotter dwarfs, \& 1 giant star) were marked as possible eclipsing binaries, 9~stars (5\%; 5~cool dwarfs \& 4 hotter dwarfs) are in close proximity to stars revealed by follow-up imaging, and 5~stars (3\%; 1 cool dwarf \& 4 hotter dwarfs) are candidate EBs that also have candidate stellar companions.  We include flags in Table~\ref{tab:classifications} to identify possible stellar binaries.

The 5~cool dwarfs with candidate stellar companions are EPIC~206061524, EPIC~220187552, EPIC~220555384, EPIC~201119435, and EPIC~202071401. The candidate companion to EPIC~206061524 is roughly $0\farcs5$ away from the target and detected in a Palomar/PHARO image obtained by D.~Ciardi. EPIC~220187552 is a candidate eclipsing binary and is roughly $0\farcs3$ away from a  star that is approximately $0.68$~mag fainter at 832nm. Those estimates were determined from the WIYN speckle observations acquired by M.~Everett but the companion was also detected in AO images obtained by D.~Ciardi with Keck/NIRC2 and Palomar/PHARO. EPIC~220555384 also has an extremely close companion (separation $\approx 0.\farcs2$) revealed by AO~imaging at Lick, Gemini-N, and Palomar (images uploaded to ExoFOP by D.~Ciardi) as well as speckle imaging at WIYN (image uploaded by M.~Everett). The candidate companions to EPIC~201119435 and EPIC~202071401 were detected in Gaia~DR2. 

There are no posted follow-up images for most of the cool dwarfs with multiple companions in Gaia~DR2, but the candidate companions to EPIC~201119435\footnote{\url{https://exofop.ipac.caltech.edu/k2/edit_target.php?id=201119435}} and EPIC~202071401\footnote{\url{https://exofop.ipac.caltech.edu/k2/edit_target.php?id=202071401}} are visible in AO images obtained by D.~Ciardi with Gemini-N/NIRI, Palomar/PHARO, and Keck/NIRC2. The companion to EPIC~202071401 was not detected in the WIYN speckle image obtained by M.~Everett, but the star may have been outside the $2\farcs8 \times 2\farcs8$ field-of-view. 

\subsubsection{Absolute Magnitudes}
\label{sss:absmag}
For all stars except EPIC~210693462 (the close binary), we calculated absolute $Ks$ magnitudes\footnote{We specifically chose to calculate $M_{Ks}$ because the empirically-determined mass-magnitude relations exhibit lower scatter in redder bands \citep{delfosse_et_al2000}.} from the distance estimates determined by \citet{bailer-jones_et_al2018}. For our full target sample, the parallaxes reported in Gaia~DR2 range from 0.07 mas to 35.8 mas \citep{gaia_et_al2018}, which corresponds to a distance range of 28--8546 pc \citep{bailer-jones_et_al2018}. The 86~cool dwarfs have parallaxes of 1.7--35.8 mas and estimated distances of 28--589~pc with a median distance of 127~pc. Our distance estimates are drawn from \citet{bailer-jones_et_al2018} and carry a Bayesian transformation of the parallax probability distribution function into a distance probability distribution function, using Bayesian priors selected for each star.

Next, we used the absolute magnitudes to place our targets on the color-magnitude diagram shown in Figure~\ref{fig:cmd} and confirm our stellar classifications. For the cool dwarfs with parallaxes in Gaia DR2, we then used the photometric relations described in Sections~\ref{sssec:photo_lum}--\ref{sssec:photo_teff} to estimate stellar radii, masses, and effective temperatures. We list the resulting parameters in Table~\ref{tab:photometric}.

In Figure~\ref{fig:cmd}, we indicate which stars have been flagged as possible EBs on the ExoFOP-K2 website and which stars have possible stellar companions revealed by ground-based follow-up images or Gaia astrometry. The three cool dwarfs that are flagged as possible EBs and are clearly above the main sequence in Figure~\ref{fig:cmd} are EPIC~248527514, EPIC~205996447, and EPIC~212009427.  Note that EPIC~220187552 (also flagged as suspected EB) does not appear on Figure~\ref{fig:cmd} because Gaia DR2 does not include a parallax for this star. 

The cool dwarfs that fall above the main sequence and are not flagged as likely EBs are EPIC~201663913 (1.2~magnitudes brighter than stars with similar $B_p - R_p$ colors), EPIC~246947582 (1.8~magnitudes brighter), EPIC~248890647 (0.6~magnitude brighter), and EPIC~251288417 (1.1~magnitude brighter).  There are no follow-up images of EPIC~201663913, EPIC~248890647, or EPIC~251288417 on the ExoFOP-K2 website, but EPIC~246947582 ($G = 7.10$, $B-R = 2.18$) was observed by D.~Ciardi using Keck/NIRC2 with a Br-$\gamma$ filter. Ciardi did not detect any nearby companions down to a limit of $\Delta M = 6$ at $0\farcs2$ and $\Delta M = 7$ at $0\farcs7$. EPIC~201663913 is 0.83~magnitudes brighter.

\subsubsection{Stellar Luminosities}
\label{sssec:photo_lum}
With the exception of EPIC~210693462 (the close binary), we determined photometric luminosity estimates for all cool dwarfs with adequate photometry. Following \citet{mann_et_al2017}, we began by consulting the Carlsberg Meridian Catalogue \citep{muinos+evans2014} to find $r$-band magnitudes for each star. We then inferred $L_\star$ from the 2MASS $J$ magnitudes reported in the EPIC \citep{huber_et_al2016, skrutskie_et_al2006}, $J$-band bolometric corrections determined from $r-J$ colors using the relations established by \citet{mann_et_al2015}, and the estimated stellar distances reported by \citet{bailer-jones_et_al2018}.

\begin{figure*}[tbp]
\centering
\includegraphics[width=0.32\textwidth]{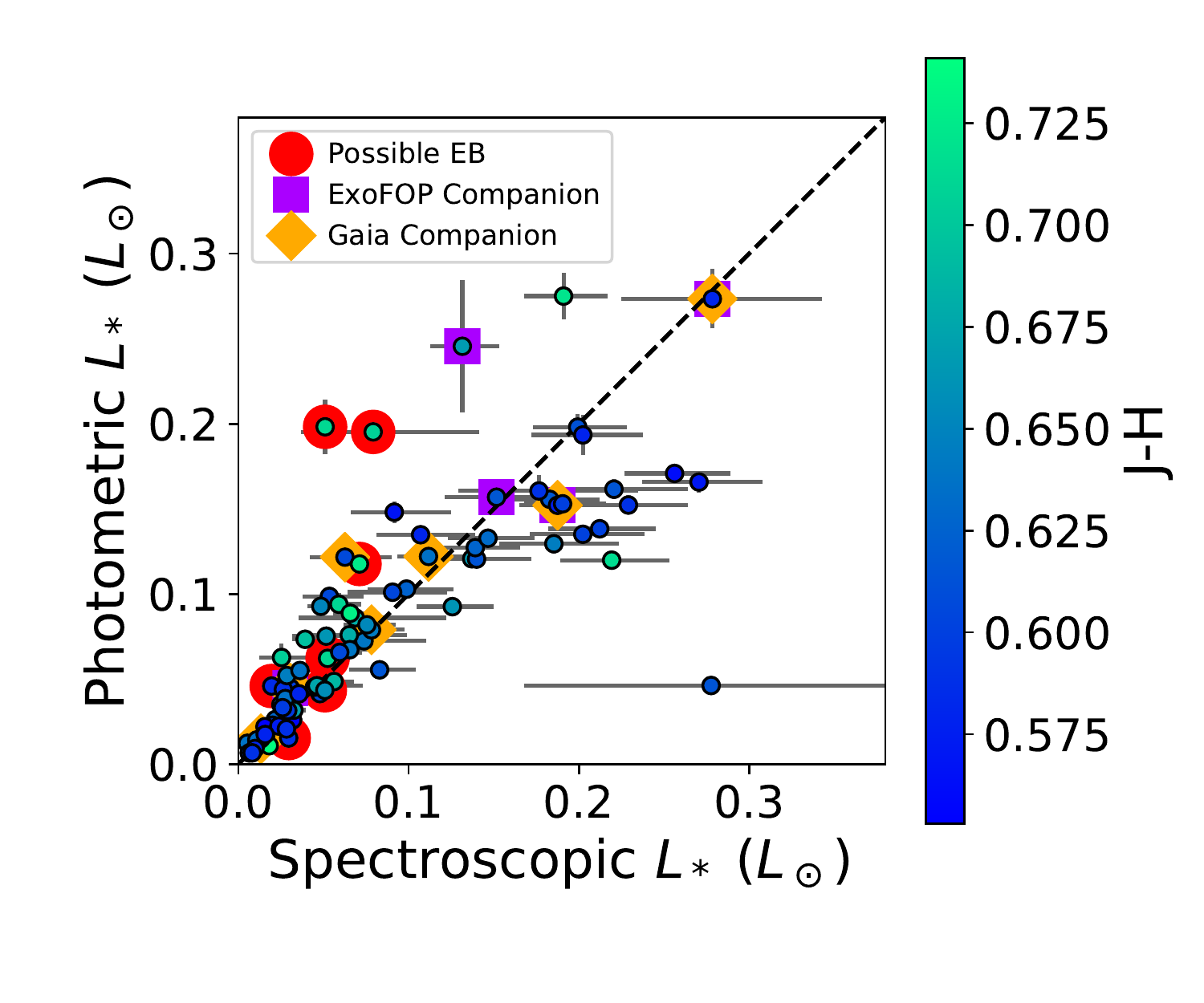}
\includegraphics[width=0.32\textwidth]{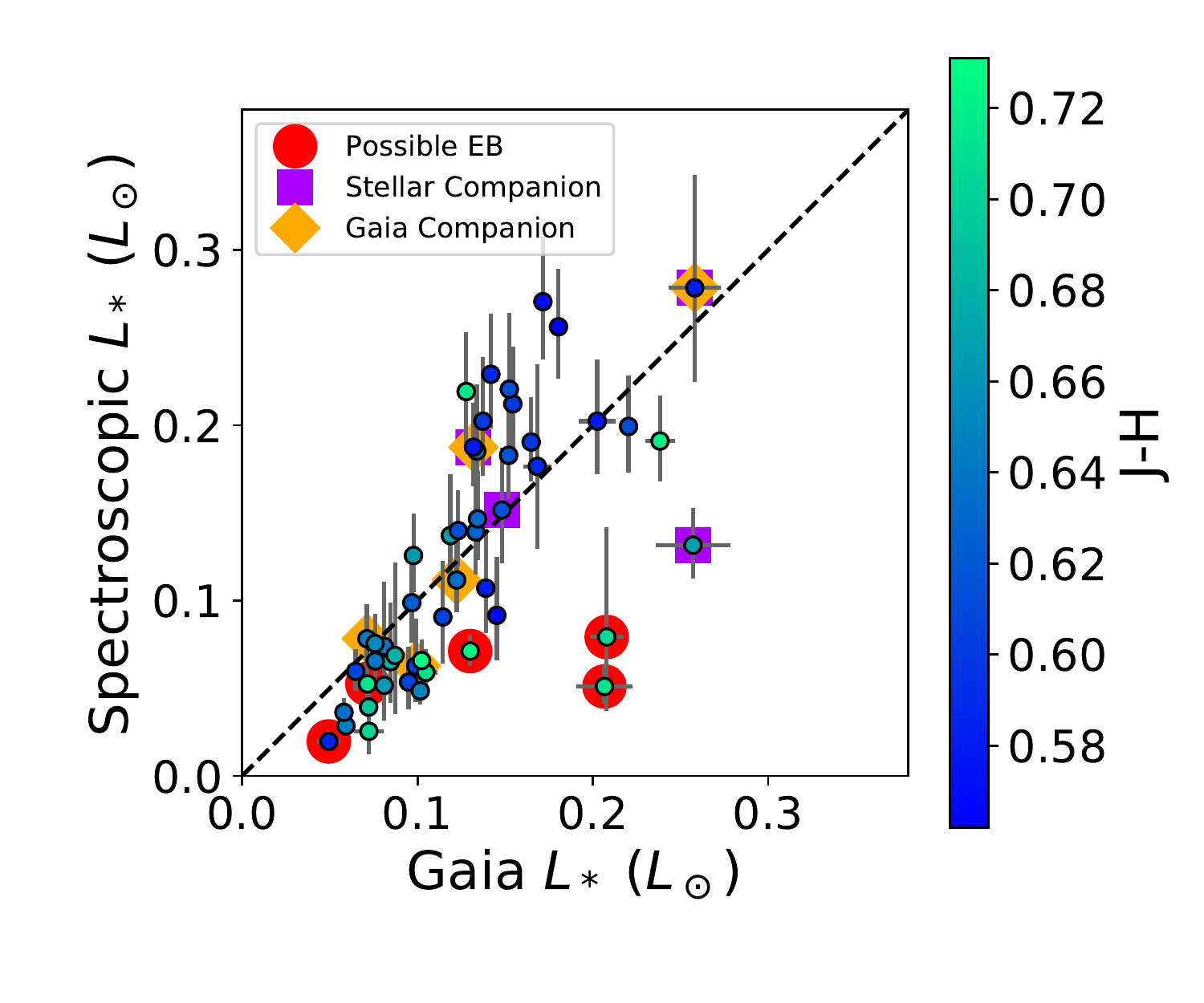}
\includegraphics[width=0.32\textwidth]{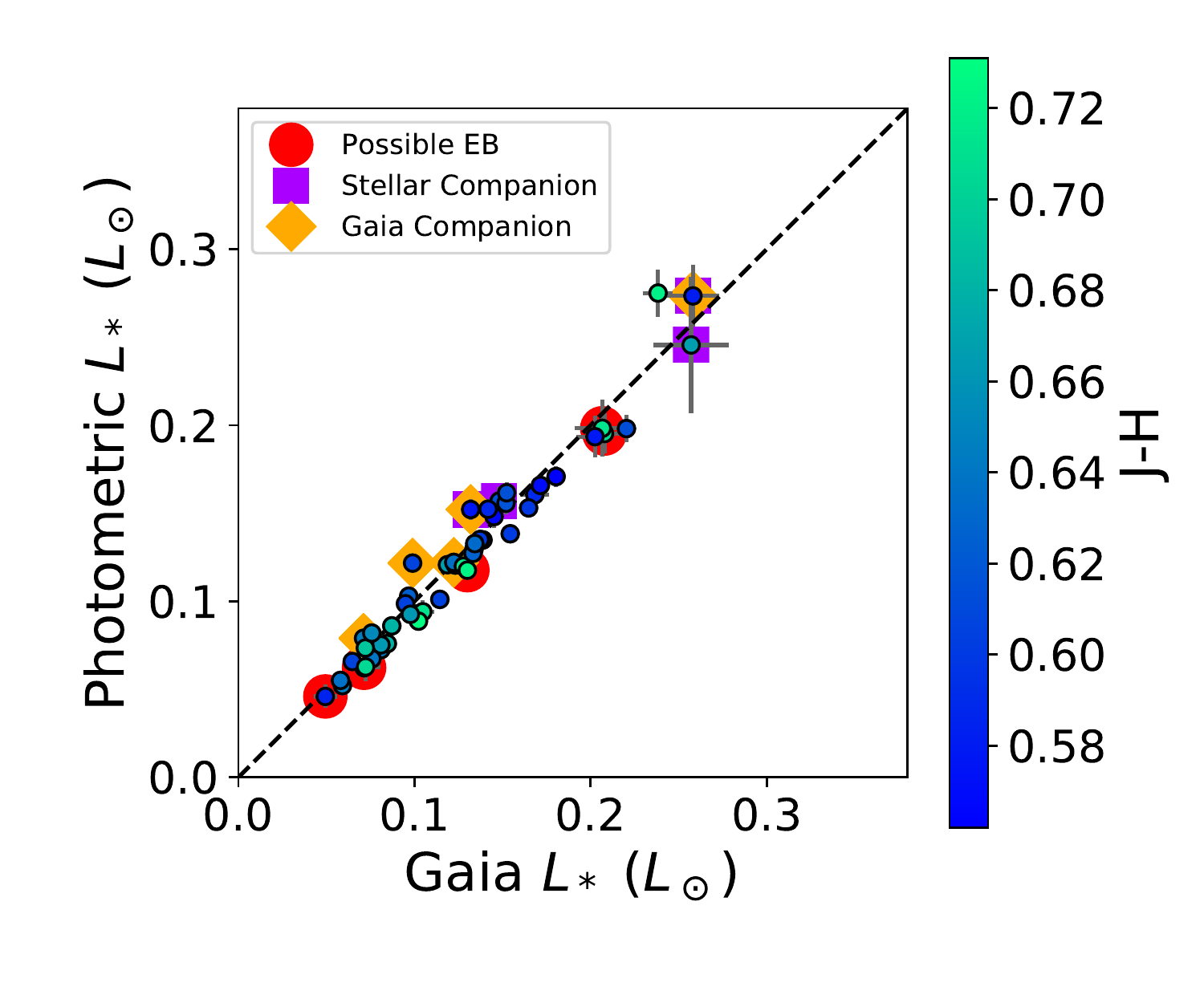}\\
\includegraphics[width=0.32\textwidth]{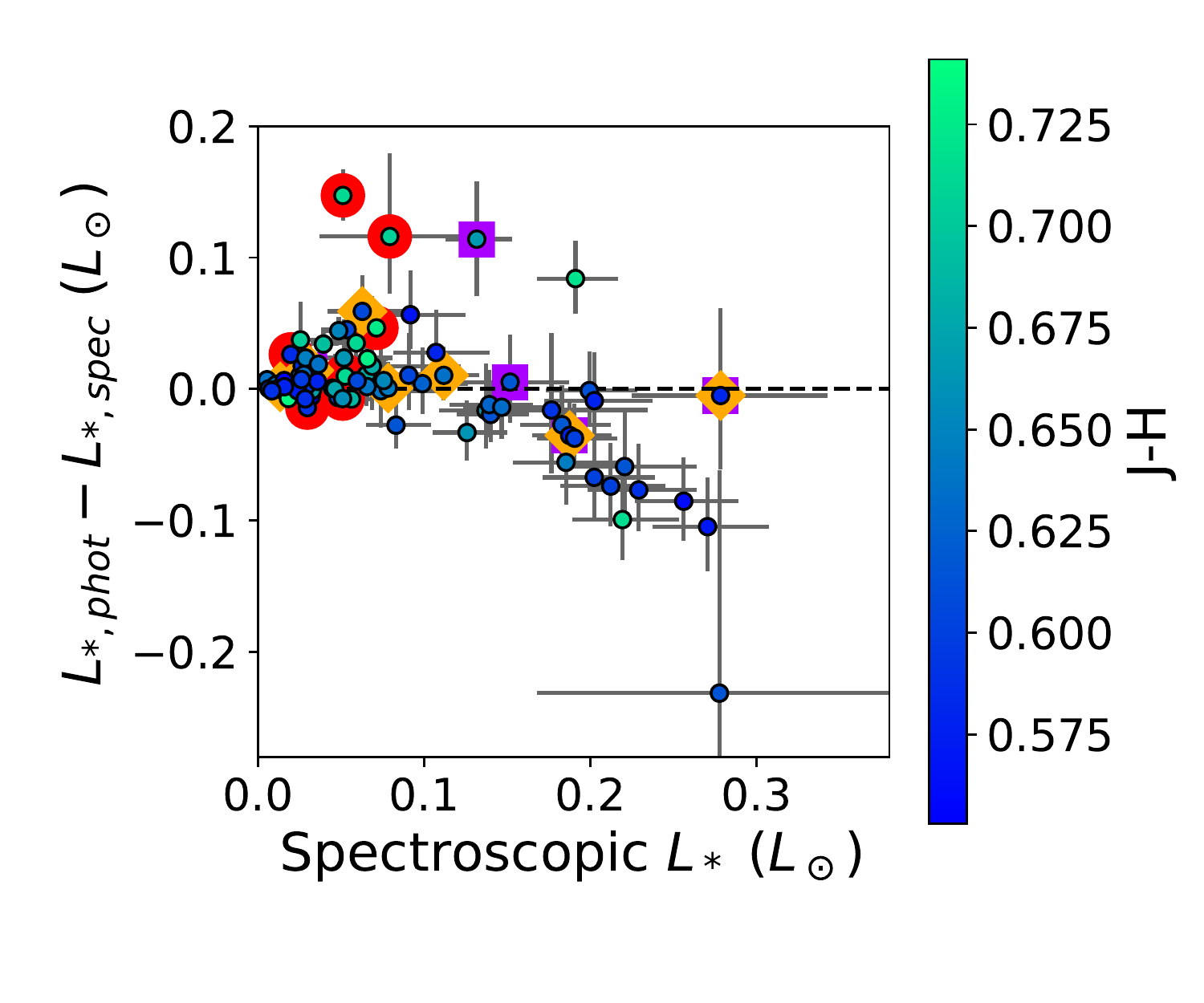}
\includegraphics[width=0.32\textwidth]{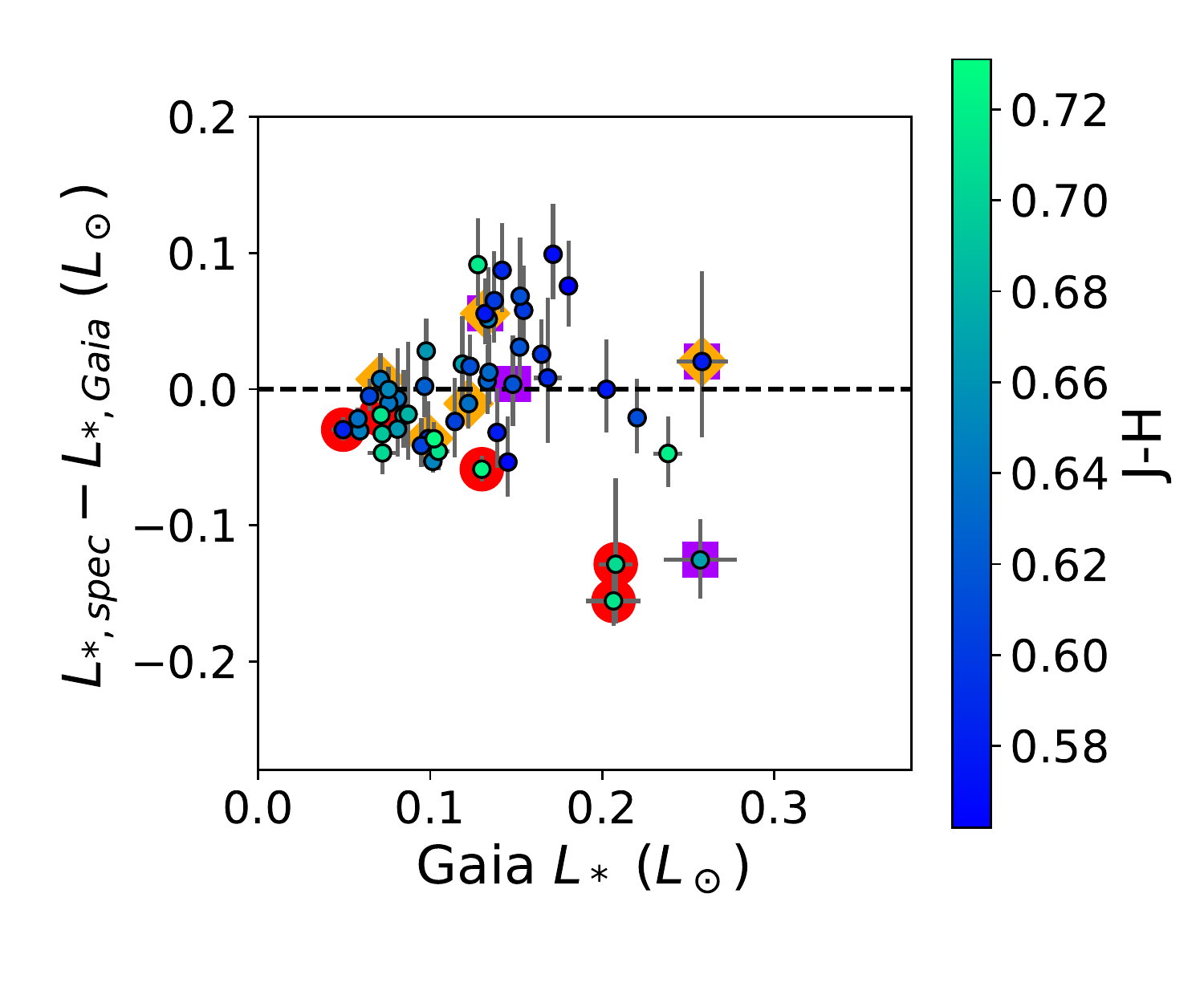}
\includegraphics[width=0.32\textwidth]{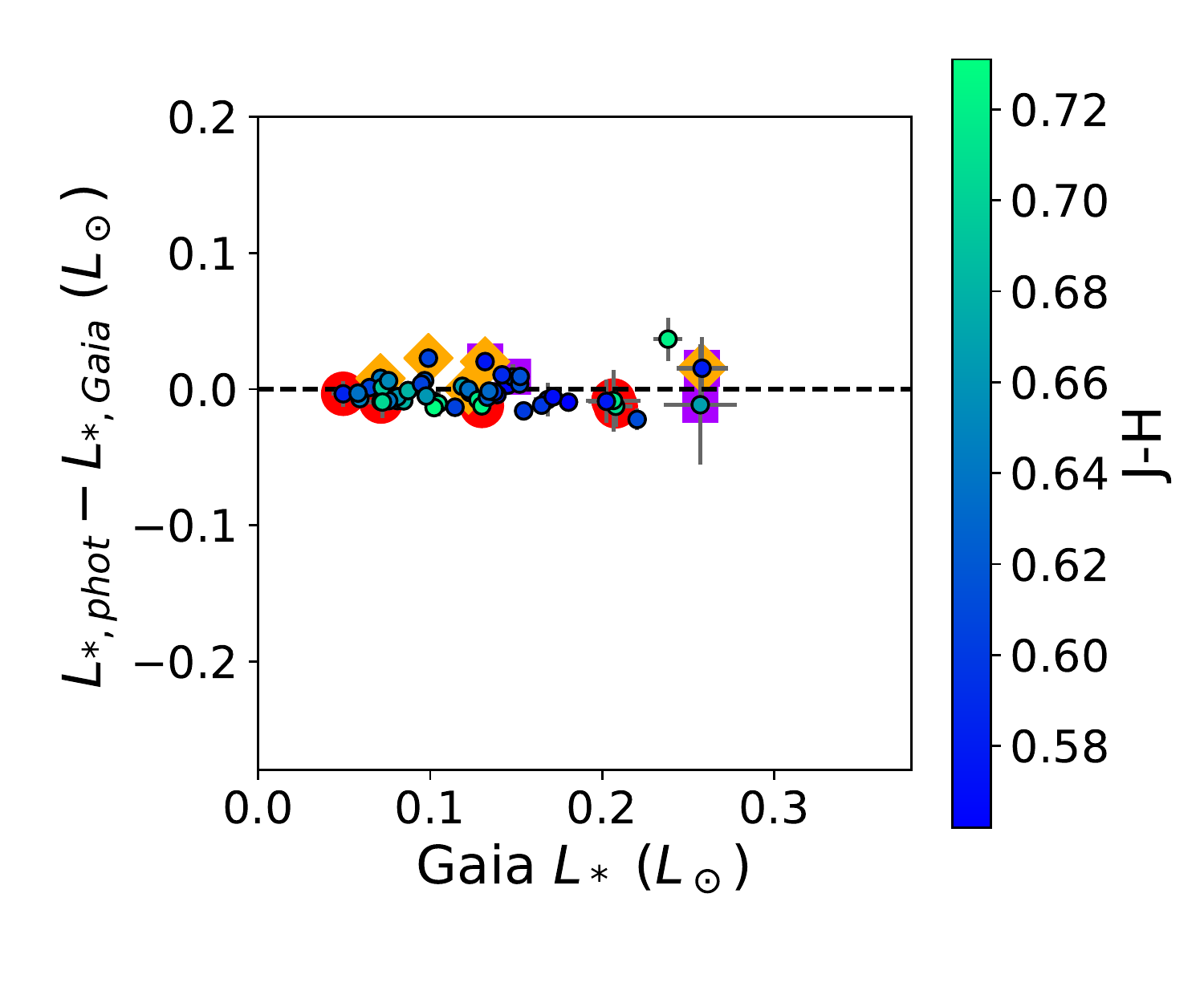}\\
\includegraphics[width=0.49\textwidth]{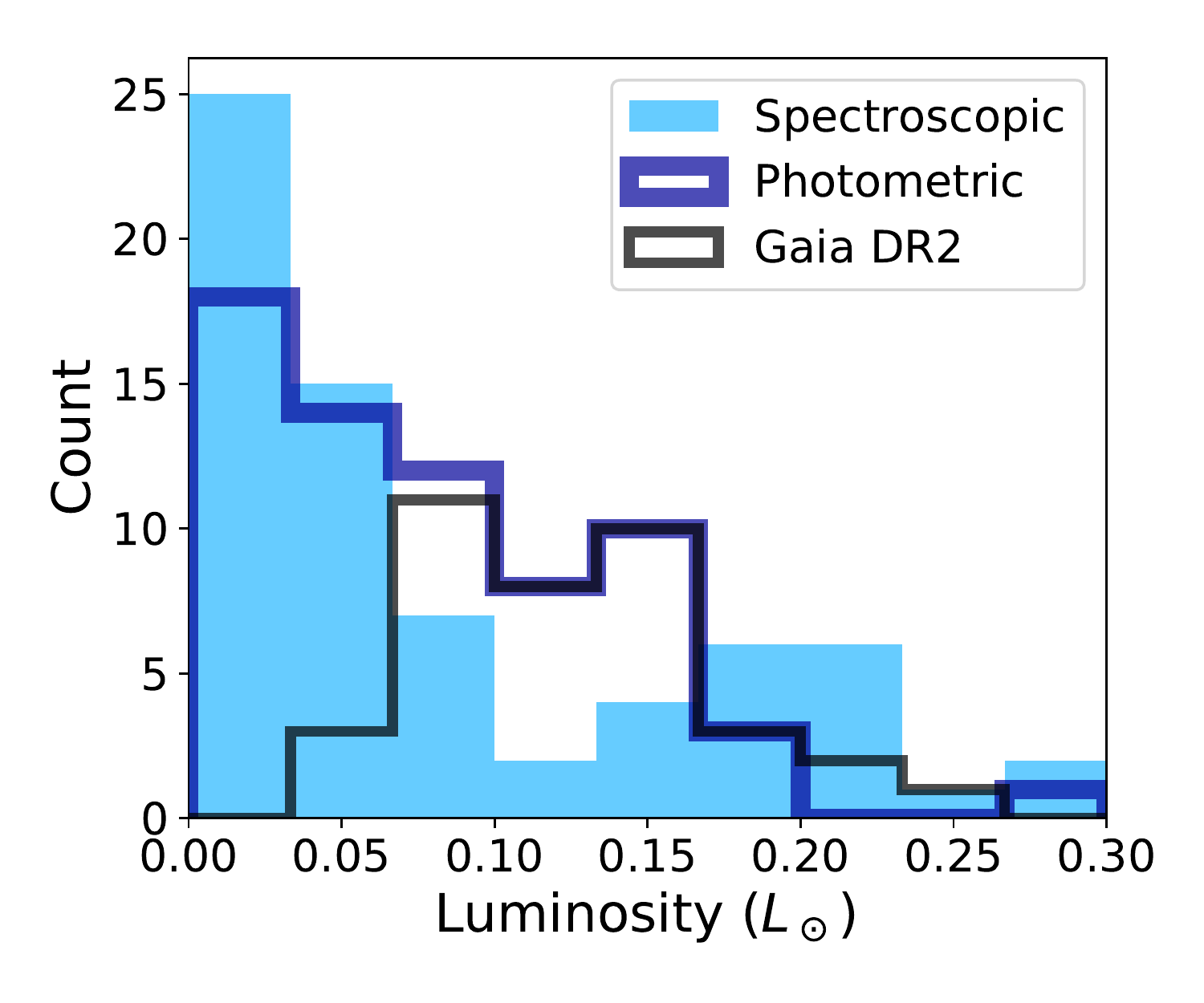}
\includegraphics[width=0.49\textwidth]{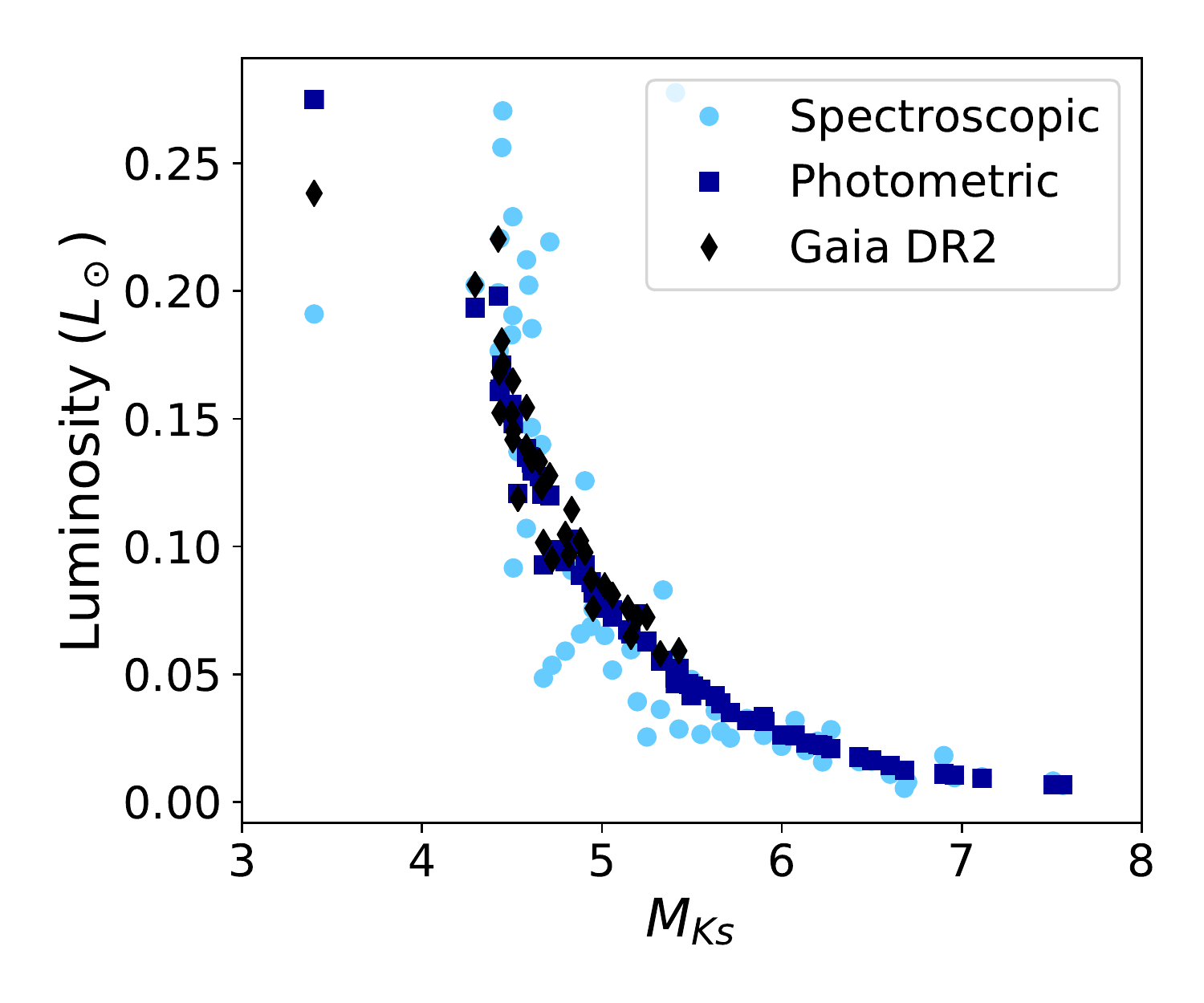}
\caption{Comparison of stellar luminosities and radii estimated from spectroscopy and photometry. We denote the values estimated in this paper as ``spectroscopic'' if they are primarily determined from our NIR spectra or ``photometric'' if they are determined from the combination of broadband photometry and Gaia parallaxes. We also compare our estimates to those determined by the Gaia team; those values are also photometric, but they are marked here as ``Gaia'' to avoid confusion. The points are color-coded by $J-H$ color. In panels displaying individual stars, we indicate possible EBs by wrapping the points with red circles. We also use orange diamonds to mark stars with companions in Gaia~DR2 and purple squares to flag stars with nearby stellar companions detected in AO or speckle images. The black dashed lines mark a 1:1 correlation. \emph{Top Left:} Photometric luminosity estimates versus spectroscopic luminosity estimates.   \emph{Top Center:} Spectroscopic luminosity estimates versus  luminosity estimates from Gaia~DR2.   \emph{Top Right:}  Photometric luminosity estimates versus Gaia luminosity estimates. \emph{Middle Left:} Difference between photometric and spectroscopic luminosity estimates versus spectroscopic estimates. \emph{Middle Center:} Difference between spectroscopic luminosity estimates and those from Gaia~DR2 versus Gaia~DR2 estimates. \emph{Middle Right: } Difference between photometric luminosity estimates and those from Gaia~DR2 versus Gaia~DR2 estimates.\emph{Bottom Left: } distribution of luminosities for stars in the cool dwarf sample that have not been classified as likely EBs and do not have stellar companions (i.e., ``single cool dwarfs''). \emph{Bottom Right: } Luminosity versus absolute $Ks$ magnitude for single cool dwarfs. The extremely bright star is EPIC~246947582; see Section~\ref{sss:absmag}.\label{fig:phot_spec_lum}}
\end{figure*}

We compare these photometric luminosity estimates to our spectroscopic estimates in the top left panel of Figure~\ref{fig:phot_spec_lum}. We find that the spectroscopic and photometric estimates agree well for single stars with spectroscopic luminosity estimates $L_{\star, spec} < 0.025 \lsun$ but that there is a systematic difference between the spectroscopic and photometric estimates for brighter stars. The photometric estimates are brighter than the spectroscopic estimates for stars with intermediate brightness ($ 0.025 \lsun < L_{\star, spec} < 0.13 \lsun$) and fainter than the spectroscopic estimates for the brightest stars \mbox{($L_{\star, spec} > 0.13 \lsun$)}.

Gaia DR2 includes luminosity estimates for 50 of the cool dwarfs in our sample. \citet{andrae_et_al2018} determined stellar parameters by using the Final Luminosity Age and Mass Estimator (FLAME) and Priam algorithms to infer stellar luminosities, radii, and effective temperatures from Gaia parallaxes and three-band photometry ($G$, $G_{\rm BP}$, $G_{\rm RP}$). Both modules are part of the larger Gaia astrophysical parameter inference system \citep[Apsis,][]{bailer-jones_et_al2013}. 

As shown in the top middle panel of Figure~\ref{fig:phot_spec_lum}, the Gaia luminosity estimates follow the same trend as the spectroscopic luminosities we estimated from the \citet{newton_et_al2015} relations in Section~\ref{ssec:details}. However, the Newton estimates are slightly lower for fainter cool dwarfs and higher for brighter cool dwarfs. Note that \citet{andrae_et_al2018} do not report luminosities or radii for stars smaller than $R_\star = 0.5 \rsun$. For field-age cool dwarfs, this boundary roughly corresponds to $M_\star = 0.5 \msun$, $T_{\rm eff} = 3660$~K, and spectral types of M1 -- M2.

There are several stars with precise Gaia luminosity estimates that are significantly higher than their spectroscopic luminosity estimates. Many of these stars have already been identified as stellar binaries, some of which are eclipsing and generated transit-like signals in the \emph{K2} photometry. Figure~\ref{fig:phot_spec_lum} demonstrates that combining spectroscopic characterization with photometric characterization is an efficient way to identify close binaries even in the absence of high-resolution follow-up imaging: stars in unresolved binaries appear overly luminous to photometric surveys but stellar spectroscopy enables independent estimates of stellar luminosities. As would be expected for unresolved binaries, the Gaia luminosities calculated for the stars identified as possible EBs are larger than our spectroscopic estimates. 

Neglecting the five stars with nearby companions (four of which have Gaia luminosity estimates)  and the seven stars flagged as likely eclipsing binaries (five of which have Gaia luminosity estimates), the median difference between the luminosity estimates is $\Delta L_\star = L_{\star, spec} - L_{\star,Gaia} = -0.003 \lsun$ and the standard deviation of the difference is $\Delta L_\star$ is $0.043\lsun$. However, while the median difference is small, Figure~\ref{fig:phot_spec_lum} reveals that the difference between the Gaia luminosity estimates and the spectroscopic luminosity estimates is luminosity-dependent. The Gaia estimates are lower than our spectroscopic estimates for stars with $L_{\star,Gaia} < 0.12\lsun$ and higher for brighter stars. The differences are roughly $0.02\lsun$ at the low luminosity end ($ L_{\star,Gaia} < 0.12\lsun$) and $0.03\lsun$ at the high luminosity end \mbox{($ L_{\star,Gaia} > 0.12\lsun$).} 

The top right panel of Figure~\ref{fig:phot_spec_lum} reveals that our photometric luminosity estimates are consistent with the Gaia luminosity estimates. All of the stars are tightly near the one-to-one relation but our photometric estimates are slightly lower than the Gaia luminosity estimates. The median difference between the luminosity estimates is $\Delta L_\star = L_{\star, phot} - L_{\star,Gaia} = -0.005 \lsun$ and the standard deviation of the difference is $\Delta L_\star$ is $0.010\lsun$. For the closest stars ($d < 75$~pc), the Gaia estimates are roughly $0.01\lsun$ larger than the photometric estimates. This difference decreases with increasing distance for distances between 29 pc and 130 pc. For intermediate distances of 130 -- 200 pc, the Gaia estimates are roughly $0.003\lsun$ smaller than the photometric estimates. Finally, for distances of 200 --  500 pc, the Gaia estimates are roughly $0.008\lsun$ larger than the photometric estimates.

As discussed in Section~\ref{ssec:comp}, for our final stellar catalog, we adopt the photometric luminosities when possible and the spectroscopic luminosities for stars without parallaxes in Gaia~DR2. We favor the photometric luminosities over the spectroscopic luminosities because the relations from \citet{newton_et_al2015} that we use to calculate spectroscopic luminosities were calibrated using a sample of only 25~stars with interferometrically-determined radii while the photometric luminosities are derived directly from photometry, precisely determined parallaxes from Gaia~DR2 \citep{gaia_et_al2018}, and established bolometric corrections \citep{mann_et_al2015}.

\subsubsection{Stellar Radii}
\label{sssec:photo_rs}
We estimated stellar radii using the empirical equations from Table~1 of \citet{mann_et_al2015, mann_et_al2016}. For the 66~cool dwarfs with Gaia parallaxes and spectral types of K7 or later, we calculated stellar radii by employing the $R_\star - M_{Ks}- $[Fe/H] relation given in their Equation~5; for the metallicity-dependent term, we used the [Fe/H] values calculated in Section~\ref{ssec:details}. For the 17~K5~dwarfs with Gaia parallaxes, we dropped the metallicity dependence because the stars were too hot for our selected metallicity relation and used the simpler $R_\star -M_{Ks}$ relation described by their Equation~4. The systematic errors on the $R_\star - M_{Ks}- $[Fe/H] and $R_\star -M_{Ks}$ relation are 2.70\% and 2.89\%, respectively \citep{mann_et_al2015, mann_et_al2016}. Our quoted errors on the stellar radii incorporate both these systematic errors and the uncertainties on $M_{Ks}$ and [Fe/H].

The relations are valid for K7 -- M7 dwarfs with \mbox{$4.6 < M_{Ks} < 9.8$} and \mbox{$-0.6 < {\rm [Fe/H]} < 0.5$}. Most of the cool dwarfs with parallaxes fall within those limits (59 stars; 69\%), but one is too metal-rich ([Fe/H] = 0.63) and 23 are too bright. We do not report photometric radius estimates for the six brightest stars, but we used the equations from \citet{mann_et_al2015} to extrapolate the relations slightly to cover $4.3 < M_{Ks} < 9.8$ and $-0.6 < {\rm [Fe/H]} < 0.65$ so that we can estimate radii for 18~stars that are only slightly outside the calibration range. We have included a flag in Table~\ref{tab:photometric} to indicate which stars have absolute magnitudes or metallicities outside the range recommended by \citet{mann_et_al2015}. 

\begin{figure*}[tbp]
\centering
\includegraphics[width=0.32\textwidth]{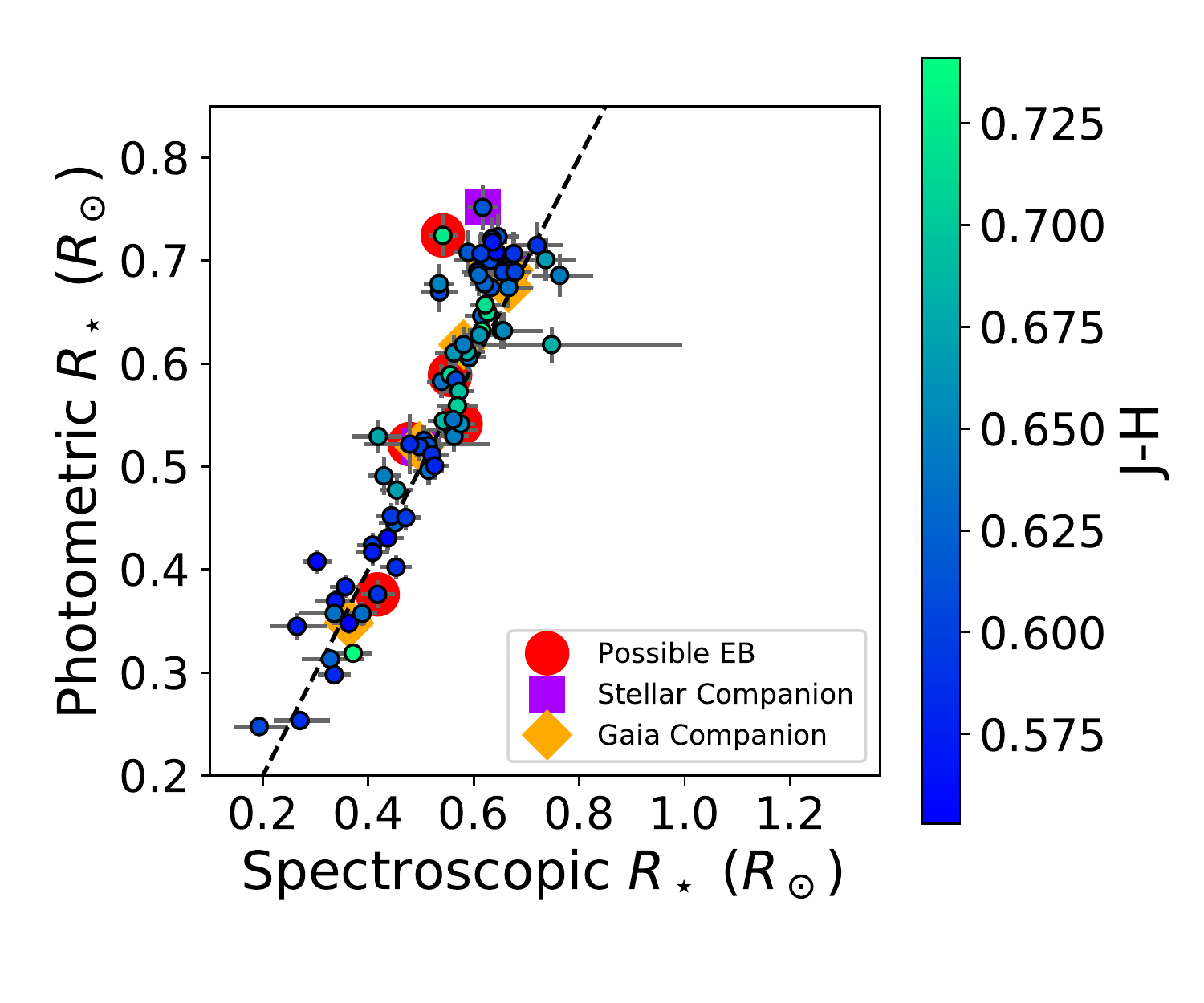}
\includegraphics[width=0.32\textwidth]{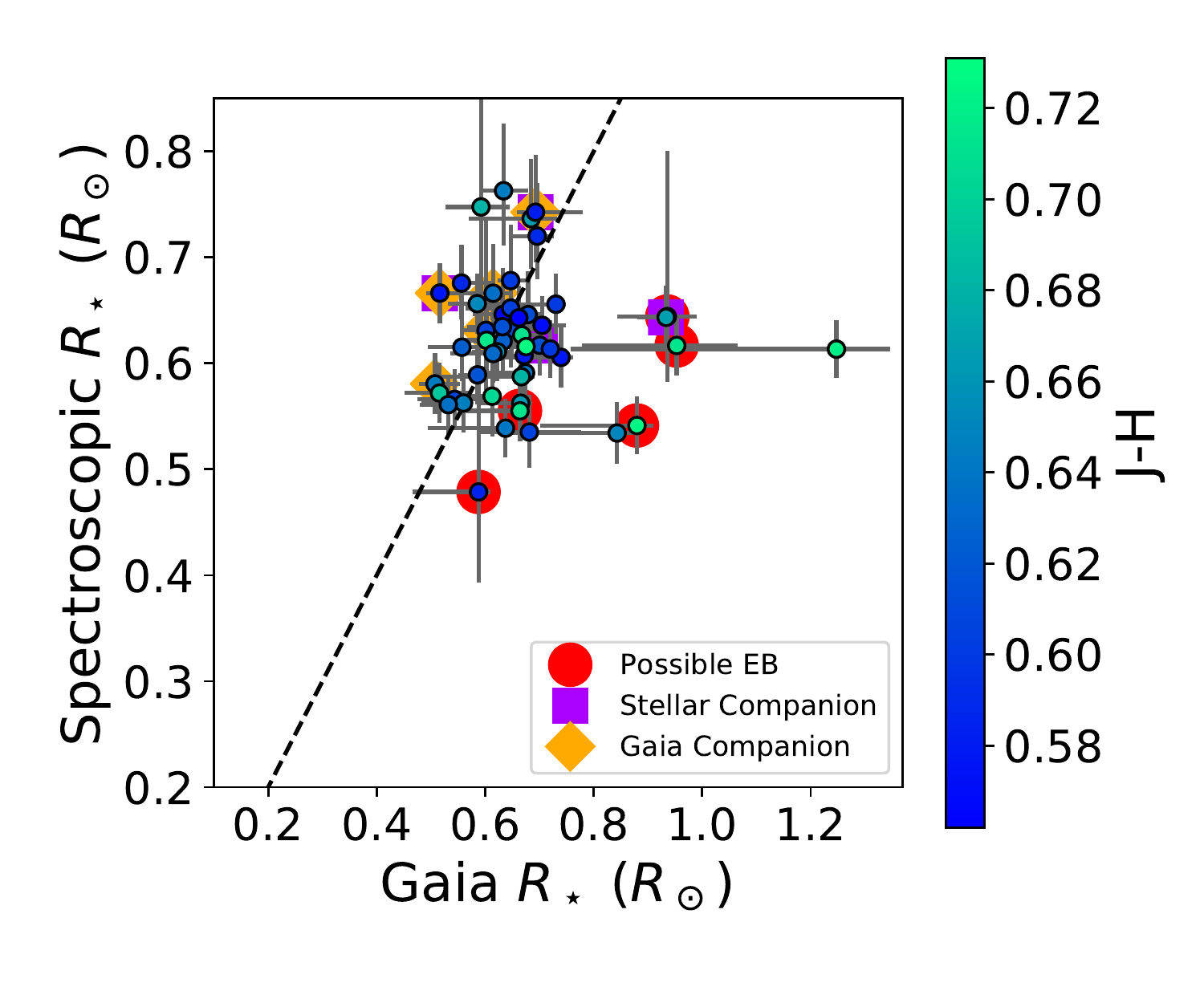}
\includegraphics[width=0.32\textwidth]{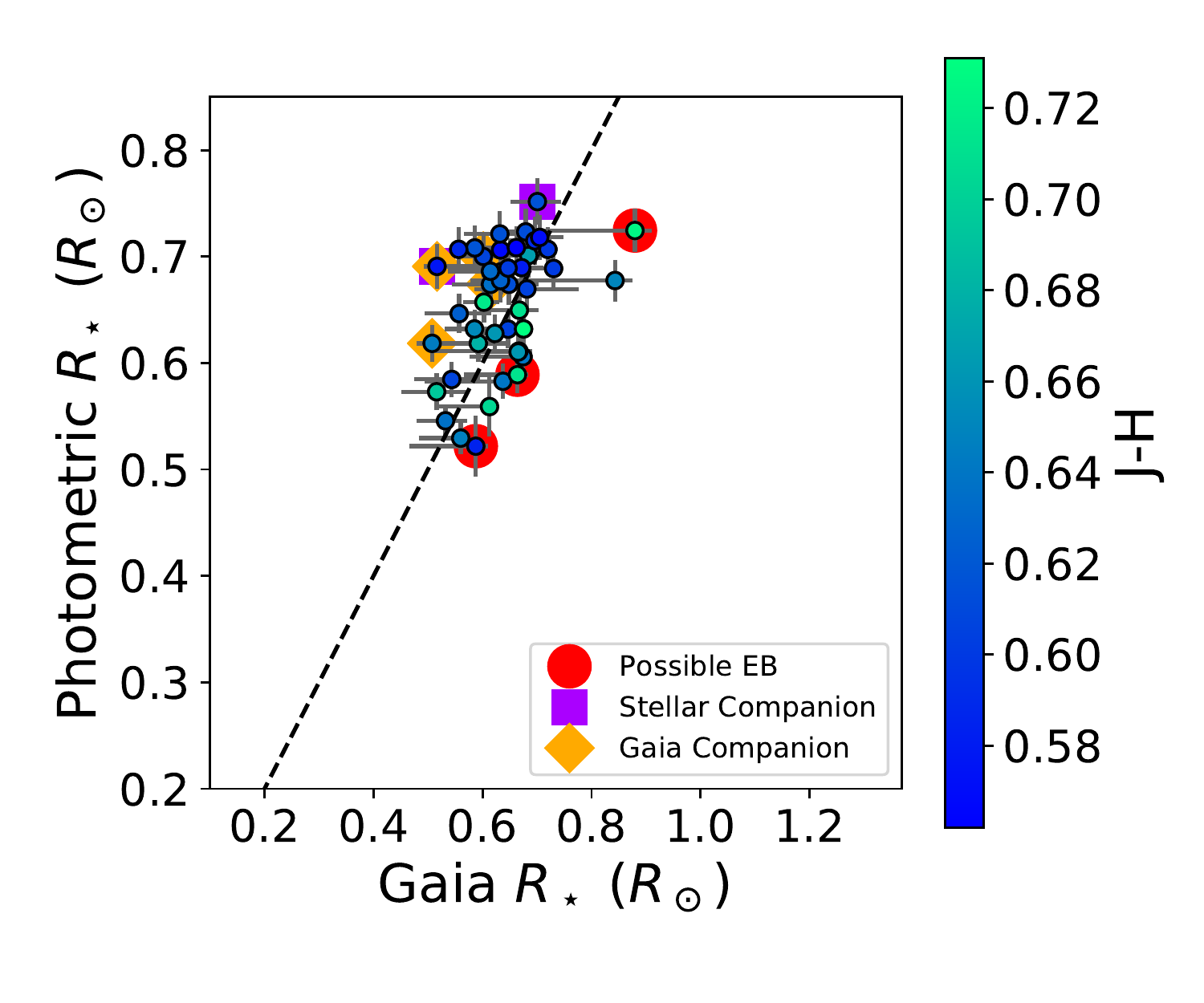}\\
\includegraphics[width=0.32\textwidth]{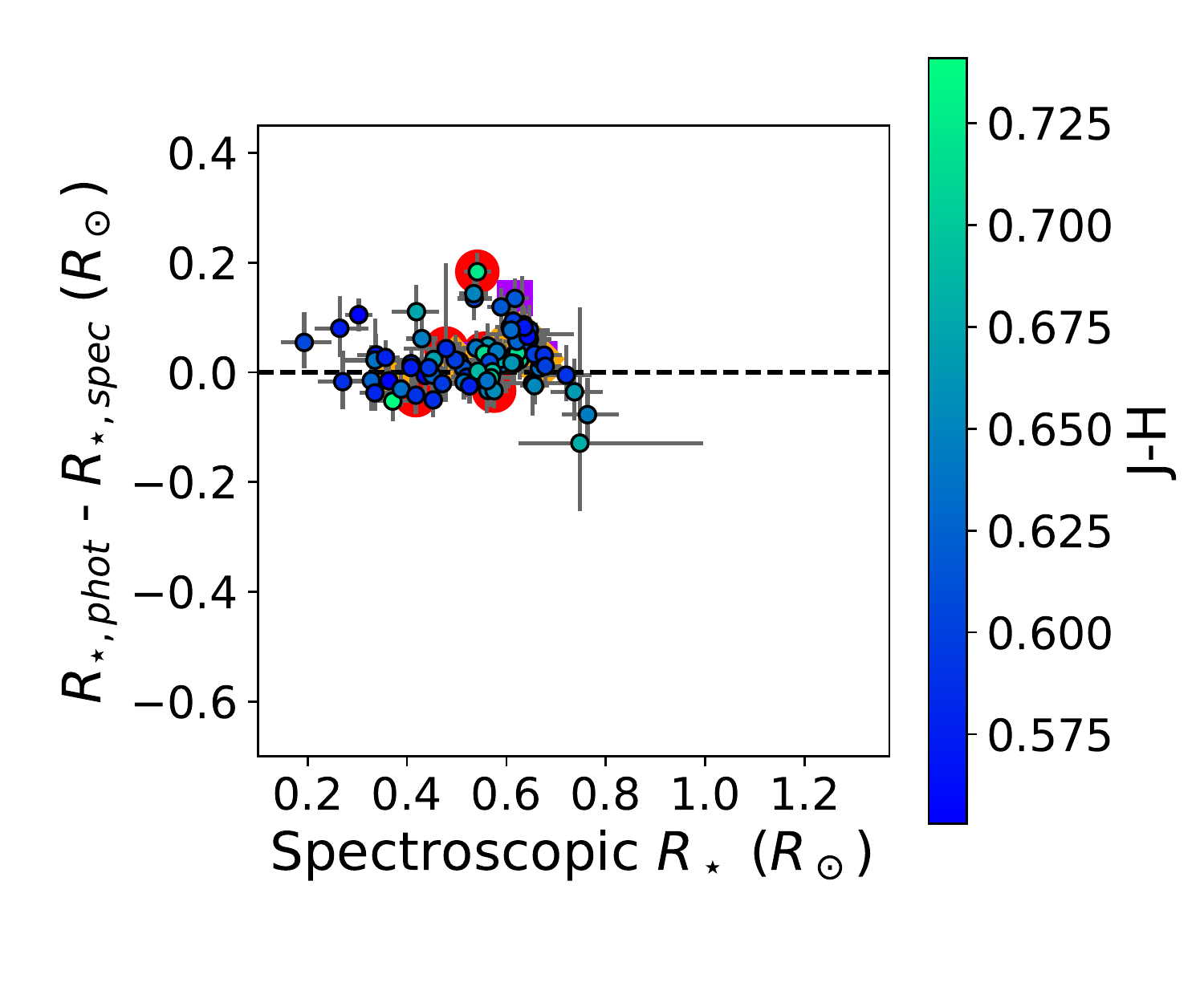}
\includegraphics[width=0.32\textwidth]{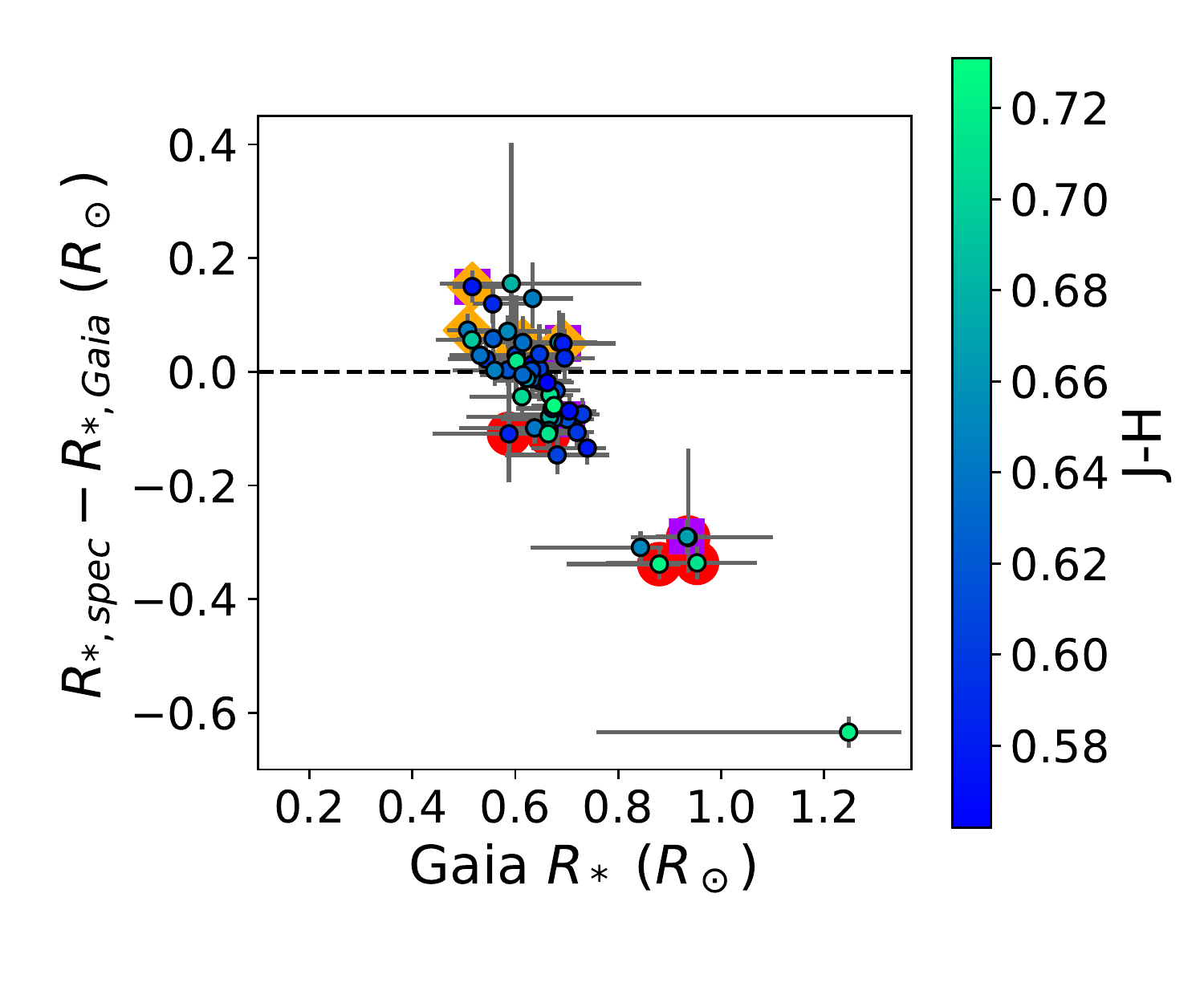}
\includegraphics[width=0.32\textwidth]{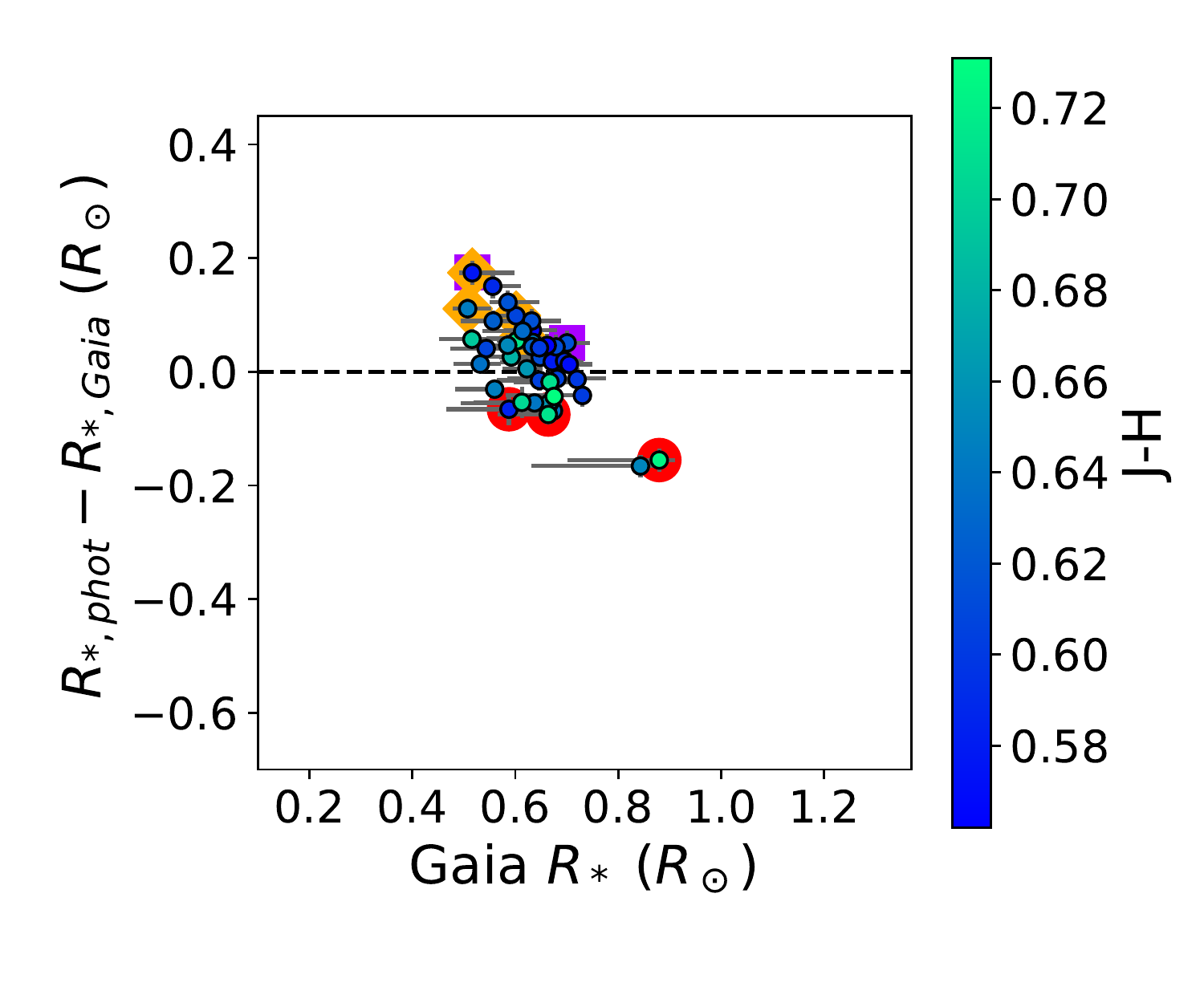}\\
\includegraphics[width=0.49\textwidth]{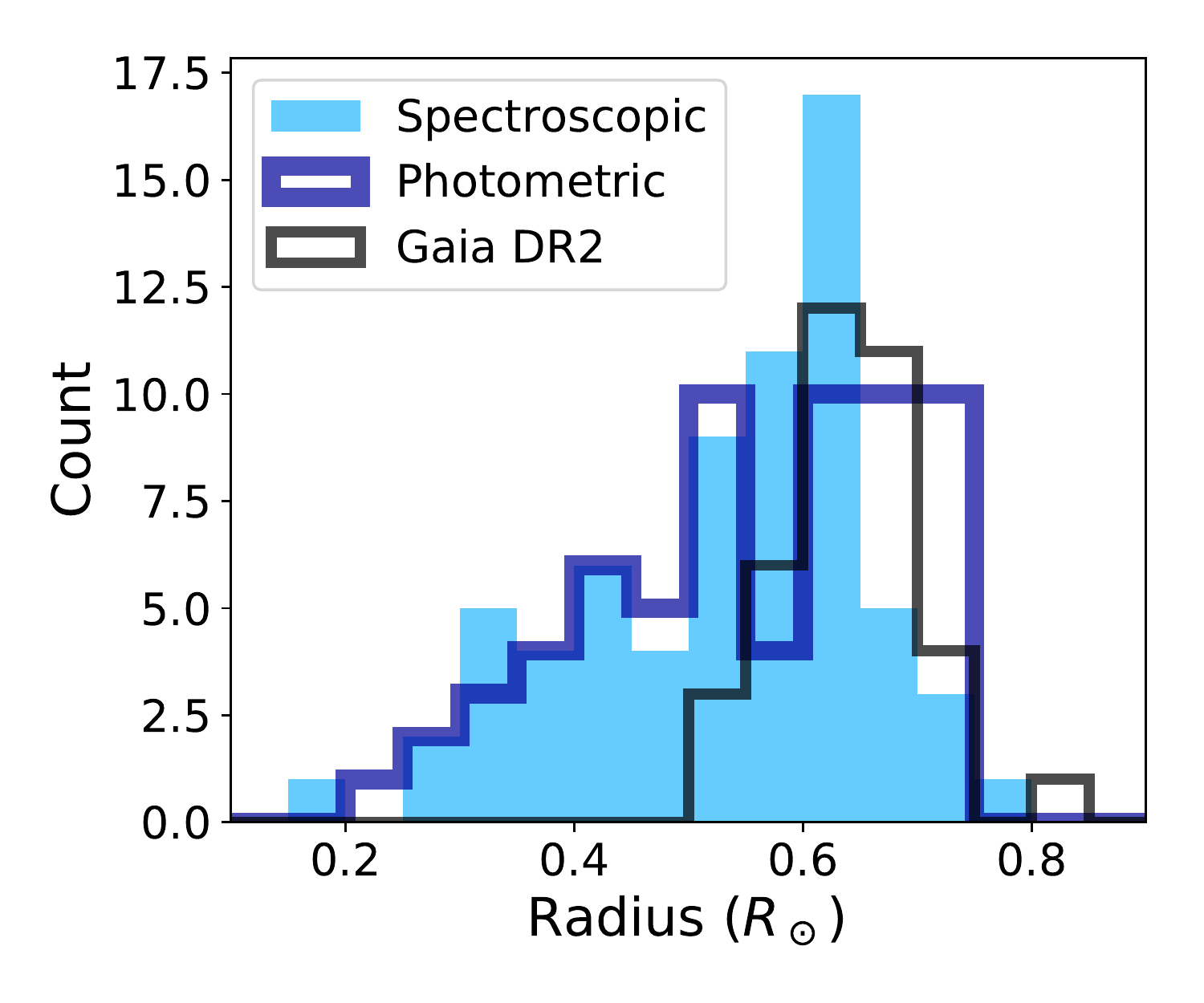}
\includegraphics[width=0.49\textwidth]{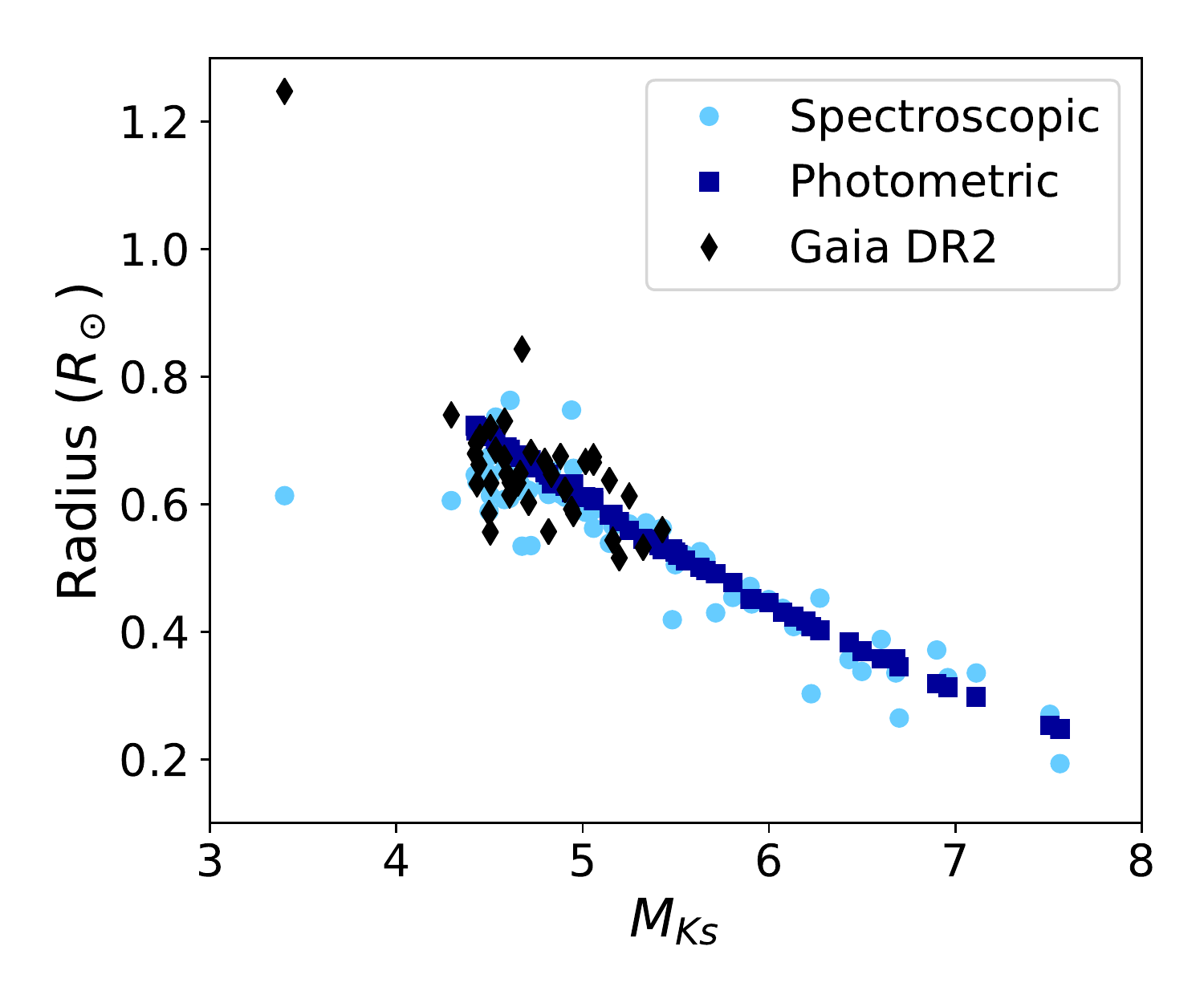}
\caption{Same as Figure~\ref{fig:phot_spec_lum}, but for stellar radii. The Gaia sample is restricted to stars larger than $0.5\rsun$. \label{fig:phot_spec_rs}}
\end{figure*}

As shown in the upper two left panels of Figure~\ref{fig:phot_spec_rs}, the photometric radius estimates agree well with the spectroscopic estimates found in Section~\ref{ssec:details}. Considering only the 65~stars that appear to be single and have radius estimates from both methods, the median difference $\Delta R_\star = R_{\star,phot} -  R_{\star,spec} = 0.02 \rsun$ (3\%) and the standard deviation in the distribution of $\Delta R_\star$ is $\sigma_{\Delta R_\star} = 0.05 \rsun$. \citet{mann_et_al2017b} previously noted that the spectroscopic radii estimated by \citet{newton_et_al2015} are consistent with the photometric radii estimated by their $R_\star - M_{Ks}- $[Fe/H] relation \citep{mann_et_al2015}, so this result is not surprising. 

The top two center panels of Figure~\ref{fig:phot_spec_rs} contrast our spectroscopic radius estimates to those estimated by the Gaia team using Apsis-FLAME \citep{bailer-jones_et_al2013, andrae_et_al2018}. Only 50 cool dwarfs have radius estimates in DR2 and all of those stars are at least $0.5\rsun$ because \citet{andrae_et_al2018} do not report luminosity or radius estimates for smaller stars. The median difference for the 38~supposedly single stars is $\Delta R_\star = R_{\star,spec} -  R_{\star,Gaia} = -0.01 \rsun$ (-2\%) and the standard deviation of the differences is $\sigma_{\Delta R_\star} = 0.1 \rsun$. Although these differences are small, there is a noticeable trend between the radius discrepancy and the estimated radius. Our spectroscopic estimates tend to be larger than the value estimated by the Gaia team for stars with $R_{\star,Gaia} > 0.6\rsun$ and lower than the Gaia estimates for stars with $R_{\star,Gaia} > 0.6\rsun$ . 

Predictably, the possible eclipsing binaries have larger radius estimates from Gaia than from spectroscopy because the added light from a companion star causes them to appear overluminous in Gaia. Two purportedly single stars also have large radius discrepancies: the M1 dwarf EPIC~201663913 has a Gaia radius estimate of $R_{\star,Gaia} = 0.84\rsun$ and a spectroscopic estimate of $R_{\star,spec} = 0.53\rsun$ while the K7 dwarf EPIC~246947582 has $R_{\star,Gaia} = 1.25\rsun$ and $R_{\star,spec} = 0.61\rsun$. These stars were assigned radii of $0.403\rsun$ and $0.428\rsun$, respectively, in the Ecliptic Plane Input Catalog \citep{huber_et_al2016}. As noted in Section~\ref{sss:absmag}, there are no follow-up observations of EPIC~201663913 posted to the ExoFOP website, but D.~Ciardi acquired a high-resolution image of EPIC~246947582 using NIRC2 on Keck 2 and did not detect any companions.

Finally, the upper two right panels of Figure~\ref{fig:phot_spec_rs} compare our photometric radius estimates to the Gaia radius estimates  \citep{bailer-jones_et_al2013, andrae_et_al2018}. Six of the 50~stars with Gaia radius estimates are too bright for the \citet{mann_et_al2015, mann_et_al2016} relations.  The remaining 44 cool dwarfs have both Gaia radius estimates and photometric radius estimates from this paper and 36 are supposedly single. The median difference for those 36~stars is $\Delta R_\star = R_{\star,phot} -  R_{\star,Gaia} = 0.02 \rsun$ (3\%) and the standard deviation of the differences is $\sigma_{\Delta R_\star} = 0.06 \rsun$. As in the center panels, we note that the radius difference is correlated with the radius estimated by the Gaia team. Specifically, our photometric estimates tend to be larger than the Gaia estimates for the 18~stars with $R_{\star,Gaia} > 0.64\rsun$. For the 18~larger stars, there is still scatter in the relation but the median difference is closer to zero ($-0.013\rsun$ versus $0.049\rsun$ for smaller stars). 

In our stellar catalog, we select the photometric radii when possible and default to spectroscopic radii for the nine stars without photometric estimates. Three of the stars with spectroscopic radius estimates lack parallaxes in Gaia~DR2 and the remaining six are too bright for the empirical relations from \citet{mann_et_al2015, mann_et_al2016}. The spectroscopic sample contains a high fraction of likely eclipsing binaries (three stars) and stars with candidate stellar companions (three stars).

\begin{figure*}[tbp]
\centering
\includegraphics[width=0.49\textwidth]{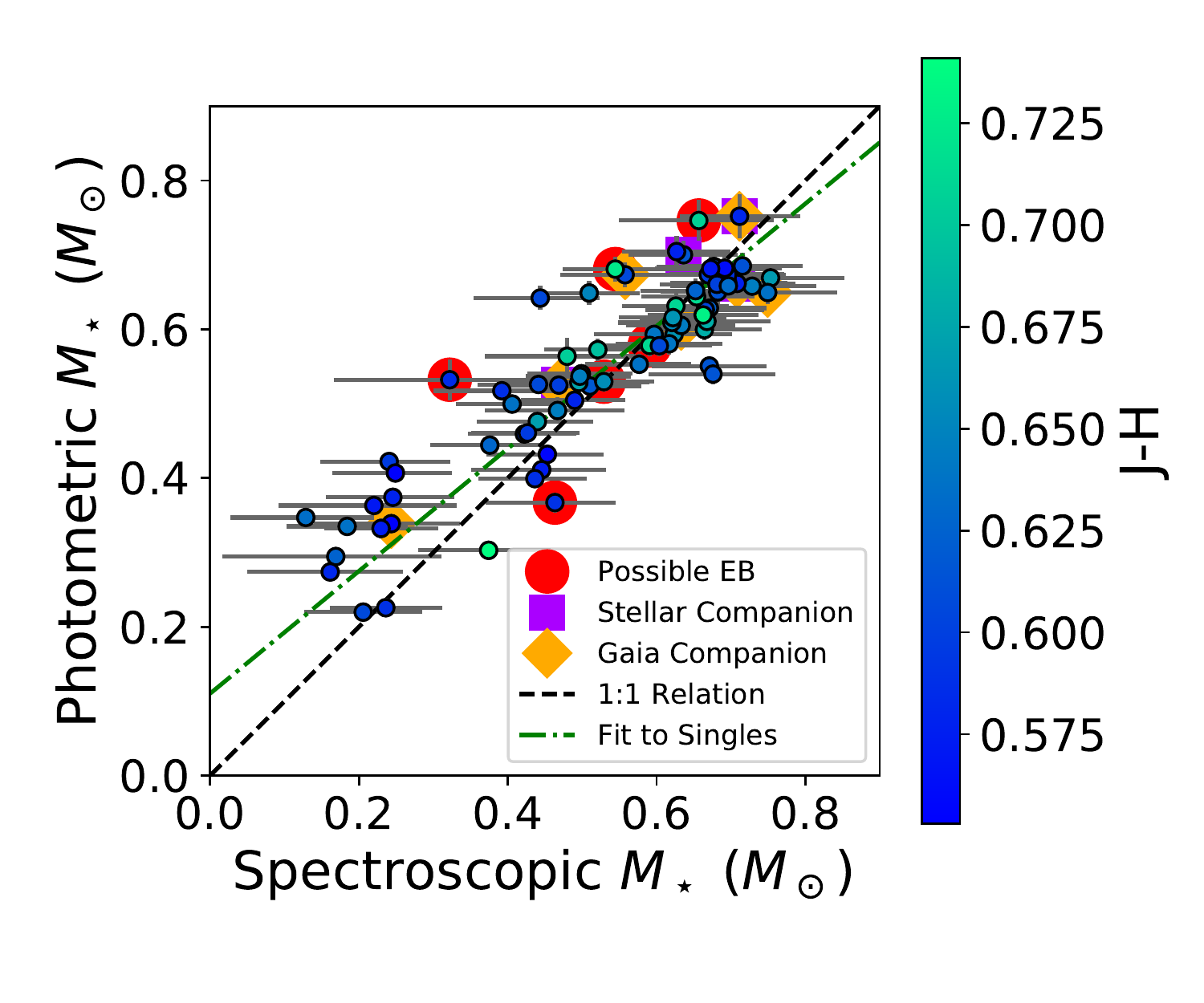}
\includegraphics[width=0.49\textwidth]{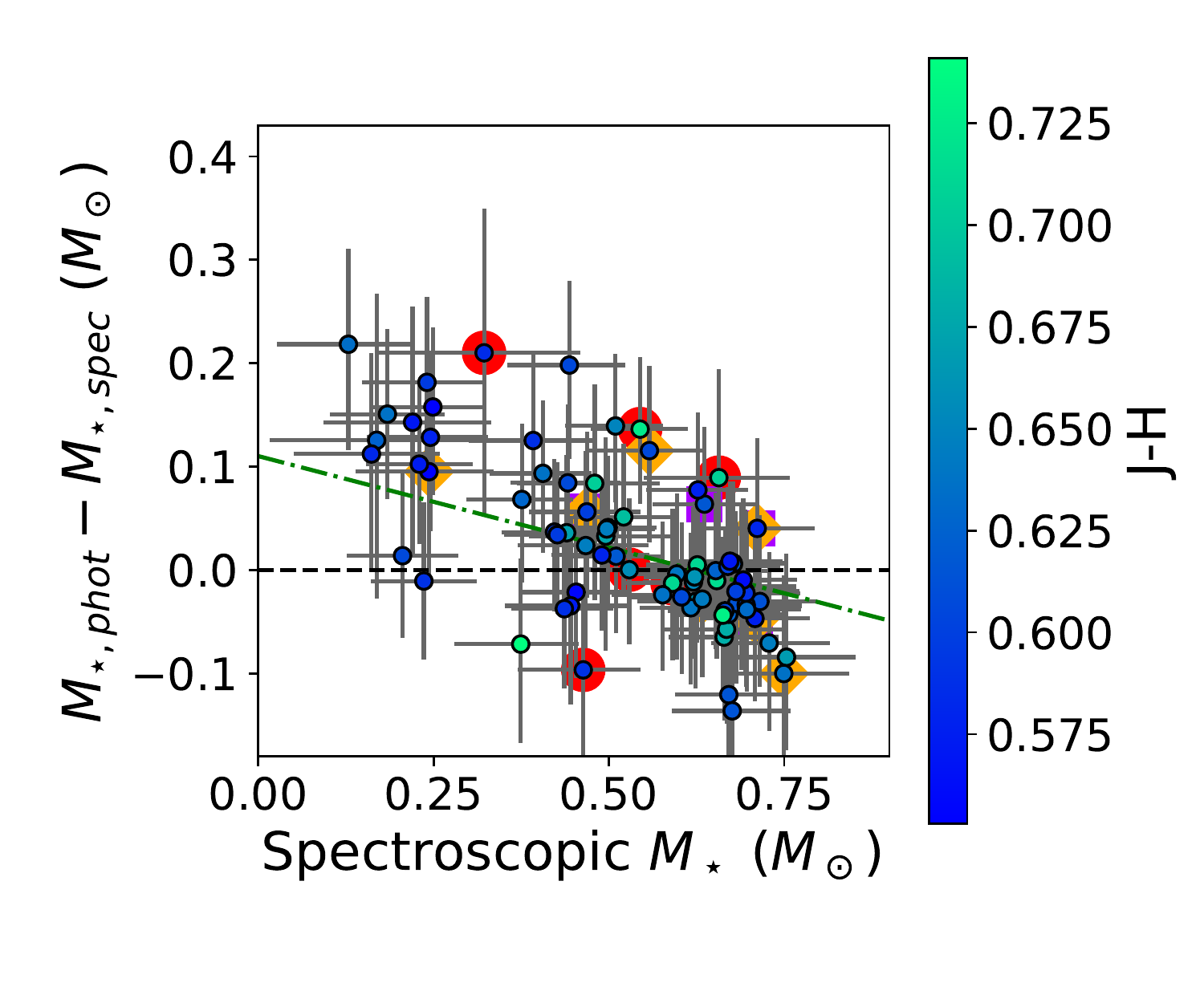}\\
\includegraphics[width=0.49\textwidth]{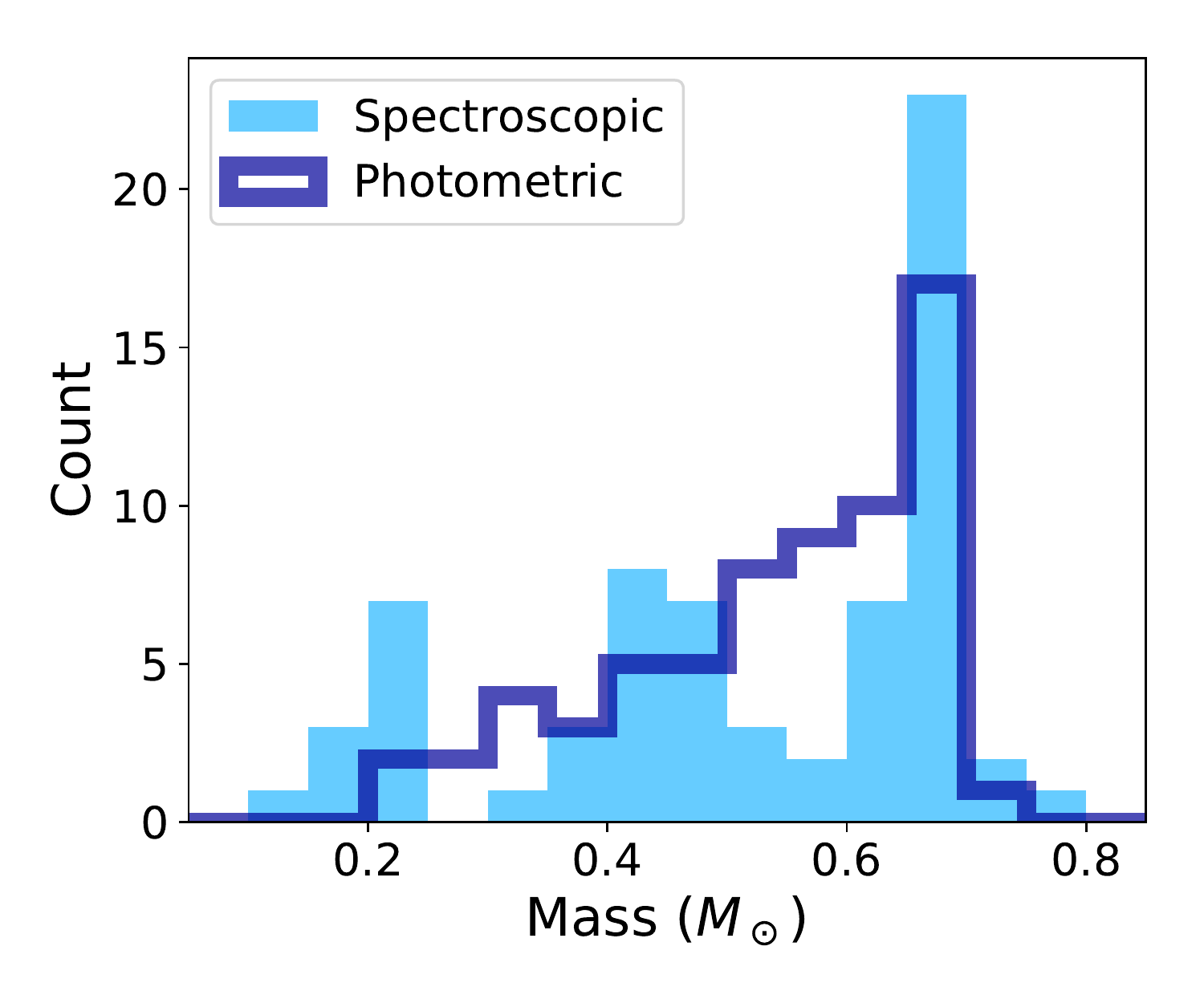}
\includegraphics[width=0.49\textwidth]{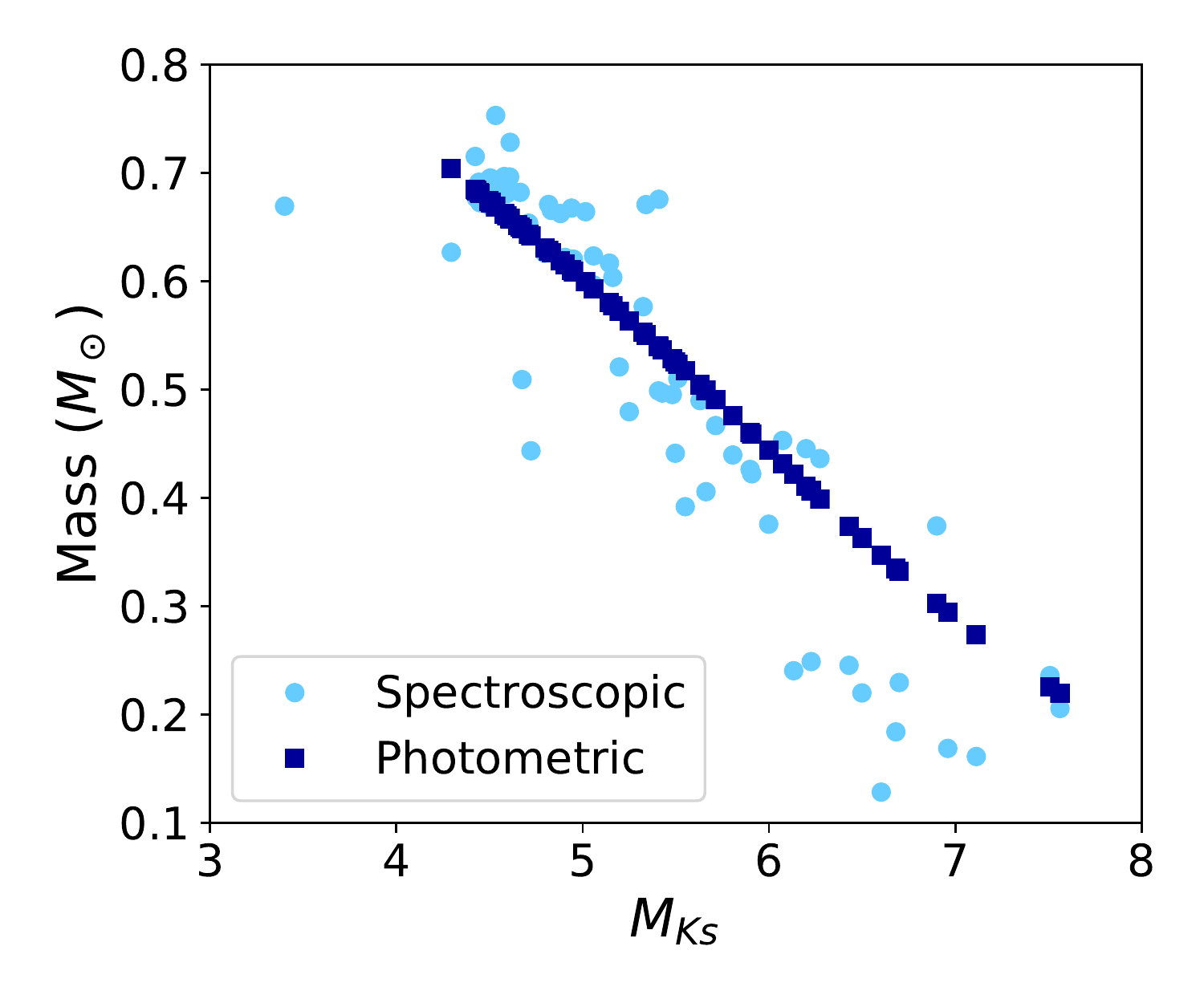}
\caption{Same as Figure~\ref{fig:phot_spec_lum}, but for stellar masses. The Gaia panels are missing because Gaia~DR2 does not include estimates of stellar mass. In the top row,} the green dotted-dash lines are a linear fit to all of the purportedly single stars and the black dashed lines mark a 1:1 correlation. \label{fig:phot_spec_mass}
\end{figure*}

\begin{figure*}[tbp]
\centering
\includegraphics[width=0.32\textwidth]{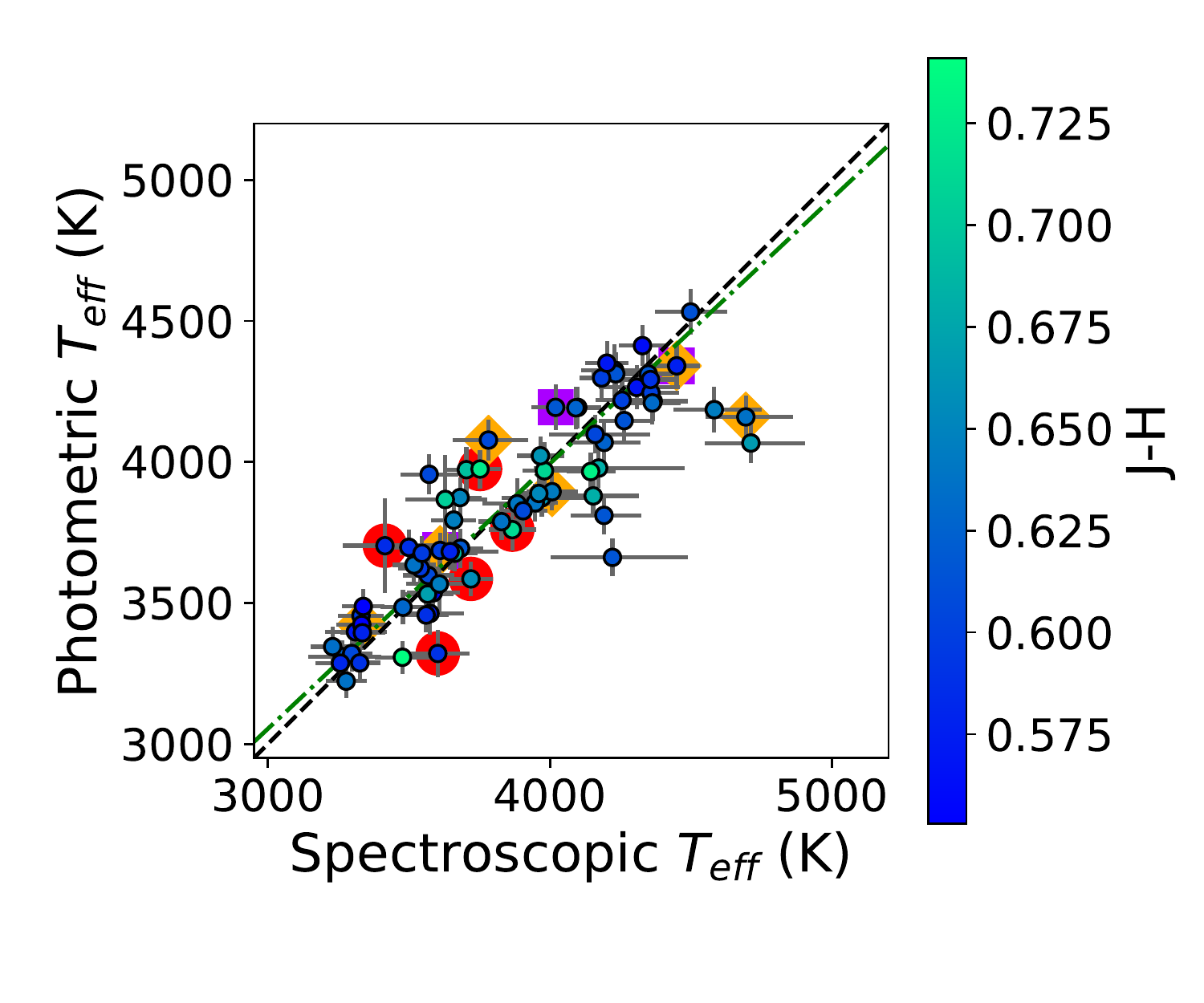}
\includegraphics[width=0.32\textwidth]{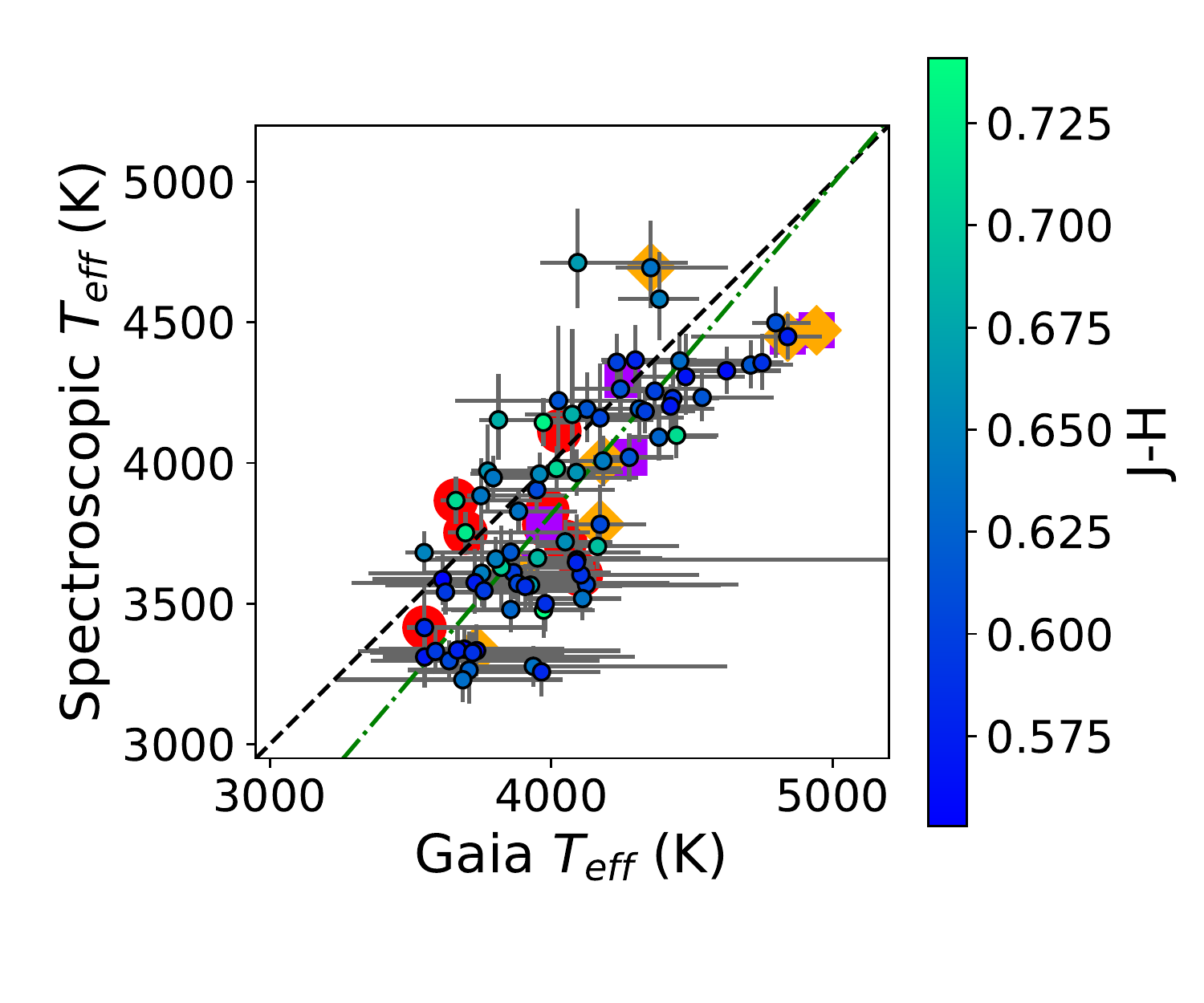}
\includegraphics[width=0.32\textwidth]{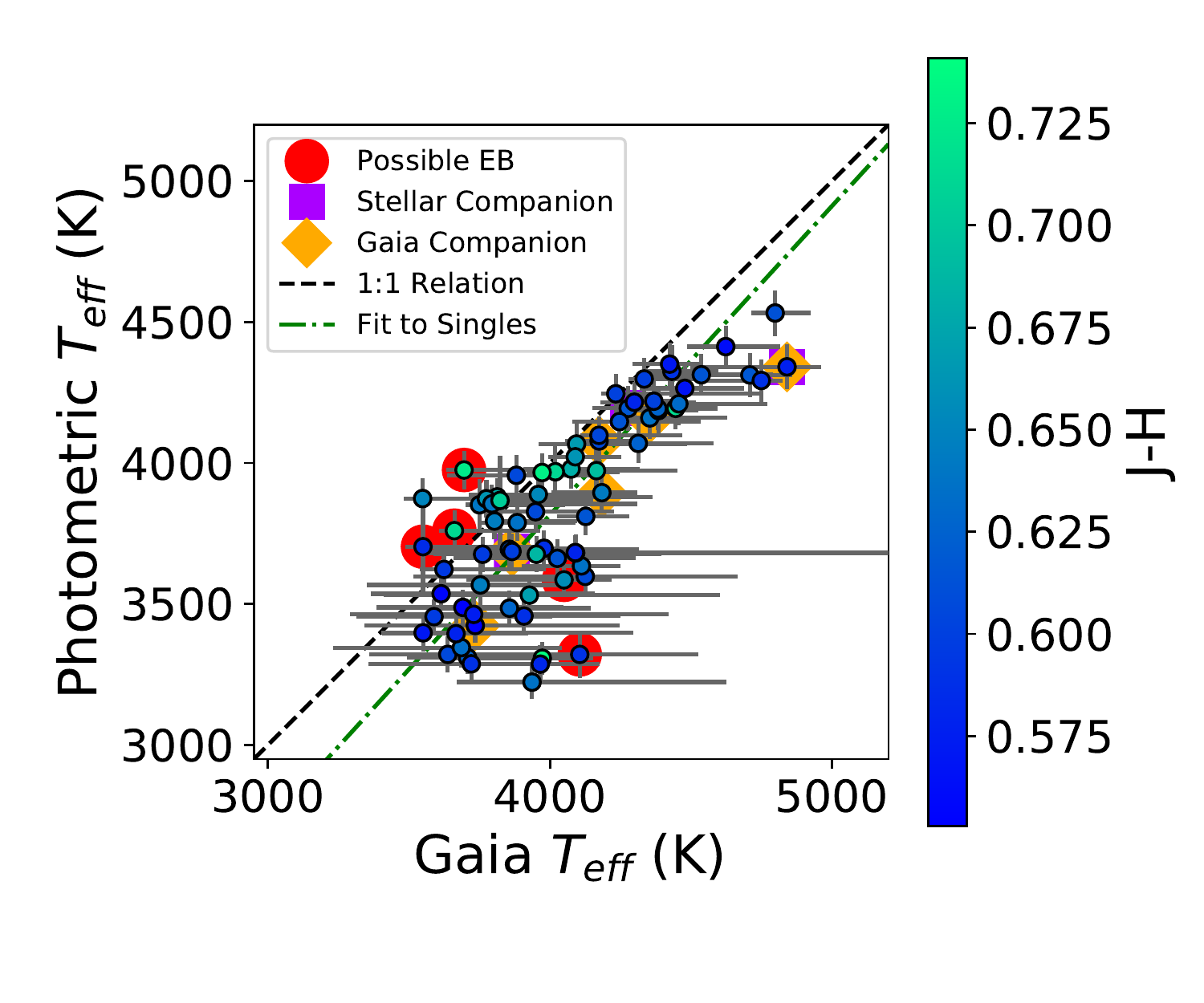}\\
\includegraphics[width=0.32\textwidth]{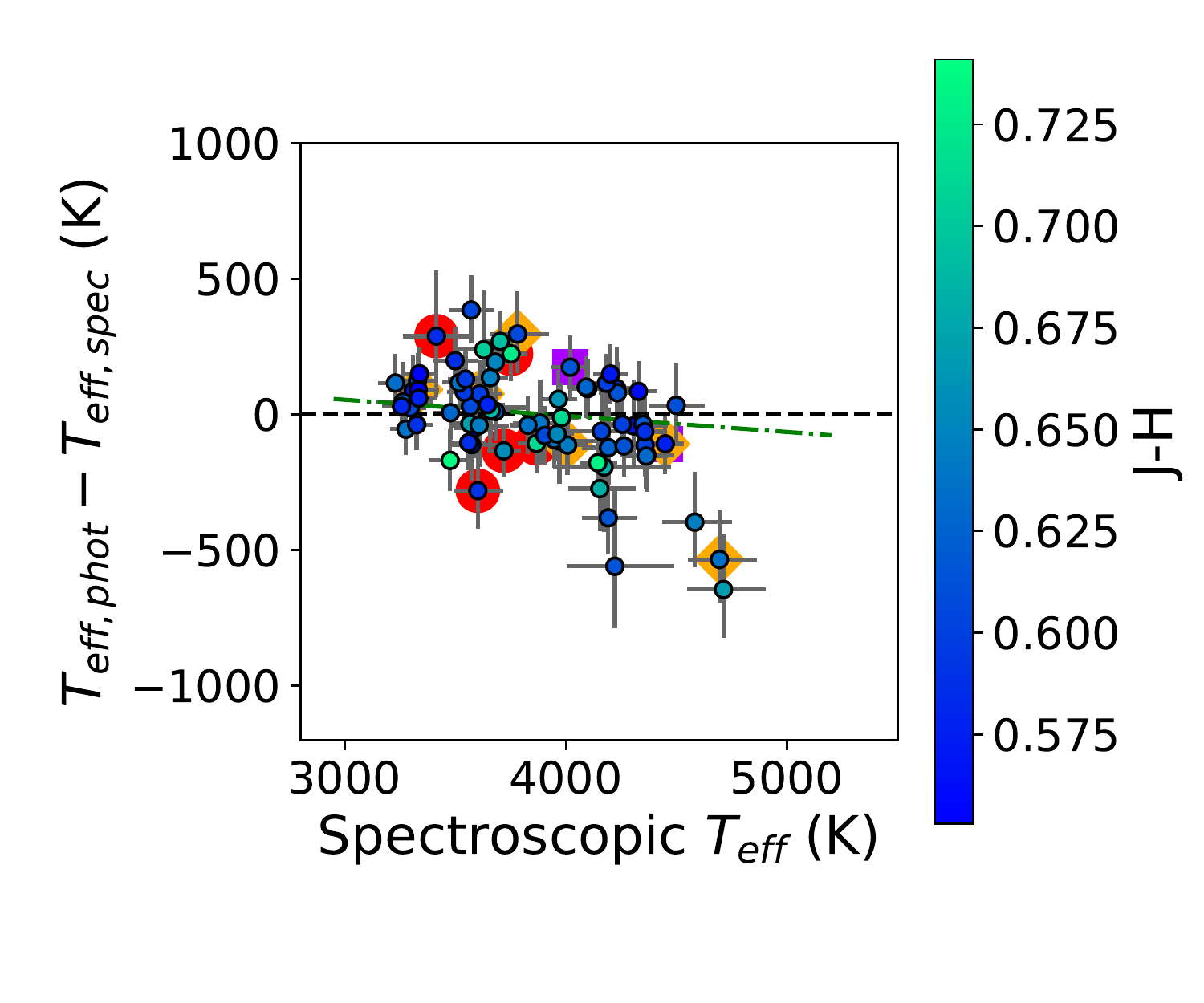}
\includegraphics[width=0.32\textwidth]{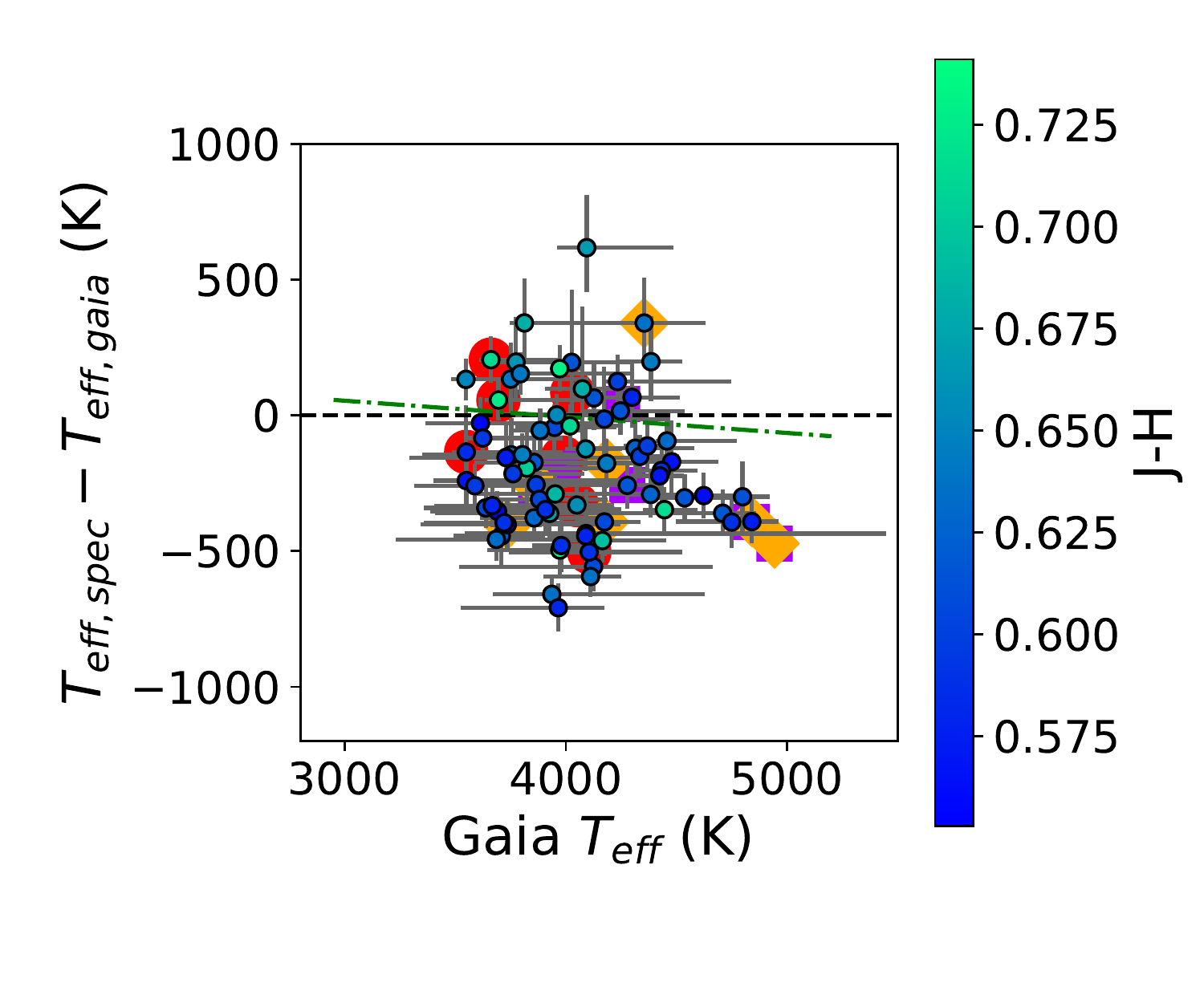}
\includegraphics[width=0.32\textwidth]{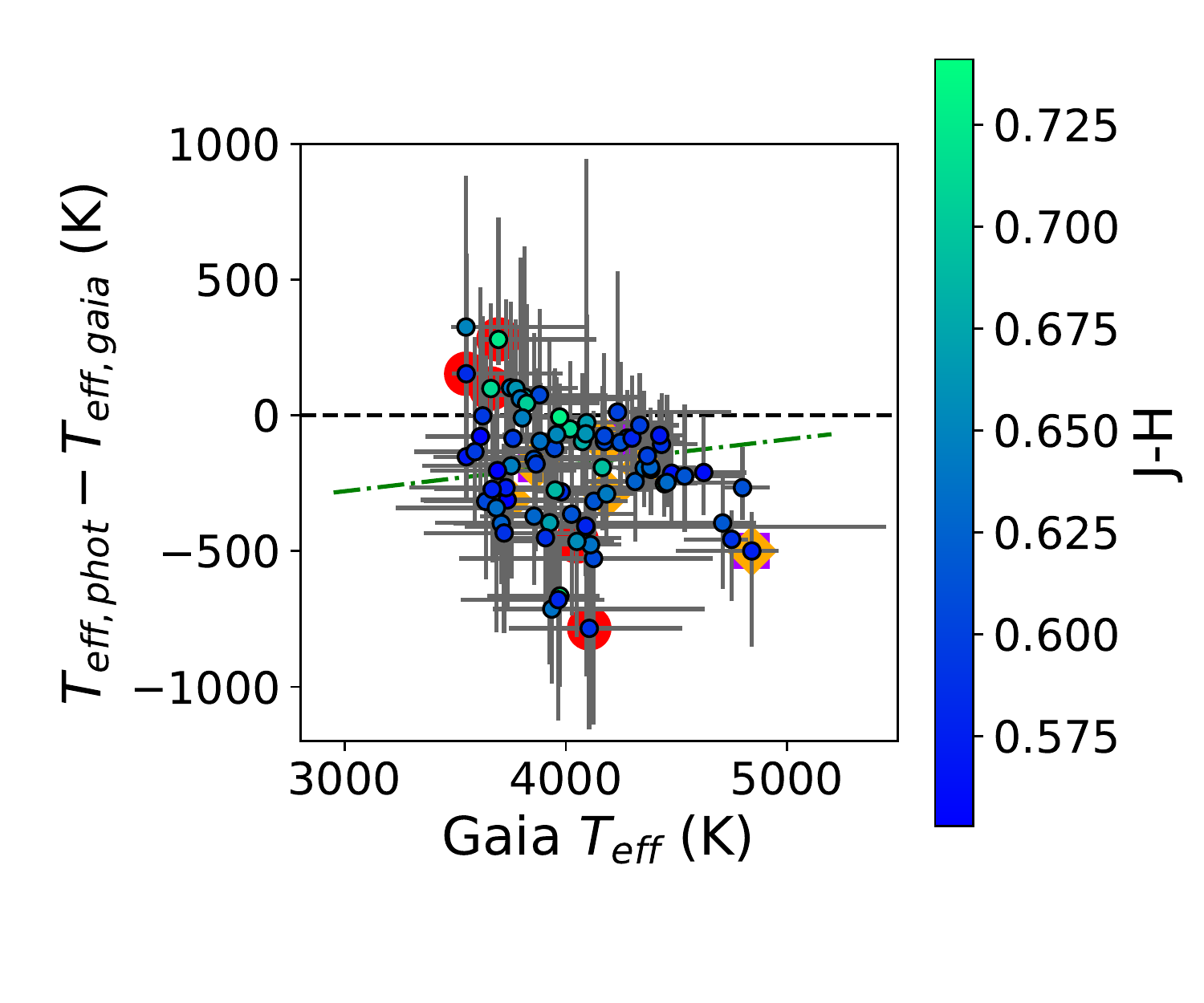}\\
\includegraphics[width=0.49\textwidth]{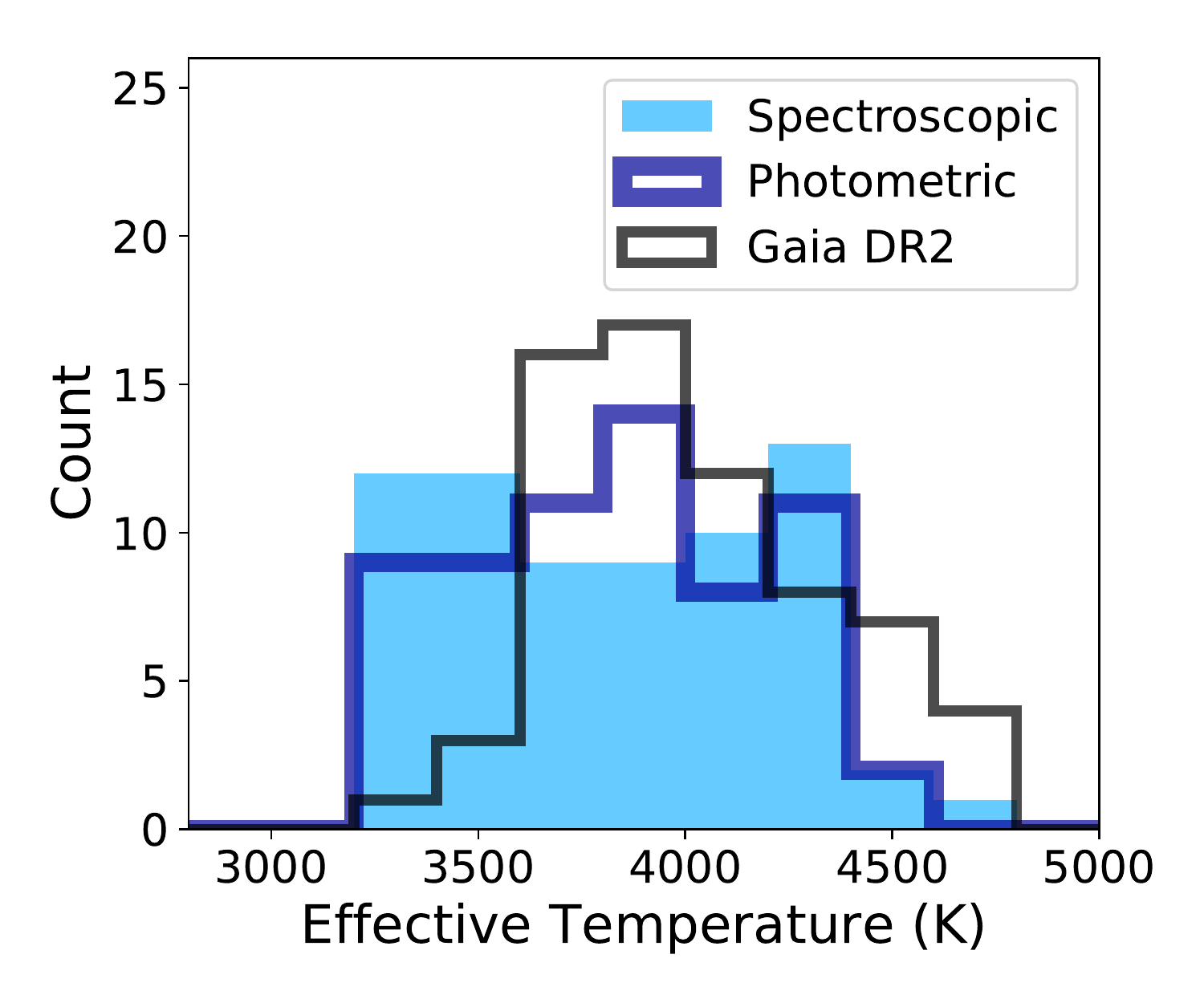}
\includegraphics[width=0.49\textwidth]{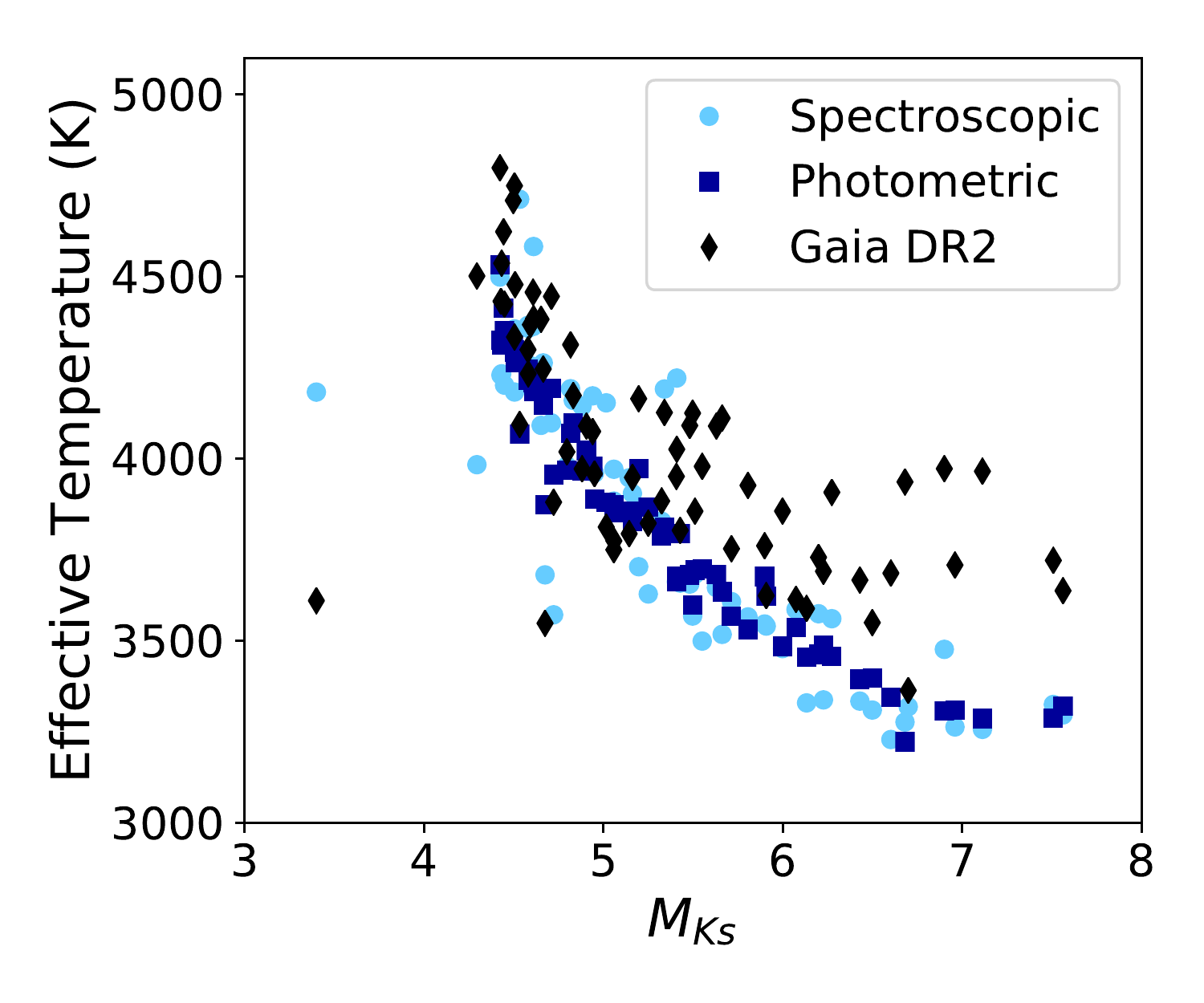}
\caption{Same as Figure~\ref{fig:phot_spec_lum}, but for stellar effective temperatures. \label{fig:phot_spec_teff}} 
\end{figure*}

\subsubsection{Stellar Masses}
\label{sssec:photo_ms}
We estimated masses by employing the $M_\star - M_{Ks}$ empirical relation from \citet{mann_et_al2019}, which was constructed by using parallaxes, imaging, and astrometry to constrain the orbits and masses of 62 nearby stellar binaries. \citet{mann_et_al2019} present six different $M_\star - M_{Ks}$ relations ranging in complexity from fourth to sixth order in $M_{Ks}$. Half of the relations incorporate a metallicity-dependent term while the remaining three are independent of metallicity. Following the advice in the paper, we used the fifth order fit and did not incorporate metallicity because the current sample of cool dwarfs with precisely determined masses is too small to warrant the addition of a metallicity-dependent term \citep{mann_et_al2019}.

The new $M_\star - M_{Ks}$ relation from \citet{mann_et_al2019} agrees well (within 5\%) with the earlier \citet{delfosse_et_al2000} relation for stars with masses $0.15 \msun < M_\star < 0.5 \msun$ and predicts masses that are roughly 10\% higher for more massive stars where the \citet{delfosse_et_al2000} sample was sparse. For stellar masses of $0.09 \msun - 0.25 \msun$, the $M_\star - M_{Ks}$relation from \citet{mann_et_al2019} also agrees well with the relation from \citet{benedict_et_al2016}, but for stars with $M_\star > 0.3\msun$, \citet{mann_et_al2019} find masses that are 10\% lower than those predicted by \citet{benedict_et_al2016} relation. \citet{mann_et_al2019} attribute this discrepancy to the inclusion of eclipsing binaries and stars with poor $M_{Ks}$ estimates in the stellar sample used by \citep{benedict_et_al2016}. For this paper, we opted to use the  relation from \citet{mann_et_al2019} because it is the most recent $M_\star - M_{Ks}$ relation available in the literature for cool dwarfs and based upon a well-vetted sample of stars with precisely and accurately determined properties.

The $M_\star - M_{Ks}$ relation from \citet{mann_et_al2019} is valid for stars with activity levels and metallicities similar to those of nearby stellar neighbors and absolute magnitudes of $4.0 < M_{Ks} < 11.0$, which corresponds to masses of $0.075\msun < M_\star < 0.75 \msun$. Our cool dwarf sample includes 80~stars within this absolute magnitude range, three~brighter stars with $3.4 < M_{Ks} < 3.9$, and three stars without parallaxes in Gaia~DR2. We do not estimate photometric stellar masses for the three~brighter stars. These targets are EPIC~220555384 (which has a nearby stellar companion reported on ExoFOP), EPIC~205947214 (which is flagged as a likely EB on ExoFOP), and EPIC~246947582.

The top left panel of Figure~\ref{fig:phot_spec_mass} demonstrates that the photometric mass estimates are systematically offset from the spectroscopic mass estimates. Ignoring the stars with nearby companions or those flagged as likely eclipsing binaries, the median mass difference for the 66~purportedly single stars is $\Delta M_\star = M_{\star,phot} -  M_{\star,spec} = 0.002 \msun$ (0.2\%) with a standard deviation of $0.08\msun$. Our stellar sample is relatively small, but the offset between the spectroscopic and photometric estimates seems to be larger at the low-mass end.  The realization that the discrepancy is largest for the lowest stellar masses is particularly problematic because even a small difference can be a large fraction of the total stellar mass for the coolest stars. 

Subdividing the sample by spectroscopic mass estimate, $\Delta M_\star = 0.11 \msun$ (34\%) for the 11~stars with \mbox{$M_{\star, phot} < 0.4 \msun$}, $\Delta M_\star = 0.03 \msun$ (6\%) for the 27~stars with \mbox{$ 0.4 \leq \msun < M_\star < 0.6 \msun$}, and  $\Delta M_\star = -0.02 \msun$ (-3\%) for the 28~more massive stars. Fitting a line to the apparently single stars and accounting for the errors in both $M_{\star, spec}$ and $M_{\star, phot}$, we find that $M_{\star, phot}$ can be estimated from $M_{\star, spec}$ using a linear fit with slope $m = 0.82 \pm 0.05$ and y-intercept  $b = 0.11 \pm 0.03$.

For our final stellar catalog, we adopt the photometric mass estimates because those are calculated directly from the absolute magnitudes of our target stars rather than indirectly by applying the older $T_{\rm eff} - M_\star$ relation derived by \citet{mann_et_al2013c} to our spectroscopic temperature estimates. The new $M_\star - M_{Ks}$ relation from \citet{mann_et_al2019} is based on a larger and more comprehensively scrutinized sample of cool dwarfs than the earlier $T_{\rm eff} - M_\star$ relation. For the two stars without parallaxes reported in Gaia~DR2, we estimate masses by using the mass-radius relation found for the photometric sample to predict the masses of stars with radii equal to our spectroscopic radius estimates. Adopting this strategy accounts for the discrepancy between photometric and spectroscopic masses (see Figure~\ref{fig:phot_spec_mass}).

\subsubsection{Stellar Effective Temperatures}
\label{sssec:photo_teff}
We determined photometric temperature estimates for all cool dwarfs with adequate photometry using the same procedure as \citet{mann_et_al2017}. We began by estimating stellar luminosities as described in Section~\ref{sssec:photo_lum}. We then combined our luminosity estimates with the photometric radii estimated in Section~\ref{sssec:photo_rs} and calculated stellar effective temperatures from the Stefan-Boltzmann relation. 

In the top left two panels of Figure~\ref{fig:phot_spec_teff}, we compare these photometric temperature estimates to the spectroscopic estimates determined in Section~\ref{ssec:details}. Overall, the photometric estimates agree well with the spectroscopic estimates. For the 64~presumedly single stars with both spectroscopic and photometric estimates, the median difference $\Delta T_{\rm eff} = T_{{\rm eff}, phot} -   T_{{\rm eff}, spec} = -3$K and the standard deviation of the differences is $\sigma_{\Delta T_{\rm eff}} = 172$K. 

\citet{mann_et_al2017} conducted a similar comparison of spectroscopic and photometric temperature estimates. They found that the temperatures estimated via the Stefan-Boltzmann relation were consistent with the spectroscopic estimates determined by \citet{newton_et_al2015}, but that the \citet{newton_et_al2015} estimates displayed more scatter. As shown in the bottom right panel of Figure~\ref{fig:phot_spec_teff}, our photometric temperature estimates also exhibit a slightly tighter relation with $M_{Ks}$ than our spectroscopic estimates.

In the top center two panels of Figure~\ref{fig:phot_spec_teff}, we investigate the similarity between our spectroscopic temperature estimates and those estimated by the Gaia team using Apsis-Priam \citep{bailer-jones_et_al2013, andrae_et_al2018}. At all spectroscopic temperatures, our estimates tend to be lower than those estimated by the Gaia team. Specifically, we find that the median difference $\Delta T_{\rm eff} = T_{{\rm eff}, Gaia} -   T_{{\rm eff}, spec} = 198$K (5\%) for the 68~supposedly single stars with temperature estimates in Gaia~DR2. The standard deviation of the difference distribution is  $\sigma_{\Delta T_{\rm eff}} =266K$.

The top right two panels of Figure~\ref{fig:phot_spec_teff} compare our photometric temperature estimates to those estimated by the Gaia team using Apsis-Priam \citep{bailer-jones_et_al2013, andrae_et_al2018}. In the Apsis-Priam framework, $T{\rm eff}$ is estimated from observed brightness of the target star in the three \emph{Gaia} photometric bands assuming zero extinction. The estimates are determined using a machine-learning algorithm training on a set of stars with known temperatures and low extinctions. The Gaia temperature estimates are noticeably larger than our own photometric temperature estimates. For the 64 stars with temperature estimates from both methods and no evidence of stellar companions, the median difference $\Delta T_{\rm eff} = T_{{\rm eff}, phot} -   T_{{\rm eff}, Gaia} = -191$K (-5\%) and the standard deviation of the differences is $\sigma_{\Delta T_{\rm eff}} = 200$K. 

\citet{andrae_et_al2018} noted a similar offset between their temperature estimates and literature values for low-mass dwarfs ($\log g \gtrsim 4.8$). They proposed that the discrepancy might be due to temperature errors induced by the presence of strong molecular absorption in the broad-band integrated photometry of cool dwarfs or to the possible tendency of Apsis-Priam to overestimate the extinction and temperatures of cool dwarfs. Apsis-Priam assigns stellar parameters by using a machine learning algorithm trained on observations of real stars, most of which are much farther away than these cool dwarfs and therefore have higher extinction. Accordingly, we provide the Gaia $T_{\rm eff}$ estimates for comparison purposes only; we do not recommend using those values for cool dwarfs.

When compiling our final cool dwarf catalog, we select the photometric temperature estimates for all stars with reported parallaxes. For stars without parallaxes, we instead adopt the spectroscopic estimates. As previously noted by \citet{mann_et_al2017}, our spectroscopic and photometric temperature estimates are in agreement but the spectroscopic estimates display more scatter.

\subsection{Overall Comparison of Spectroscopic \& Photometric Estimates}
\label{ssec:comp}
\begin{figure*}[tbp]
\centering
\includegraphics[width=0.49\textwidth]{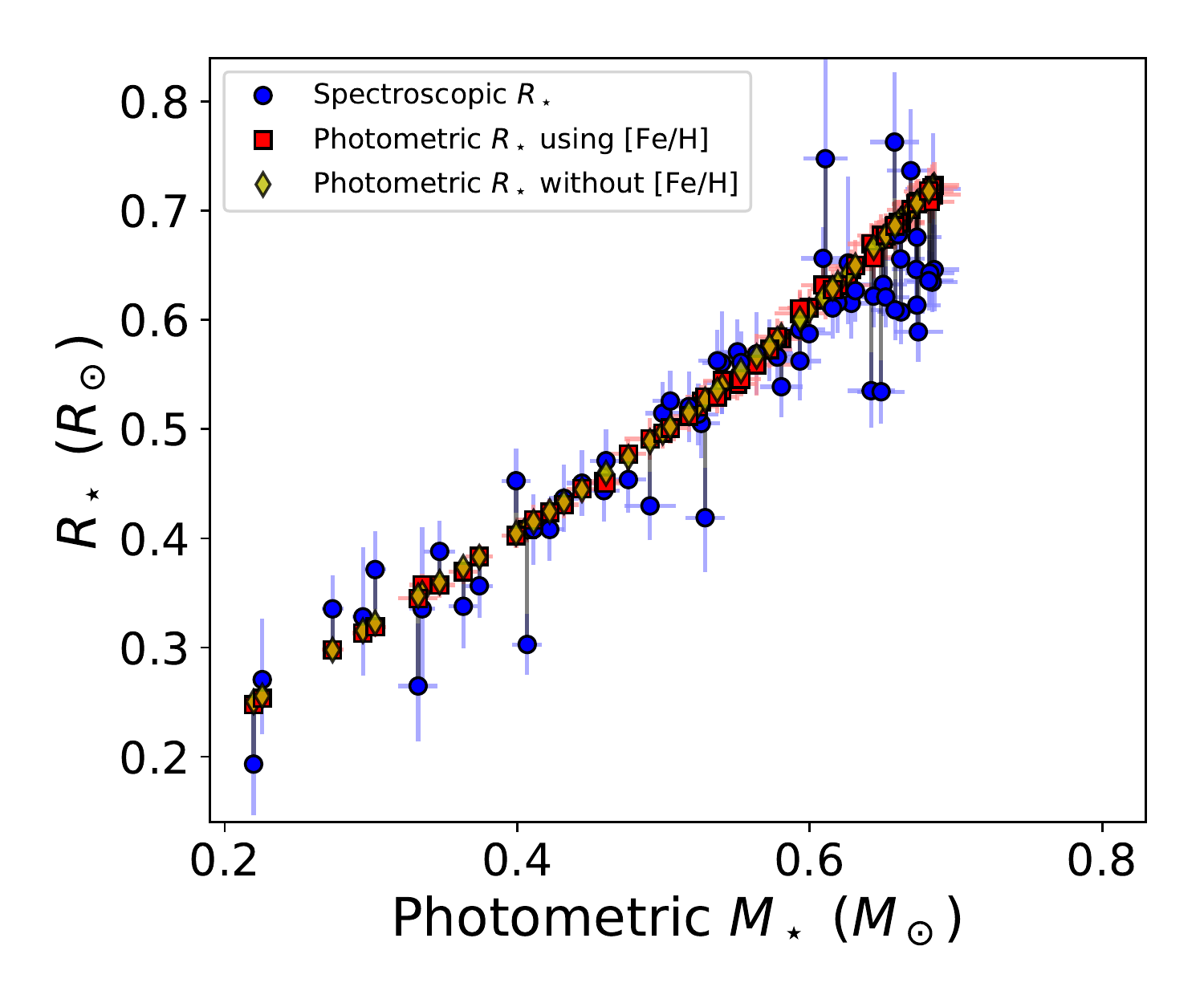}
\includegraphics[width=0.49\textwidth]{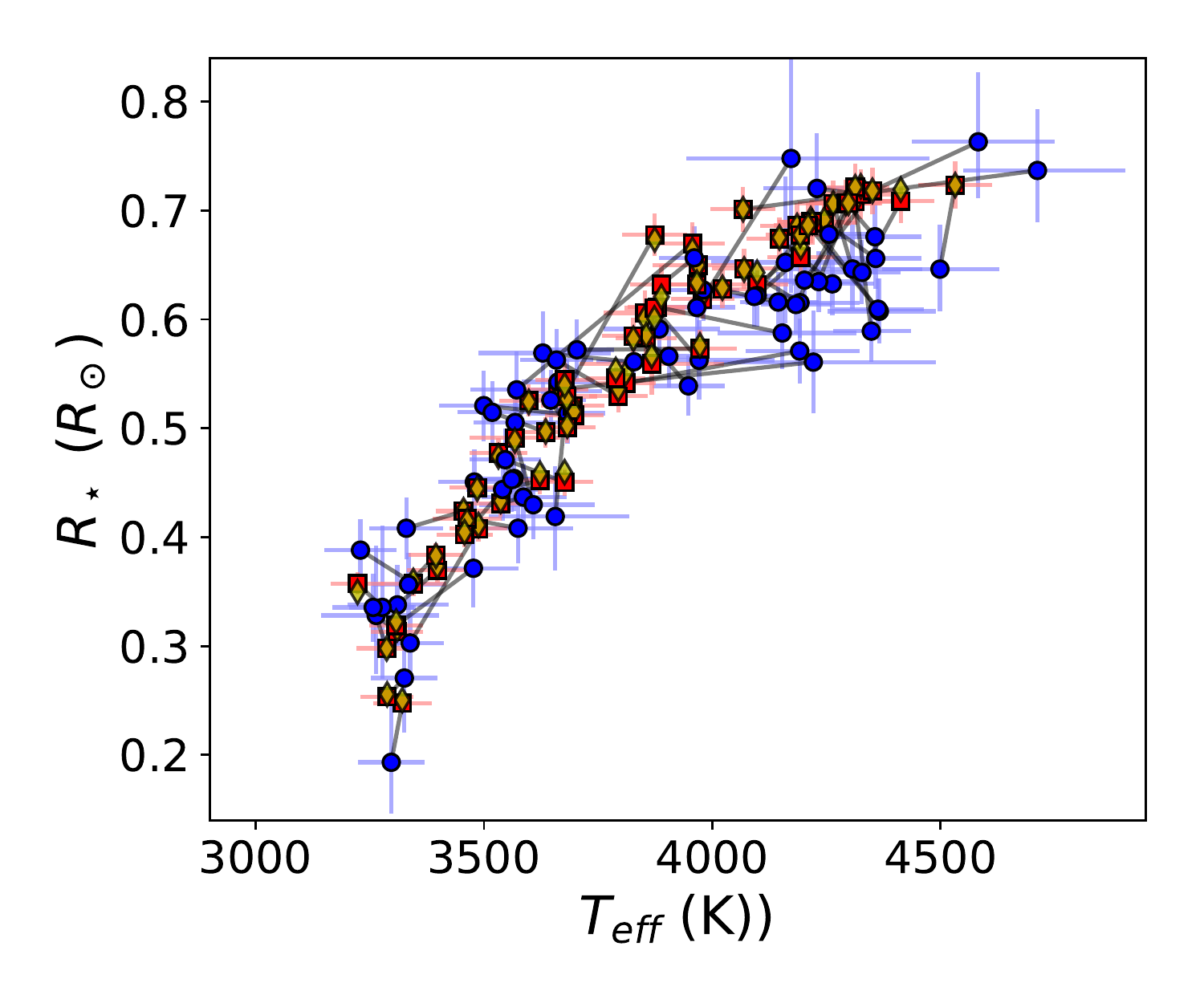}
\caption{Comparison of stellar parameters estimated from spectroscopy and photometry. \emph{Left: }  Radius versus mass for estimates based on spectroscopy (blue circles), photometry incorporating knowledge of [Fe/H] (red squares), and photometry without [Fe/H] constraints (yellow diamonds). The gray lines connect the spectroscopic and [Fe/H]-free photometric estimates for each star to the [Fe/H]-based photometric estimates. \emph{Right: } Radius versus stellar effective temperature.  \label{fig:joint_phot_spec}} 
\end{figure*}

In Figure~\ref{fig:joint_phot_spec}, we compare the stellar radii and effective temperatures resulting from the spectroscopic analysis in Section~\ref{ssec:details} and the photometric analysis in Section~\ref{ssec:gaia}. For clarity, we exclude the five stars identified as possible eclipsing binaries (EBs) and the seven stars with detected nearby companions in Gaia~DR2 or follow-up images. There are therefore 65~stars included in both the $R_\star - M_\star$ and the $R_\star - T_{\rm eff}$ panels. 

Our photometric radius and mass estimates are both primarily determined by $M_{Ks}$, leading the photometric estimates to follow a tight trend on the left panel of Figure~\ref{fig:joint_phot_spec}. In contrast, the spectroscopic estimates are more broadly dispersed. At the more massive end of the cool dwarf sample ($M_\star > 0.67\msun$), there is cluster of stars for which the spectroscopic radius estimates are roughly 10\% lower that the photometric estimates, indicating that the spectroscopic relations may systematically underestimate the radii of the most massive cool dwarfs.    

The difference between the spectroscopic and photometric estimates is less stark in the $R_\star - T_{\rm eff}$ plot displayed in the right panel of Figure~\ref{fig:joint_phot_spec}. Although our photometric estimates incorporate the spectroscopic [Fe/H] constraints when possible, the difference between [Fe/H]-dependent photometric estimates and [Fe/H]-free photometric estimates is much smaller than the overall difference between the photometric and spectrosopic estimates. The [Fe/H]-dependent radius estimates fall nearly on top of the [Fe/H]-free radius estimates: the median $\Delta R_\star =R_{\star, {\rm [Fe/H]}} - R_{\star, {no \rm [Fe/H]}}  = -0.002\rsun$ (-0.4\%) with a standard deviation of $0.005\rsun$. (This quoted difference was calculating using the 52~stars with spectral types of K5 or later; the [Fe/H]-based relations are not valid for K7 dwarfs.) 

Assuming that the [Fe/H]-dependent photometric estimates are the ``true'' values, the $M_{Ks}$-based photometric relations employed in Sections~\ref{sssec:photo_ms}--\ref{sssec:photo_teff} yield remarkably accurate and precise stellar properties even in the absence of [Fe/H] constraints. Accordingly, we are now using the combination of Gaia DR2 data \citep{gaia_et_al2018} and archival photometry from the KIC \citep{brown_et_al2011} and EPIC \citep{huber_et_al2016} to produce catalogs of updated properties for all \emph{K2} and \emph{Kepler} cool dwarfs (Schwab Abrahams, Dressing et al., \emph{in prep}).

For the remainder of the paper, we restrict the discussion to the 75~stars that have not been classified as likely eclipsing binaries and do not have candidate stellar companions within $1"$.
Nearly all of these stars (97\%) have parallaxes reported in Gaia~DR2. For the 73~stars with Gaia parallaxes, we adopt the photometric estimates as our preferred values for each star. These estimates incorporate our spectroscopic estimates of [Fe/H] for the 56~stars with Gaia parallaxes and spectral types of K7 or later and are agnostic to [Fe/H] for the 17~K5 dwarfs with Gaia parallaxes. For the remaining two~stars without Gaia parallaxes, we resort to our spectroscopic estimates, but replace the spectroscopic masses by those found by interpolating the mass - radius relation found for the photometric sample because of the discrepancy between photometric and spectroscopic masses (see Figure~\ref{fig:phot_spec_mass} and Section~\ref{sssec:photo_ms}). Even though we adopt photometric estimates when possible, our spectroscopic characterization was important for determining spectral types, estimating stellar metallicities, and identifying close stellar binaries. 

Accounting for the validity ranges of the various photometric relations, our sample includes 70 cool dwarfs with photometric radius and mass estimates and five with spectroscopic radius estimates based on the \citet{newton_et_al2014, newton_et_al2015} relations and masses estimated by placing the spectroscopic radii on the photometric mass-radius relation. We adopt the photometric luminosities for 43~cool dwarfs and report spectroscopic estimates based on \citet{newton_et_al2014, newton_et_al2015} for the remaining 32~cool dwarfs. All stars have photometric temperature estimates based on the relations from \citet{mann_et_al2015}. 

\begin{figure}[tbp]
\centering
\includegraphics[width=0.5\textwidth]{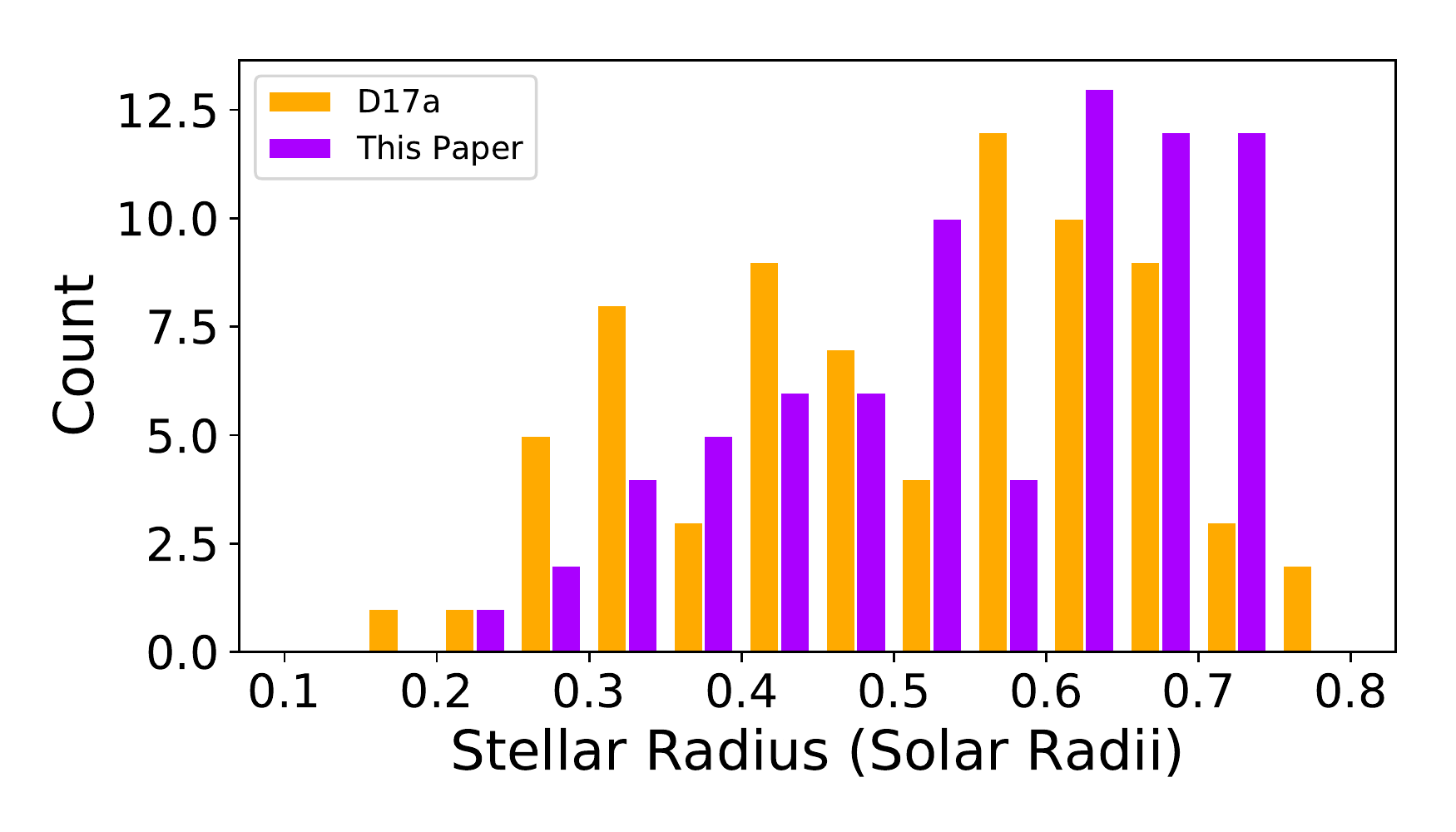}
\includegraphics[width=0.5\textwidth]{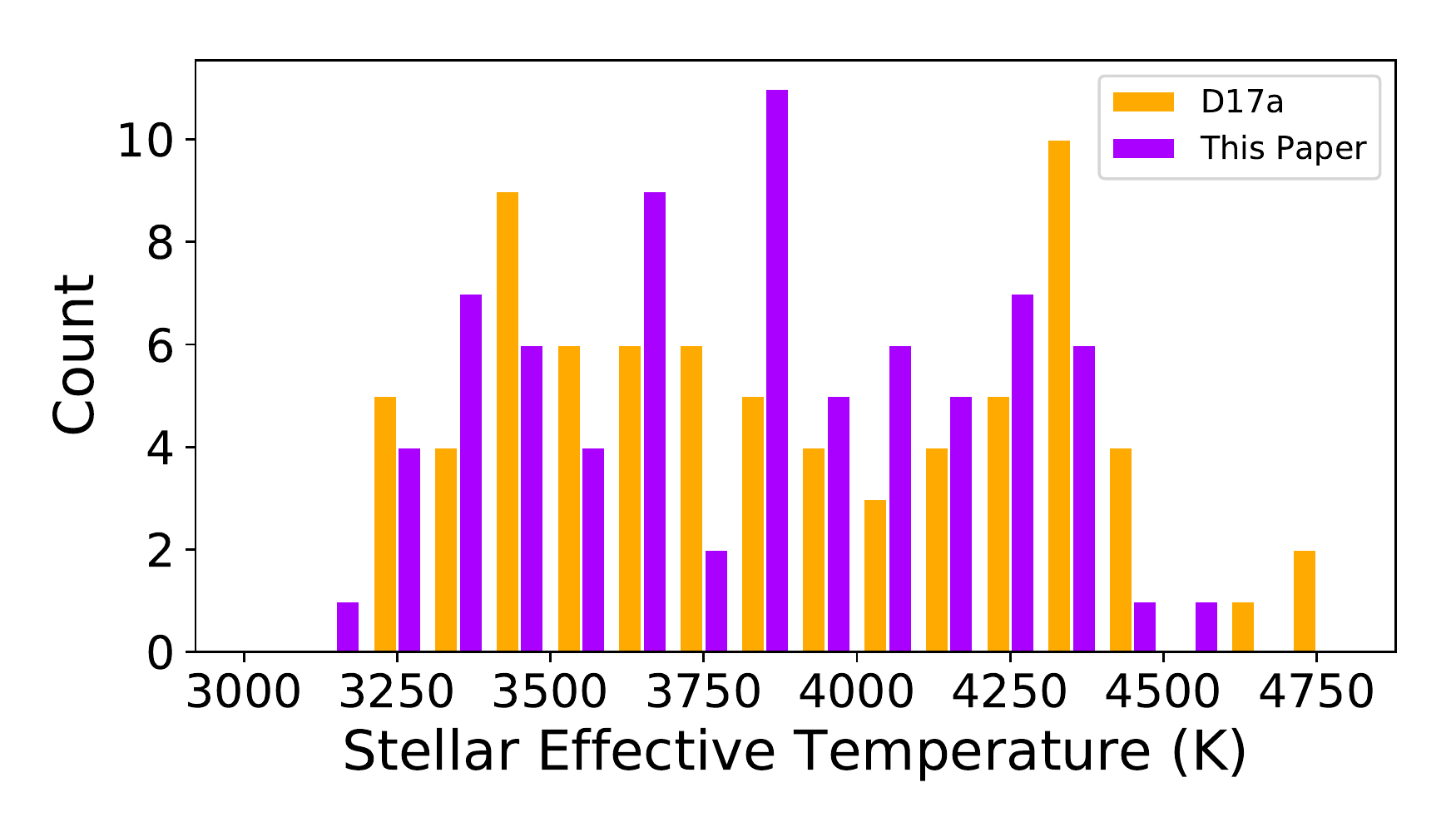}\\
\caption{Distribution of radii (top) and effective temperatures (bottom) for the cool dwarfs analyzed in this paper (purple) compared to those previously characterized in \citet[][orange]{dressing_et_al2017a}. \label{fig:rs_teff_comp}} 
\end{figure}

We list the adopted parameters for all 75~cool dwarfs presumed to be single in Table~\ref{tab:adopted} and display the resulting distribution of stellar radii and effective temperatures in Figure~\ref{fig:rs_teff_comp}. The radii range from $0.24\rsun$ to $0.74\rsun$ with a median value of $0.58\rsun$ and the stellar effective temperatures extend from 3178K to 4531K with a median value of 3851K. Compared to the sample of cool dwarfs we characterized in \citet{dressing_et_al2017a}, this sample is shifted toward higher radii and cooler stellar effective temperatures. The offset is partially due to our use of spectroscopic estimates in \citet{dressing_et_al2017a} and predominantly photometric estimates in this paper as well as sample selection effects influencing both the original \emph{K2} target lists and the sample of stars for which we obtained follow-up observations. 

In summary, we reached the following conclusions from comparing the spectroscopic and photometric stellar parameters calculated in this paper to those reported in Gaia~DR2:
\begin{itemize}
\item{Our photometric estimates of stellar luminosity are consistent with those reported in Gaia~DR2 \citep{gaia_et_al2018}.}
\item{Relative to our photometric estimates, our spectroscopic luminosities are roughly $0.03\lsun$ brighter for the brightest stars ($L_{\star, spec} > 0.13 \lsun$).}
\item{Our photometric and spectroscopic estimates of stellar radius agree well. Across our full cool dwarf sample, the median radius difference is only $0.02\rsun$, with the photometric estimates slightly larger than the spectroscopic estimates.}
\item{The stellar radii reported in Gaia~DR2 are systematically offset from our spectroscopic and photometric estimates. Compared to our photometric estimates, the Gaia estimates are roughly $0.04 \rsun$ smaller for stars with $R_{\star, phot}  > 0.67\rsun$.}
\item{Our photometric and spectroscopic mass estimates are correlated, but our spectroscopic estimates are smaller than our photometric estimates for the least massive stars and larger than our photometric estimates for the most massive stars. The discrepancy is roughly $M_{\star, phot}  - M_{\star, spec} = 0.11\msun$ for stars with \mbox{$M_{\star, phot} < 0.4 \msun$} and $-0.02\msun$ for stars with \mbox{$M_{\star, phot} > 0.6 \msun$}.}
\item{Our photometric and spectroscopic temperature estimates agree well (median difference of \mbox{$T_{{\rm eff}, phot} -   T_{{\rm eff}, spec} = -3$K}), but the temperatures reported in Gaia~DR2 are roughly 200K higher than our estimates.}
\end{itemize}

\section{Discussion}%
\label{sec:discussion}

\begin{figure*}[tbph]
\centering
\includegraphics[width=0.49\textwidth]{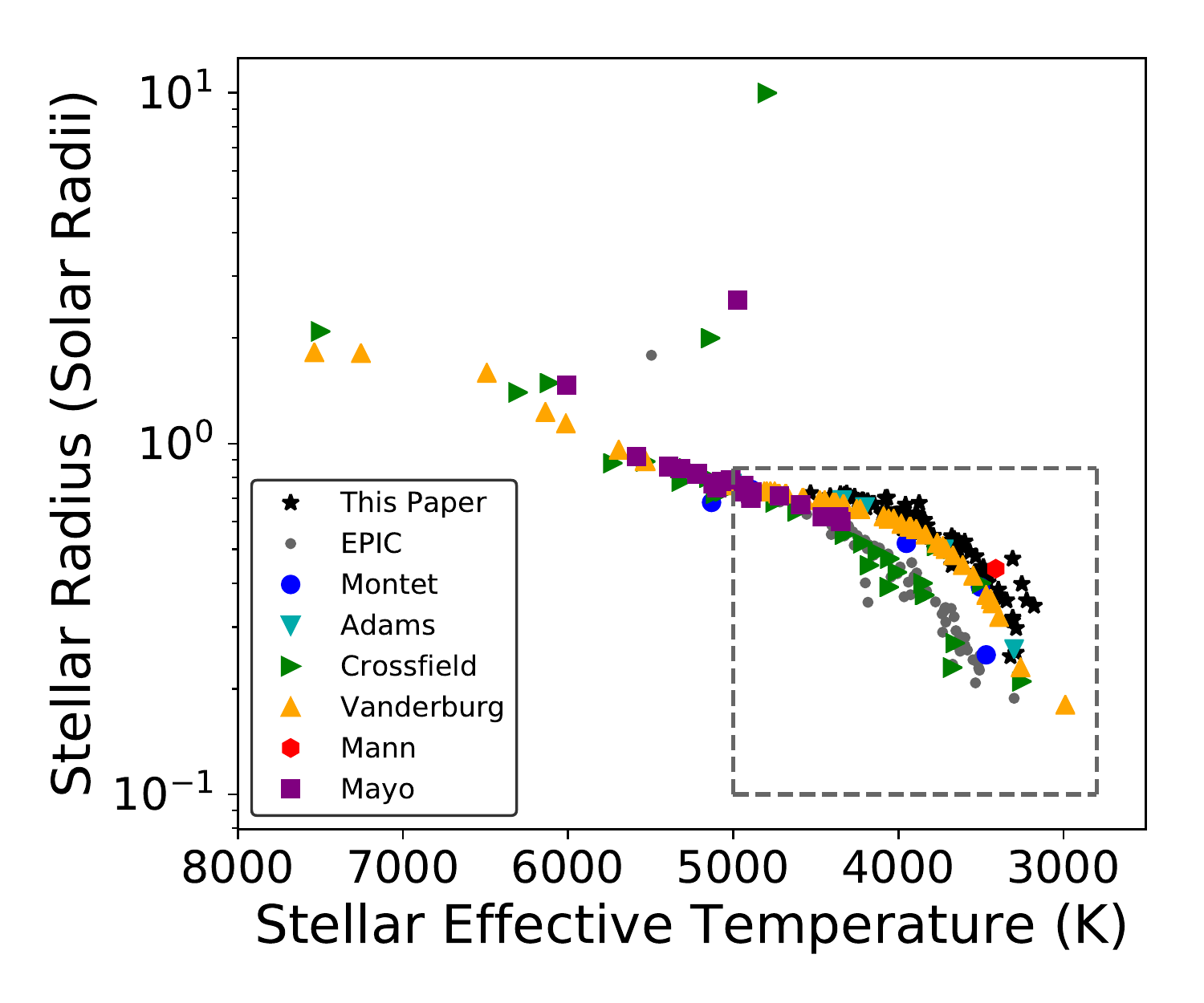}
\includegraphics[width=0.49\textwidth]{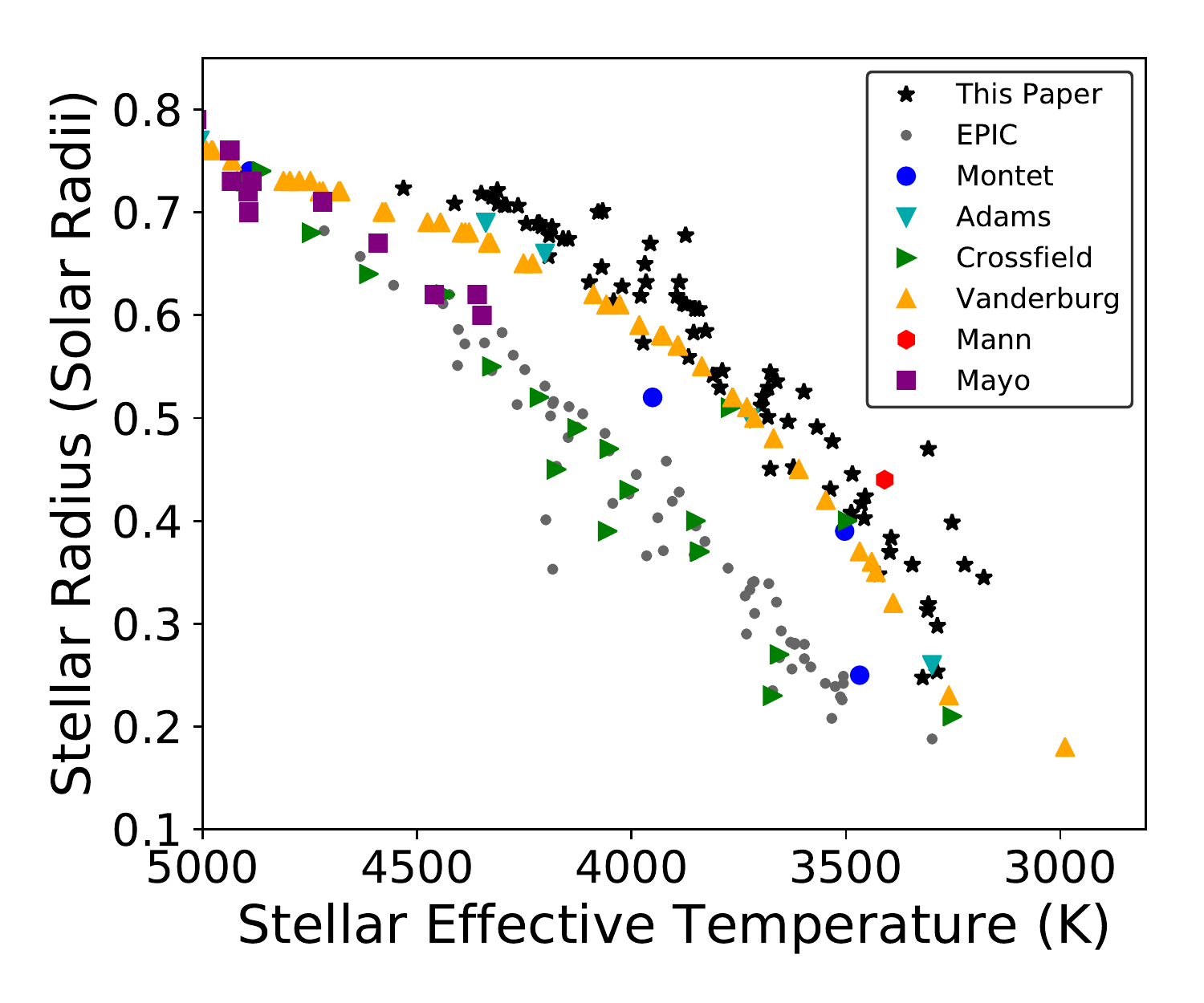}\\
\includegraphics[width=0.49\textwidth]{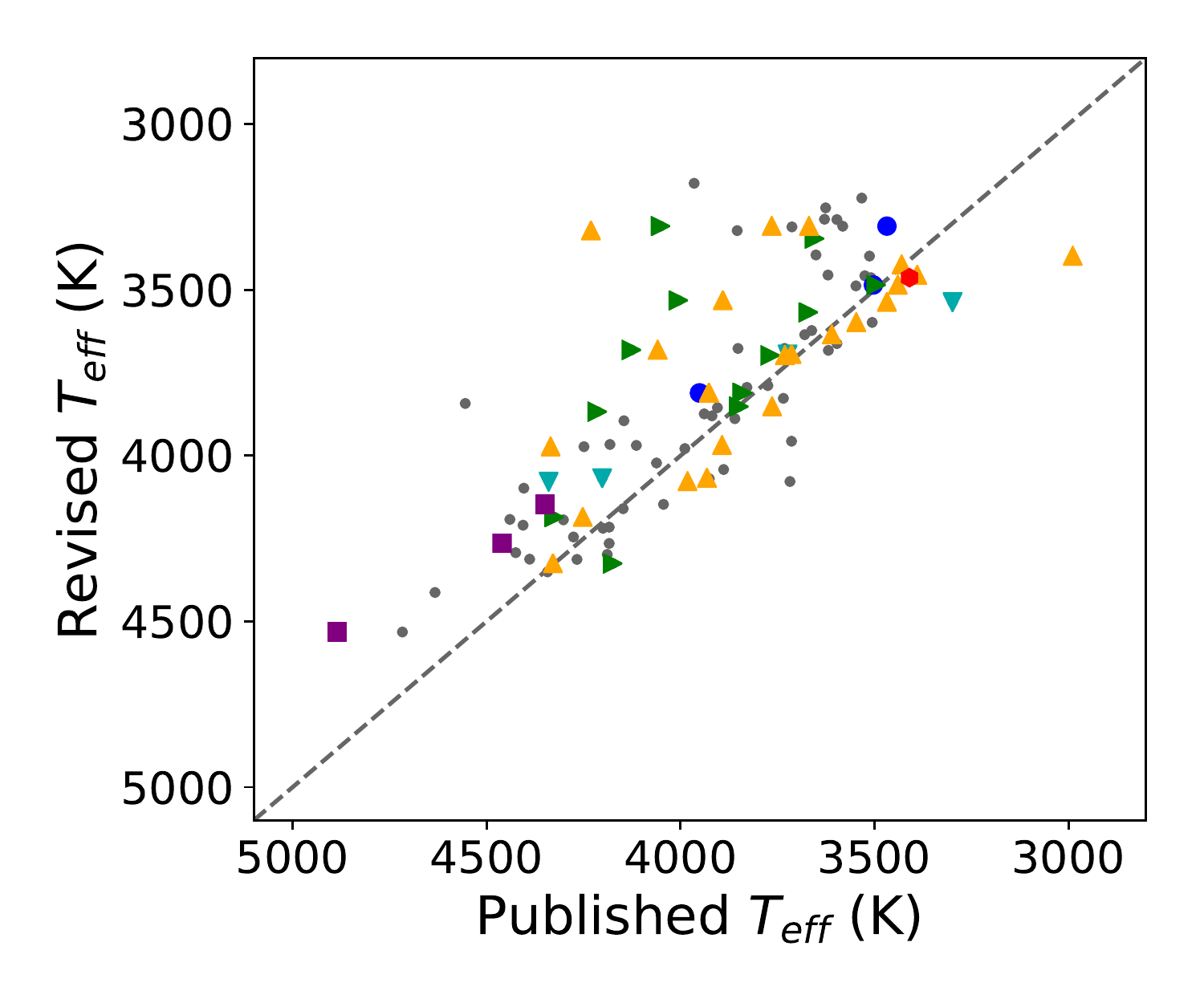}
\includegraphics[width=0.49\textwidth]{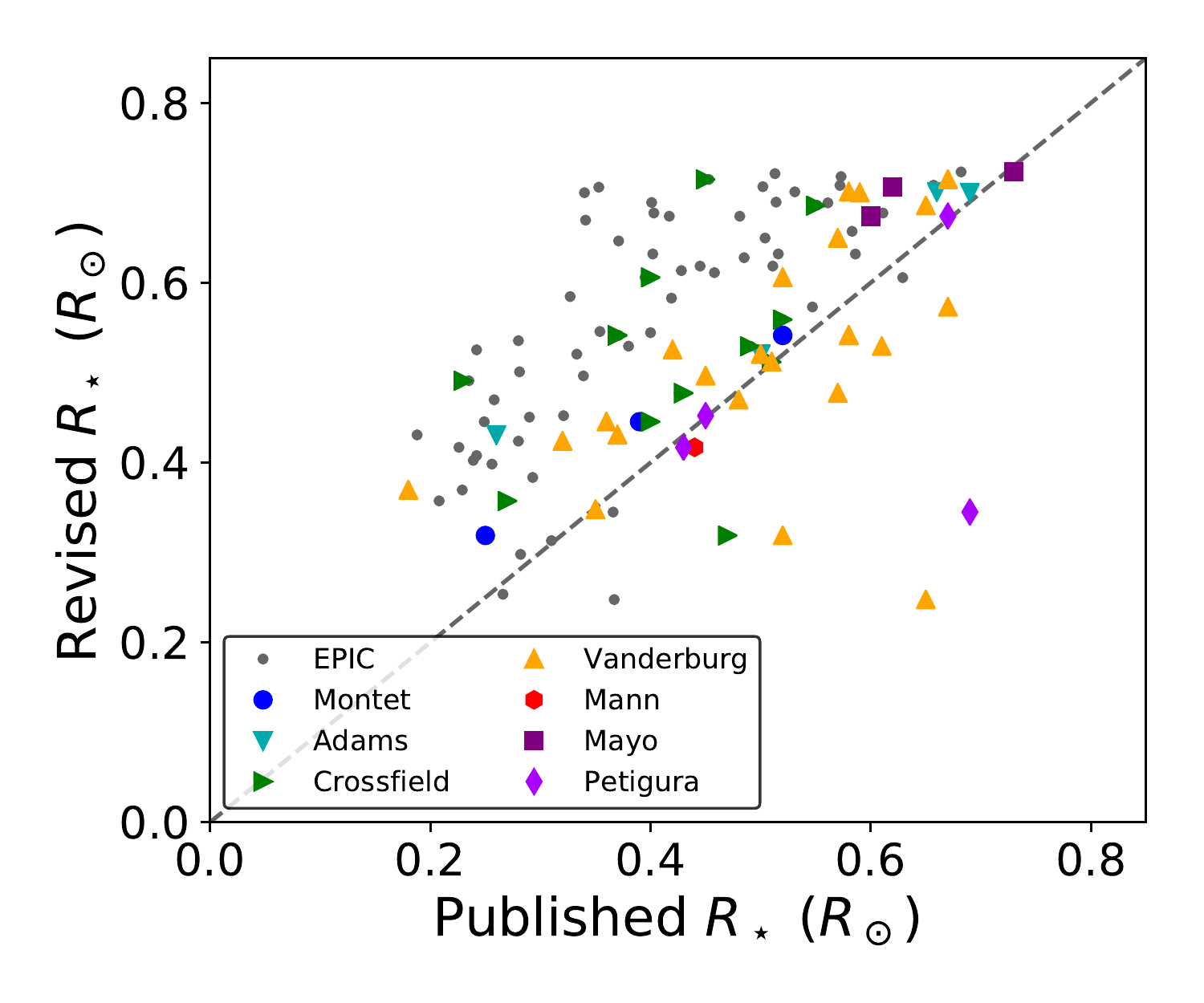}\\
\caption{Comparison of our revised stellar parameters to earlier estimates from other studies. \emph{Top:} Stellar radius versus effective temperature. Solid lines connect our revised estimates (black stars) to the the earlier estimates (colors) for each star. The right panel shows a zoomed-in view of the boxed region shown in the left panel. The errorbars are omitted from this figure for clarity; consult Figure~\ref{fig:bypaper} to see the errorbars.  \emph{Bottom Left: } Revised stellar effective temperatures versus previously published values. \emph{Bottom Right: } Revised stellar radii versus previously published values.  \label{fig:comparison}} 
\end{figure*}

\begin{figure*}[tbph]
\centering
\includegraphics[width=0.49\textwidth]{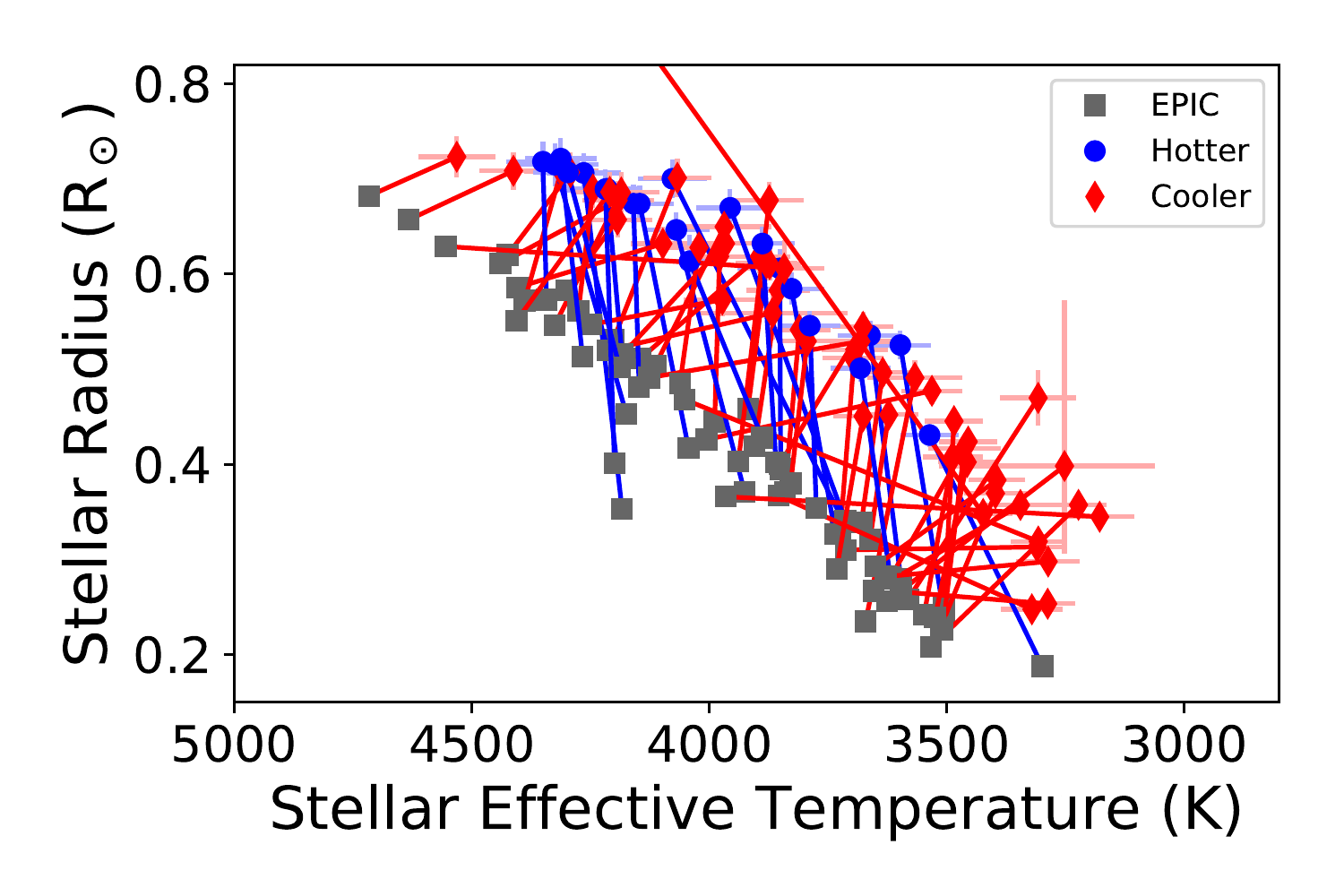}
\includegraphics[width=0.49\textwidth]{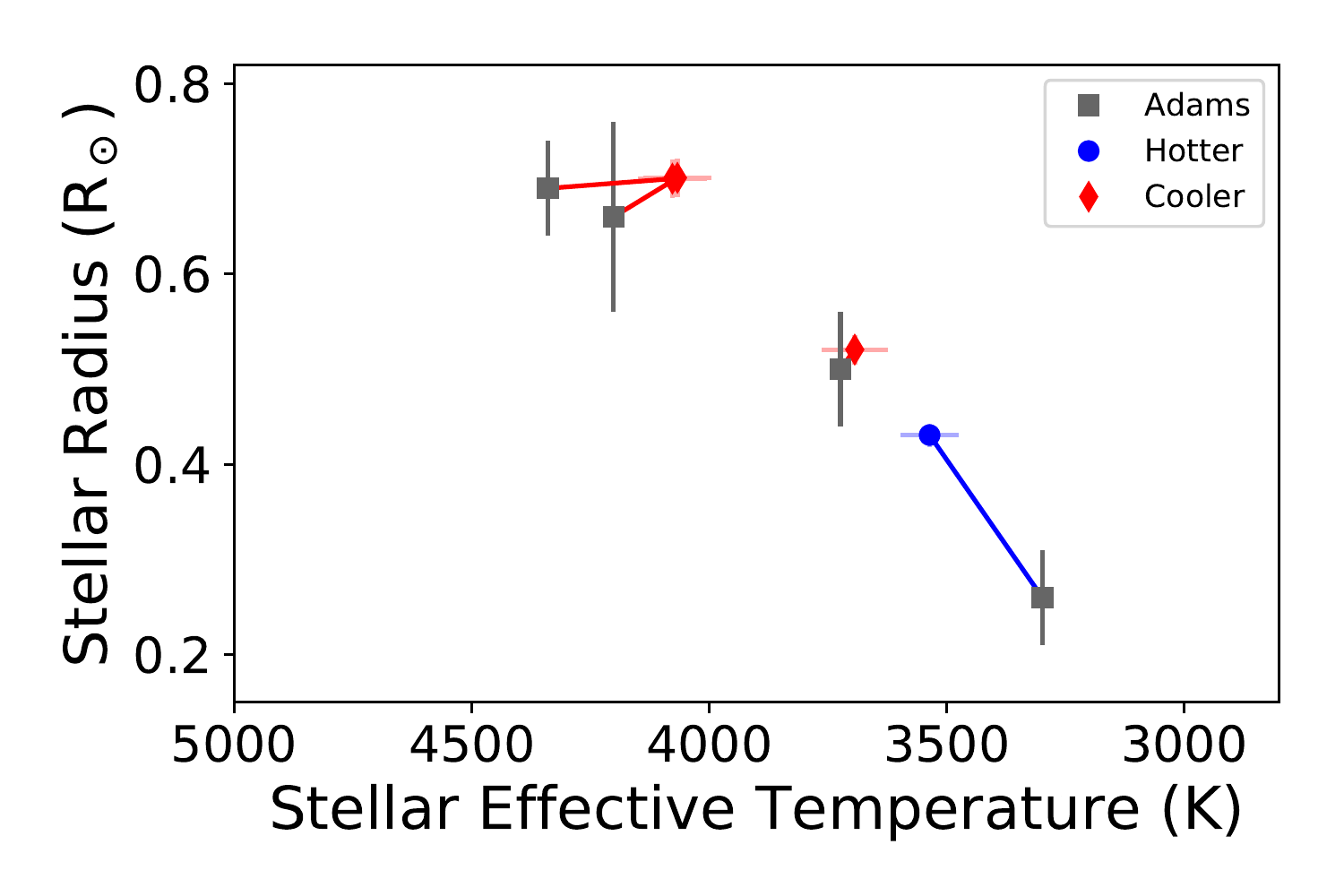}
\includegraphics[width=0.49\textwidth]{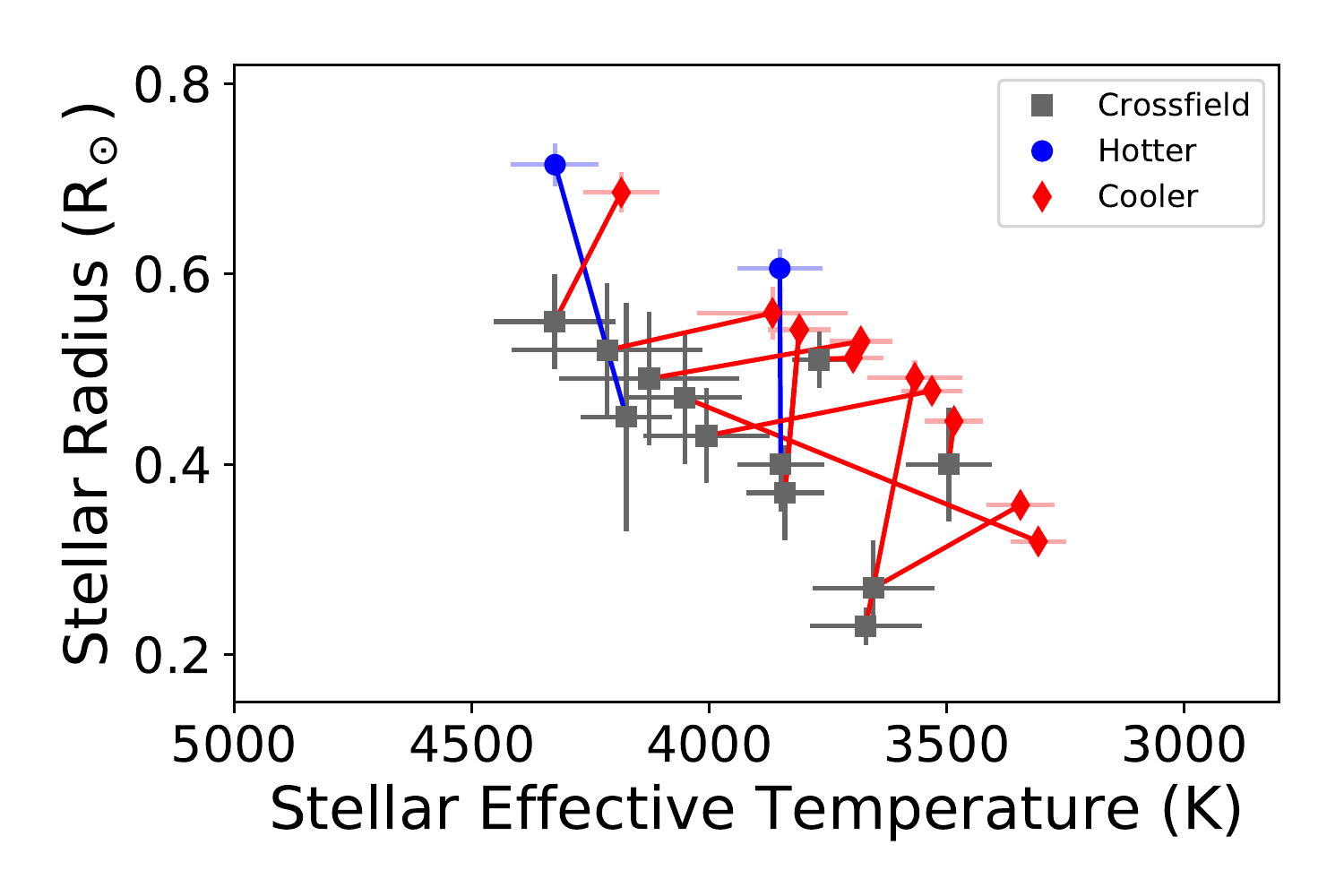}
\includegraphics[width=0.49\textwidth]{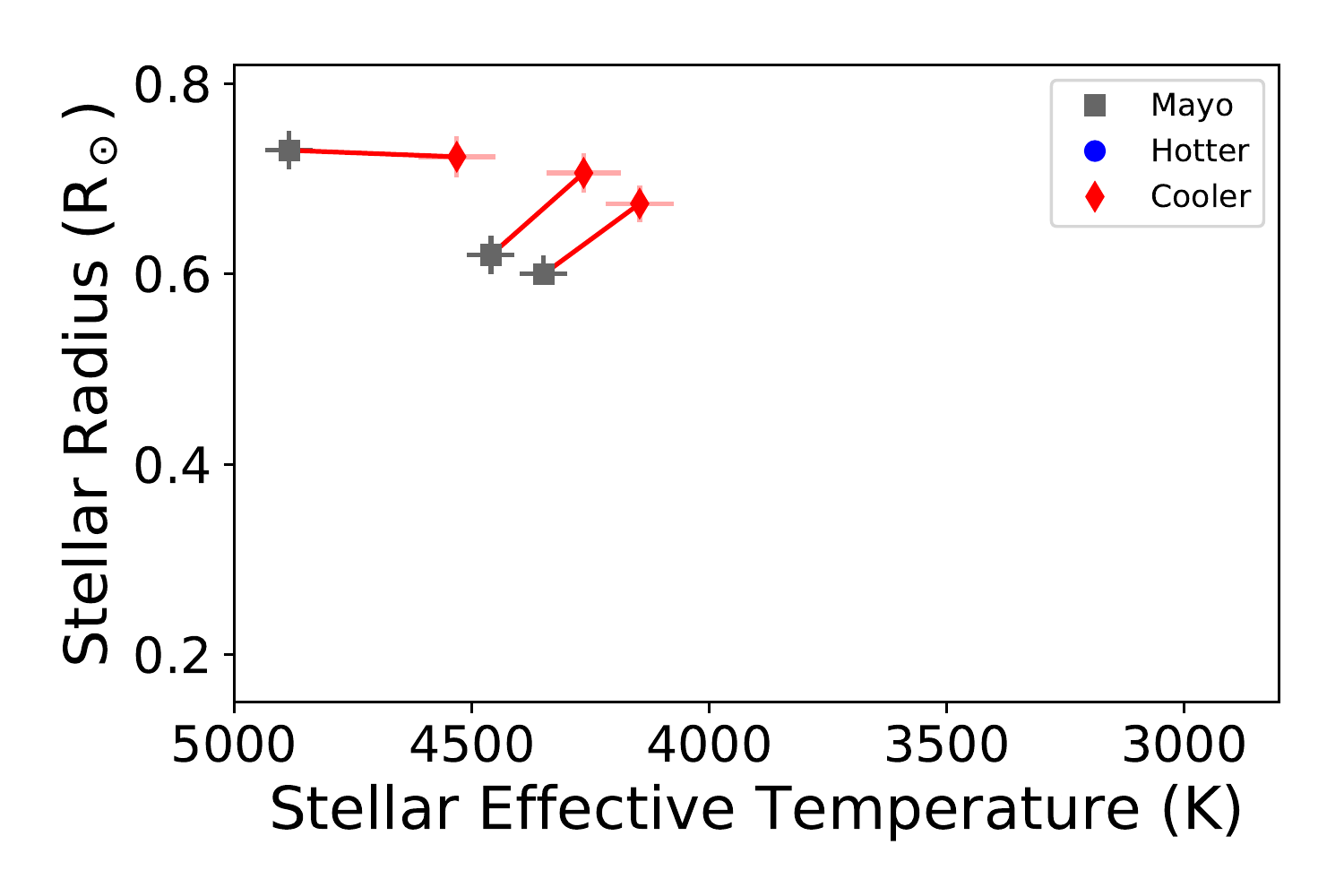}
\includegraphics[width=0.49\textwidth]{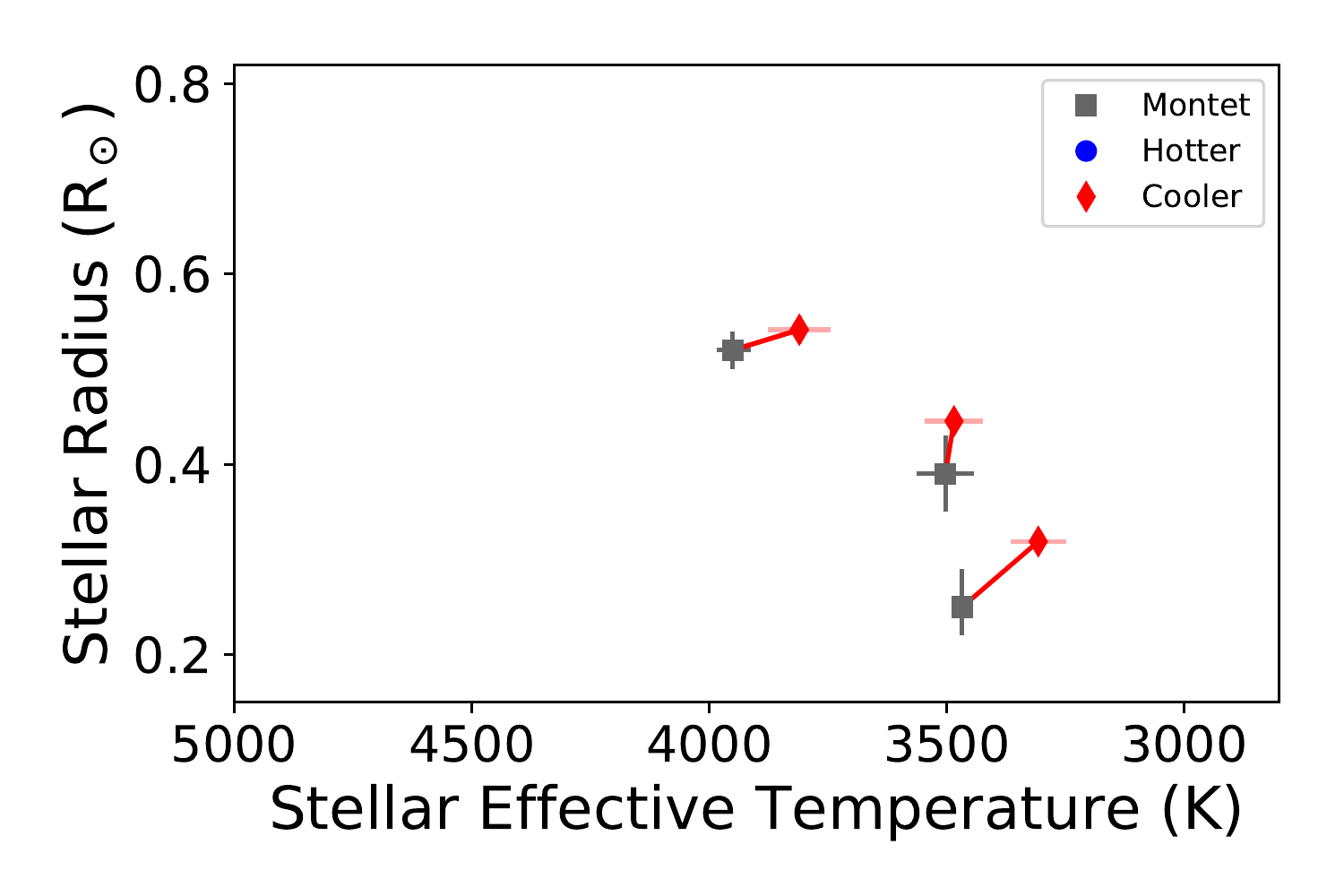}
\includegraphics[width=0.49\textwidth]{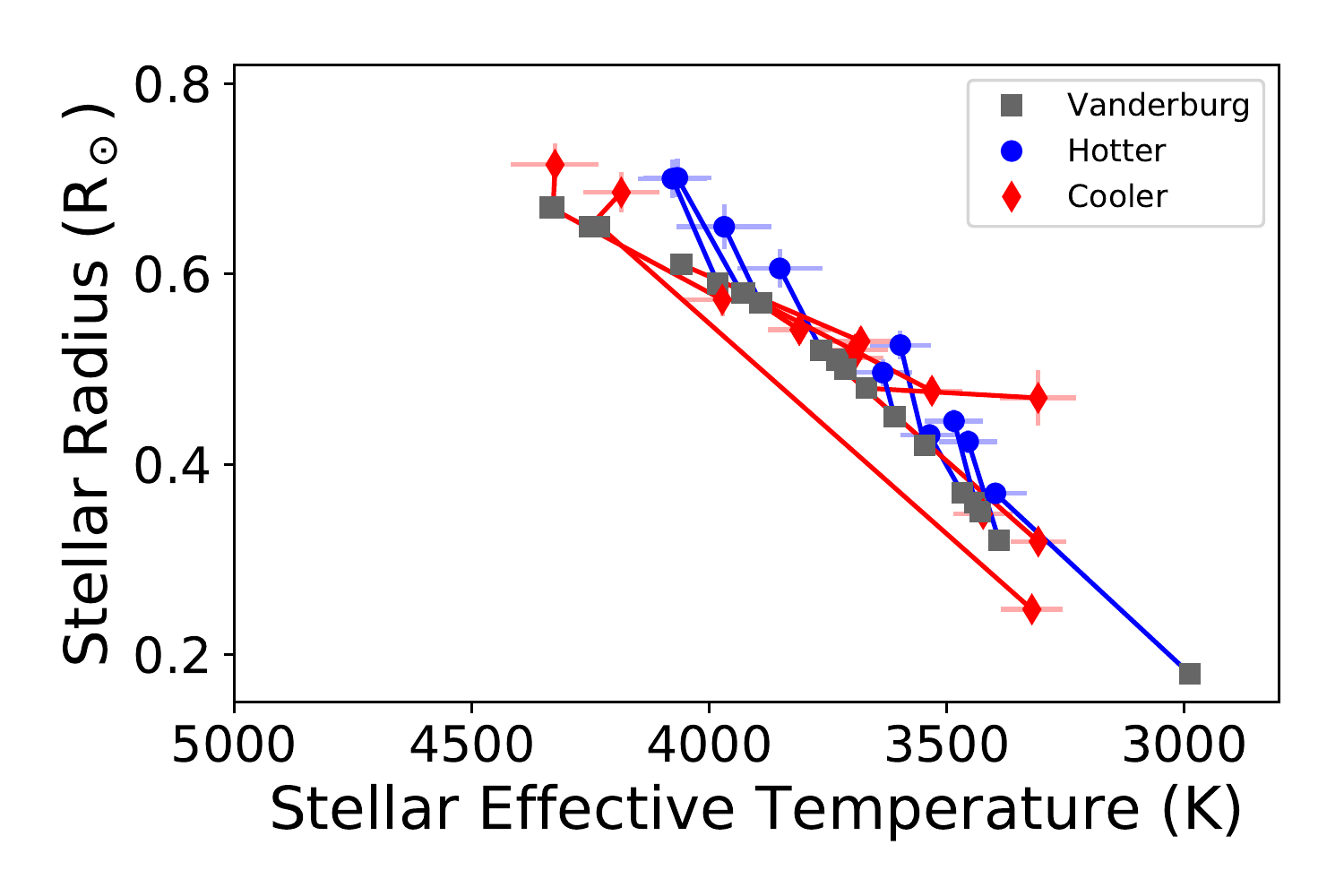}
\caption{Comparison of our revised stellar parameters to earlier estimates from other studies.  Solid lines connect our revised estimates (black stars) to the the earlier estimates (colors) for each star. The revised values are blue if the temperature estimate has increased and red if the temperature estimate has decreased. \emph{Top Left:} Comparison to values published in the Ecliptic Plane Input Catalog \citep[EPIC][]{huber_et_al2016}. \emph{Top Right:} Comparison to \citet{adams_et_al2016}. \emph{Middle Left:} Comparison to \citet{crossfield_et_al2016}. \emph{Middle Right:} Comparison to \citet{mayo_et_al2018}. \emph{Bottom Left:} Comparison to \citet{montet_et_al2015}. \emph{Bottom Right:} Comparison to \citet{vanderburg_et_al2016}.  \label{fig:bypaper}} 
\end{figure*}

As mentioned in Section~\ref{sec:targets}, we observed these stars because they were initially identified as candidate cool dwarfs. In Figures~\ref{fig:comparison} and \ref{fig:bypaper}, we compare our revised stellar parameters to earlier estimates from the Ecliptic Plane Input Catalog \citep[EPIC, ][]{huber_et_al2016} and previous studies. Several of the earlier planet catalogs did not estimate host star parameters \citep{barros_et_al2016, pope_et_al2016, schmitt_et_al2016, rizzuto_et_al2017}. The top and left panels of Figure~\ref{fig:comparison} contrast our new estimates of the stellar effective temperature with those previously estimated by \citet{montet_et_al2015, adams_et_al2016, crossfield_et_al2016, vanderburg_et_al2016, mann_et_al2017b, mayo_et_al2018}. For the stellar radius comparison (bottom right panel), we include past estimates from those six studies as well as \citet{petigura_et_al2018}. 

Figure~\ref{fig:comparison} clearly shows that our estimated stellar radii are significantly larger than the radii estimated in previous studies. The $R_\star - T_{\rm eff}$ relation traced out by our revised parameters has a similar shape to the relations assumed by \citet{crossfield_et_al2016}, \citet{huber_et_al2016}, and \citet{vanderburg_et_al2016}, but our results are shifted toward larger radii and cooler temperatures. The temperature offset is readily apparent in the $T_{\rm eff}-T_{{\rm eff}, pub}$ plot of Figure~\ref{fig:comparison}: the majority of the previous estimates are roughly 200K hotter than our revised estimates. The scatter is larger on the accompanying $R_\star-R_{\star, pub}$ plot, but there is a clear excess of stars with previously underestimated radii. The tendency for models to underpredict the radii of cool stars has been well-established in past studies \citep[e.g.,][]{boyajian_et_al2012, zhou_et_al2014, newton_et_al2015, mann_et_al2017b} and is unsurprising. 

For instance, \citet{boyajian_et_al2012} found that cool dwarf radii predicted by the Dartmouth models are roughly 10\% too small at a given temperature and \citet{mann_et_al2017b} found that the model radius of Kepler-42 (a 3269K cool dwarf hosting three transiting planets) was 6\% too small. Similarly, \citet{zhou_et_al2014} found tentative evidence that stellar models underpredict the radii of cool dwarfs by roughly 5\%. In addition, \citet{newton_et_al2015} measured the radii of \emph{Kepler} cool dwarfs using the spectroscopic methods employed in Section~\ref{ssec:details}. \citet{newton_et_al2015} found that their spectroscopic radius estimates were typically $0.09\rsun$ larger than the radii determined by \citet{dressing+charbonneau2013} by fitting photometry to Dartmouth models.

For the 69~purportedly single stars in our cool dwarf sample, Figure~\ref{fig:bypaper} reveals that nearly all of our revised radius estimates are larger than those published by \citet{huber_et_al2016} in the EPIC. Overall, the median change in the estimated stellar radius is $0.15\rsun$ (40\%). In addition, our revised temperature estimates are typically 65K cooler than the EPIC values. The difference between our estimated radii and the EPIC radii is larger than the discrepancy reported in previous studies, but the bottom right panel of Figure~\ref{fig:comparison} reveals that we measure smaller offsets of roughly 5\% between our estimates and those reported by other previous studies \citep[e.g., ][]{adams_et_al2016, vanderburg_et_al2016}.

For their \emph{K2} catalog, \citet{montet_et_al2015} estimated stellar properties by using the {\tt isochrones}\footnote{\url{http://github.com/timothydmorton/isochrones}} Python module \citep{morton2015} to identify the stellar models in the Dartmouth Stellar Evolution Database \citep{dotter_et_al2008} that were most consistent with the archival photometry for each star. \citet{dressing_et_al2017a} contained five~stars from \citet{montet_et_al2015}; this cool dwarf sample includes three~stars from \citet{montet_et_al2015}. Compared to the values published by \citet{montet_et_al2015}, we find that our revised radii are fairly consistent but that our temperature estimates differ by 18--160K. Our estimated $T_{\rm eff}$ are 140~K cooler for EPIC~201367065, 160~K cooler for EPIC~201465501, and 18~K cooler for EPIC~201912552.

\citet{adams_et_al2016} adopted stellar effective temperatures from the \emph{K2}-TESS Stellar Properties Catalog for most candidates. For candidates identified in Campaign~4, \citet{adams_et_al2016} estimated temperatures from spectra they acquired using the Tull Coud\'{e} spectrograph  \citep{tull_et_al1995} at the Harlan J. Smith \mbox{2.7-m} telescope at McDonald Observatory. They estimated the radii of their targets using the radius-$T_{\rm eff}$ relations established by \citet{boyajian_et_al2012}. Four of the stars in our cool dwarf sample were previously published by \citet{adams_et_al2016}. Compared to the estimates published by \citet{adams_et_al2016}, our estimated stellar effective temperatures are between 240K hotter and 260K cooler with a median temperature difference of 82~K cooler. Our estimated radii are $0.02\rsun  - 0.17\rsun$ larger with a median difference of $0.03\rsun$ (5\%).

For most targets, \citet{crossfield_et_al2016} determined $T_{\rm eff}$, log~$g$, and [Fe/H] by using {\tt SpecMatch} to analyze spectra they obtained using the HIRES echelle spectrometer \citep{vogt_et_al1994} on the 10-m Keck~I telescope, the Levy spectrograph  \citep{vogt_et_al2014} at the Automated Planet Finder, and the FEROS echelle spectrograph \citep{kaufer+pasquini1998} at the 2.2-m MPG telescope. They then determined masses and radii using by the {\tt isochrones} Python package \citep{morton2015}. A subset of the stars in the \citet{crossfield_et_al2016} catalog did not have {\tt SpecMatch} parameters. Those stars were assigned the stellar parameters from \citet{huber_et_al2016} if they were included in the EPIC or from {\tt isochrones} fits to broadband photometry from APASS, 2MASS, and \emph{WISE} for stars not in the EPIC. 

Like \citet{montet_et_al2015}, \citet{crossfield_et_al2016} used Dartmouth stellar models \citep{dotter_et_al2008} for the {\tt isochrones} analysis and therefore also underestimated the radii of cool dwarfs. In \citet{dressing_et_al2017a}, we found that our revised radius estimates were typically  28\% ($0.10\rsun$) larger than the radii reported by \citet{crossfield_et_al2016}. The cool dwarf sample in this paper includes 12~stars from \citet{crossfield_et_al2016}. As in our 2017 paper, our radius estimates are typically $0.11\rsun$ (28\%) larger. In addition, our temperature estimates are roughly 87K cooler than the \citet{crossfield_et_al2016} estimates. The radius and temperature changes across the sample are relatively uniform, with nearly all stars moving upward and toward the right to larger radii and cooler temperatures. 

 \citet{vanderburg_et_al2016} report a mix of spectroscopic and photometric parameters for their targets. For stars with spectroscopic estimates, \citet{vanderburg_et_al2016} obtained optical spectra with the Tillinghast Reflector Echelle Spectrograph (TRES) on the 1.5-m telescope at Fred L. Whipple Observatory and analyzed the spectra using the Stellar Parameter Classification \citep[SPC][]{buchhave_et_al2012, buchhave_et_al2014} method. For stars without TRES spectra, \citet{vanderburg_et_al2016} estimated stellar effective temperatures using a variety of color-temperature relations. Their preferred relation was the $V-K$ relation from \cite{boyajian_et_al2013} but they defaulted to the $B-V$ or $g-r$ relations from \citet{boyajian_et_al2013} or the $J-K$ relation from \citet{gonzalez-hernandez+bonifacio2009} when necessary. For stars with colors beyond the validity range of those relations, \citet{vanderburg_et_al2016} instead estimated temperatures by consulting the spectral type tables published by \citet{pecaut+mamajek2013} or by applying the $V-K$ color-temperature relation from \citet{casagrande_et_al2008} for the reddest stars. For stars cooler than 5778K, \citet{vanderburg_et_al2016} then estimated stellar radii by applying the temperature-radius relationships from \citet{boyajian_et_al2012}. \citet{dressing_et_al2017a} contained 9~stars from \citet{vanderburg_et_al2016}; we found that our revised radius estimates were 8\% ($0.05\rsun$) larger than the radii reported by \citet{vanderburg_et_al2016}. 
 
The cool dwarf sample in this paper contains 22~stars from \citet{vanderburg_et_al2016}. The median changes between our revised parameters and those published by \citet{vanderburg_et_al2016} are that our radius estimates are $0.03\rsun$ (5\%) larger and our stellar effective temperatures are 92~K hotter. Although most stars slightly upward to larger radii and moderately different temperatures, three stars (EPIC~201465501 (K2-9), EPIC~206032309, and EPIC~206215704) have extremely different parameter estimates in this paper than in \citet{vanderburg_et_al2016}.
 
 The M3~dwarf EPIC~201465501 (K2-9) was previously estimated by \citet{vanderburg_et_al2016} to have \mbox{$R_{\star, pub} = 0.52\rsun$} and \mbox{$T_{{\rm eff}, pub} = 3765$K}, but the revised radius is 40\% smaller \mbox{($R_\star = 0.32 \pm 0.01 \rsun$)} and the revised temperature \mbox{$T_{\rm eff} = 3308 \pm 58$K} is 457K cooler. These revised estimates are consistent with the earlier classification by \citet{schlieder_et_al2016} of K2-9 as an M2.5V$ \pm 0.5$ star with $T_{\rm eff} = 3390 \pm 150$K and $R_\star = 0.31 \pm 0.11 \rsun$ and close to the values of $T_{\rm eff} = 3468^{+20}_{-19}$K and $R_\star = 0.25^{+0.04}_{-0.03} \rsun$ reported by \citet{montet_et_al2015} in the discovery paper. Our revised estimates are also consistent with the constraints of $T_{\rm eff} = 3460 \pm 164$K and $R_\star = 0.366 \pm 0.053 \rsun$ published by \citet{martinez_et_al2017}. 
 
EPIC~206032309, an M2 dwarf at a distance of $161 \pm 1.8$~pc, was initially estimated to have  \mbox{$R_{\star, pub} = 0.18\rsun$} and \mbox{$T_{{\rm eff}, pub} = 2989$K}, but our analysis suggests that the star is much hotter and larger (\mbox{$R_\star = 0.37 \pm 0.01 \rsun$},  \mbox{$T_{\rm eff} = 3398 \pm 66$K}). Finally, we found that both the temperature and the radius were significantly overestimated for the M4~dwarf EPIC~206215704: the published  values were  \mbox{$R_{\star, pub} = 0.65\rsun$} and \mbox{$T_{{\rm eff}, pub} = 4231$K} while we find \mbox{$R_\star = 0.25 \pm 0.01 \rsun$} and  \mbox{$T_{\rm eff} = 3321 \pm 65$K}.
 
 \citet{mann_et_al2017b} classified their target stars by acquiring optical spectra with the SuperNova Integral Field Spectrograph \citep[SNIFS][]{aldering_et_al2002, lantz_et_al2004} on the University of Hawai'i 2.2-m telescope on Mauna Kea and near-infrared spectra with SpeX on the IRTF and the Immersion Grating Infrared Spectrometer \citep[IGRINS][]{park_et_al2014, mace_et_al2016}. They then confirmed that the stars were members of Praesepe and determined stellar effective temperatures by comparing their dereddened spectra to a grid of BT-SETTL CIFIST stellar models \citep{allard_et_al2012}. Next, they estimated bolometric fluxes of by comparing their spectra to archival photometry and determined stellar radii using the Stefan-Boltzmann relation. Finally, \citet{mann_et_al2017b} used the mass-$M_K$ relation they established in \citet{mann_et_al2015} to determine stellar masses. EPIC~211916756 (K2-95) is the only star from \citet{mann_et_al2017b} included in this paper. Our estimates of \mbox{$R_\star = 0.42 \pm 0.01 \rsun$},  \mbox{$T_{\rm eff} = 3463 \pm 76$K} agree well with the previously published estimates of \mbox{$R_\star = 0.44 \pm 0.02 \rsun$} and  \mbox{$T_{\rm eff} = 3410 \pm 65$K} from \citet{mann_et_al2017b}; \mbox{$R_\star = 0.402\pm 0.050 \rsun$} and  \mbox{$T_{\rm eff} = 3471 \pm 124$K} from \citet{obermeier_et_al2016};  \mbox{$R_\star = 0.44 \pm 0.03 \rsun$} and \mbox{$T_{\rm eff} = 3325 \pm 100$K} from \citet{pepper_et_al2017b}; and  \mbox{$R_\star = 0.42 \pm 0.09 \rsun$} and  \mbox{$T_{\rm eff} = 3704 \pm 214$K} from \citet{martinez_et_al2017}.

Like \citet{vanderburg_et_al2016}, \citet{mayo_et_al2018} estimated spectroscopic stellar parameters by obtaining TRES spectra and running SPC. Three stars from \citet{mayo_et_al2018} are in our cool dwarf sample. As shown in Figure~\ref{fig:bypaper}, our radius estimates are significantly larger for two stars (EPIC~201110617 = K2-156 and EPIC~220321605 = K2-212), but our radius estimate for EPIC~201390048 (K2-162) is consistent with that from \citet{mayo_et_al2018}. Our temperature estimates for K2-156 and K2-212 are nearly 200K cooler than those estimated by \citet{mayo_et_al2018} and our estimate for K2-162 is roughly 350K cooler. 

\section{Conclusions}
\label{sec:conclusions}
This paper is the fourth in a series of papers about cool dwarfs observed by the \emph{K2} mission. We presented NIR spectroscopy and revised classifications for 172~candidate cool dwarfs observed by \emph{K2} during campaigns~\mbox{1 -- 17}. While 86 (50\%) of our target stars were indeed cool dwarfs, our sample also included 74~hotter stars and 12~giant stars. 

For the cool dwarfs, we estimated stellar properties from our NIR spectra using empirical relations developed by \citet{newton_et_al2014, newton_et_al2015} and \citet{mann_et_al2013a, mann_et_al2013c}. We also determine photometric properties by combining parallaxes and inferred distances from Gaia DR2 with archival photometry. We found that the radius and effective temperature estimates from both methods agreed well. However, the stellar effective temperatures reported by the Gaia team were approximately 200K hotter than our photometric or spectroscopics estimates. 

The spectroscopic and photometric mass estimates are correlated, but the slope of the relation is shallower than a 1:1 line, which causes the photometric mass estimates to be larger than the spectroscopic mass estimates for the least massive stars and smaller than the spectroscopic mass estimates for the most massive stars. For the eleven stars with photometric mass estimates below $0.4\msun$, the photometric estimates were systematically $0.11\msun$ (34\%) higher than the spectroscopic mass estimates. For stars with photometric masses  
$0.4\msun < M_{\rm phot} < 0.6\msun$, the offset persists, but the difference is smaller: the photometric masses are roughly $0.03\msun$ (6\%) higher than the spectroscopic masses. Finally, for the most massive cool dwarfs ($M_{\rm phot} > 0.6 \msun$), we found that the photometric mass estimates were $0.02\msun$ (3\%) lower than the spectroscopic mass estimates. The offset between the spectroscopic and photometric mass estimates could be partially explained by unresolved binaries.

Our cool dwarf sample extended from K5 to M4. Eleven of the 86~cool dwarfs have candidate stellar companions within $1"$ revealed by AO or speckle imaging (3 stars) or were identified as possible eclipsing binaries (8 stars). For the remaining 75~stars that are presumed to be single or in wide binaries, we found that the distribution of stellar radii extends from  $0.24\rsun-0.74\rsun$ with a median value of $0.58\rsun$, the stellar masses range from $0.22\msun$ to $0.75\msun$ with a median value of $0.58\msun$, and the stellar effective temperatures span 3077 -- 4730K with a median value of 3693K. The typical star in the sample is slightly metal-poor (median [Fe/H]$ = -0.06$), but the sample extends from [Fe/H]$ = -0.42$ to [Fe/H]$ = 0.50$.

Compared to the original stellar radii published in the EPIC, our revised radii tend to be larger. The median increase in the estimated stellar radius is $0.15\rsun$ (40\%). This increase is nearly identical to the difference of $0.13\rsun$ (39\%) we found in \citet{dressing_et_al2017a} between our revised stellar radii and the original EPIC estimates for the first set of stars considered as part of this project. In addition to the change in the radius estimates, we find that the stellar effective temperatures in the EPIC are overestimated by roughly 65K relative to our revised values. 

Extending the comparison to previously published \emph{K2} planet candidate catalogs \citep{adams_et_al2016, crossfield_et_al2016, mayo_et_al2018, montet_et_al2015, vanderburg_et_al2016}, we find that other previous studies have also tended to underestimate stellar radii and overestimate stellar effective temperatures. The radii and equilibrium temperatures of transiting planets are derived from their transit depths and the properties of their host stars, so systematic errors in stellar properties will lead to corresponding errors in planetary properties. Ignoring any possible systematic over- or under-estimates of the planet/star radius ratios, we anticipate the radii of any associated planets are also 5--40\% larger than previously calculated using catalogs that relied on theoretical models to estimate stellar properties.  

\begin{acknowledgments}
Many of our targets were provided by the \emph{K2} California Consortium (K2C2). We thank K2C2 for sharing their candidate lists and vetting products. We are grateful to Michael Cushing for sharing a beta version of the Spextool pipeline designed for TripleSpec data. We thank Philip Muirhead and Juliette Becker for providing advice regarding TripleSpec data acquisition and reduction. We also acknowledge helpful conversations with Chas Beichman, Eric Gaidos, and Ellianna Schwab Abrahams. 

This work was performed in part under contract with the Jet Propulsion Laboratory (JPL) funded by NASA through the Sagan Fellowship Program executed by the NASA Exoplanet Science Institute. C.D.D. and I.J.M.C. acknowledge support from the \emph{K2} Guest Observer Program. 

This paper was motivated by data collected by the \emph{K2} mission, which is funded by the NASA Science Mission directorate. Our follow-up observations were obtained at the IRTF, which is operated by the University of Hawaii under contract NNH14CK55B with the National Aeronautics and Space Administration and at Palomar Observatory. We thank the staff at both observatories and the Caltech Remote Observing Facilities staff for supporting us during our many observing runs.We are grateful to the IRTF and Caltech TACs for awarding us telescope time. This research has made use of the NASA Exoplanet Archive, which is operated by the California Institute of Technology, under contract with the National Aeronautics and Space Administration under the Exoplanet Exploration Program. This work initially made use of the gaia-kepler.fun crossmatch database created by Megan Bedell.

The authors wish to recognize and acknowledge the very significant cultural role and reverence that the summit of Maunakea has always had within the indigenous Hawaiian community.  We are most fortunate to have the opportunity to conduct observations from this mountain. 
\end{acknowledgments}

\facilities{IRTF (SpeX), Palomar (TripleSpec), Palomar (PHARO), Keck (NIRC2), Gemini-N (NIRI), Lick-3m (ShaneAO), WIYN (NESSI)}
\software{ {\tt am\_getmetal} \citep{mann_et_al2013a}, {\tt astropy} \citep{astropy_et_al2018}, {\tt astroquery} \citep{ginsburg_et_al2016}, {\tt iPython} \citep{perez+granger2007}, {\tt matplotlib} \citep{hunter2007}, {\tt nirew} \citep{newton_et_al2014, newton_et_al2015}, {\tt numpy} \citep{oliphant2015}, {\tt pandas} \citep{mckinney2010}, {\tt Spextool}  \citep{cushing_et_al2004},   {\tt RV\_code} by Andrew Mann\footnote{\url{https://github.com/awmann/RV_code}}, {\tt scipy} \citep{jones_et_al2001}, {\tt tellrv} \citep{newton_et_al2014},  {\tt xtellcor} \citep{vacca_et_al2003}}

\begin{longrotatetable}
\begin{deluxetable*}{cccccccccccc}
\tablecolumns{12}
\tabletypesize{\small}
\tablecaption{Spectral Indices and Gaia Crossmatches for Targets Classified as Evolved Stars\label{tab:giants}}
\tablehead{
\multicolumn{3}{c}{EPIC} &
\multicolumn{5}{c}{Gaia DR2} &
\multicolumn{4}{c}{This Work} \\
\colhead{EPIC} &
\colhead{$Kp$} &
\colhead{$Ks$} & 
\colhead{Designation} & 
\colhead{PM RA} &
\colhead{PM Dec} &
\colhead{Parallax} &
\colhead{Distance} &
\colhead{Instrument} &
\colhead{EW K I} &
\colhead{EW Na I} &
\colhead{EW Ca II} 
}
\startdata
 201231064\tablenotemark{a} &  12.358 &  10.261 &   3598019894861833856 &  -0.384758 & -16.393264 &  1.062600 &   916.318517 & TSPEC & $\cdots$ &     $\cdots$ &         $\cdots$ \\
  \hline
   201516974 &  11.238 &   9.270 &   3798898062211776256 &  -5.199212 & -37.565546 &  1.233189 &   792.410202 & TSPEC & $\cdots$ &     $\cdots$ &         $\cdots$ \\
  \hline
 203776696 &  15.037 &  11.853 &   6049057713786919936 &   7.634639 &   0.802432 &  0.939381 &  1038.411156 & TSPEC & $\cdots$ &     $\cdots$ &         $\cdots$ \\
  \hline
 205084841 &  15.605 &  12.612 &   4131264587452082944 &  -3.982365 &  -2.197555 &  1.159403 &   843.771463 & TSPEC & $\cdots$ &     $\cdots$ &         $\cdots$ \\
  \hline
 206311743 &  12.922 &  10.701 &   2616298436668399616 &   7.075691 & -36.846634 &  1.494429 &   656.665138 & TSPEC & $\cdots$ &     $\cdots$ &         $\cdots$ \\
  \hline
 211333233 &   9.653 &   5.883 &   600784874384342400 &  -1.802325 &   3.770199 &  0.774226 &  1239.495584 &   SpeX & 2.295561 & -0.755824 &  13.595743 \\
 \hline
 214741009 &  14.012 &  11.788 &  4073371142719483264 &  -3.310084 &  -0.777168 &  0.138980 &  4885.770766 &  SpeX & -0.434720 & -0.465135 &   5.117540 \\
 214741009 &  14.012 &  11.788 &  4073371142719483136 &        $\cdots$ &        $\cdots$ &       $\cdots$ &          $\cdots$ & SpeX &  -0.434720 & -0.465135 &   5.117540 \\
\enddata
\tablenotetext{a}{\citet{brahm_et_al2019} classified EPIC~201231064 (K2-161) as a ``slightly evolved G~star'' with $R_\star = 1.669 \pm 0.022 \rsun$ and $M_\star = 1.105 \pm 0.019 \msun$.}
\tablenotetext{b}{Our SpeX observation of EPIC~224685166 had insufficient SNR at blue wavelengths to compute these spectral indices.}
\end{deluxetable*}
\end{longrotatetable}

%\clearpage
\begin{longrotatetable}
\begin{deluxetable*}{ccccccccccccccccc}
\tablecolumns{17}
\tabletypesize{\tiny}
\tablecaption{Candidate Stellar Companions Identified in Gaia DR2\label{tab:companions}}
\tablewidth{700pt}
\tablehead{
\multicolumn{7}{c}{Primary} &
\multicolumn{6}{c}{Secondary} &
\multicolumn{4}{c}{Analysis} \\
\cline{1-7}
\cline{14-17}
\colhead{} & 
\colhead{} & 
\colhead{RA} & 
\colhead{Dec} &
\colhead{$\mu_{\rm RA}$} &
\colhead{$\mu_{\rm Dec}$} &
\colhead{$\pi$} &
\colhead{} & 
\colhead{RA} & 
\colhead{Dec} &
\colhead{$\mu_{\rm RA}$} &
\colhead{$\mu_{\rm Dec}$} &
\colhead{$\pi$} &
\colhead{$\Delta \theta$} &
\colhead{$\Delta \mu$} &
\colhead{Cut} &
\colhead{Bound} \\
\colhead{EPIC} & 
\colhead{Designation\tablenotemark{a}} & 
\colhead{($^\circ$)} & 
\colhead{($^\circ$)} &
\colhead{mas/yr} &
\colhead{mas/yr} &
\colhead{mas} &
\colhead{Designation\tablenotemark{a}} & 
 \colhead{($^\circ$)} & 
\colhead{($^\circ$)} &
\colhead{mas/yr} &
\colhead{mas/yr} &
\colhead{mas} &
\colhead{($"$)} &
\colhead{mas/yr} &
\colhead{mas/yr} &
\colhead{Pair?} 
}
\startdata
\toprule
201119435 &   3595791498326534144 &  180.104930 &  -6.052428 &  -28.7&   -8.19 &  1.79 &   3595791498324496384 &  180.105224 &  -6.052890 &  -27.6 &   -7.31 &        -0.08 &   1.97&    1.42 &  1.0 &       no \\
\hline
 201650711 &   3812335125095532672 &  172.044400 &   2.826726 &   80.2 &  -28.8 &  10.91 &   3812335125094701056 &  172.044175 &   2.827166 &   80.1 &  -29.1 &        10.84 &   1.78 &    0.38 &  65.2 &      yes \\
 \hline
 202071401 &   3378104379464943616 &  100.378771 &  20.921884 &  178.3 & -208.6 &  7.83 &   3378104379464943104 &  100.379545 &  20.922202 &  177.8 & -208.9 &        7.82 &   2.93 &    0.58 &  3051 &      yes \\
 202071401 &   3378104379464943616 &  100.378771 &  20.921884 &  178.3 & -208.6 &  7.83 &   3378104379464943232 &  100.378110 &  20.923489 &   -0.95 &   -2.709 &       0.72 &   6.22 &  272.96 &  107 &       no \\
 \hline
  202071635 &   3373469800514670976 &   93.571792 &  18.627016 &   -0.19 &   -1.49 &  0.34 &   3373469800511625856 &   93.571783 &  18.625649 &   -0.29 &   -2.036 &        0.87 &   4.92 &    0.55 &  0.00001 &       no \\
  \hline
  202086968 &   3369361402301215616 &   97.005609 &  16.388881 &   -3.87 &   -4.41 &  1.96 &   3369361406595494016 &   97.005521 &  16.388346 &   -4.20 &   -4.15 &        1.94 &   1.95 &    0.42 &  0.002 &       no \\
 \hline
 205040048 &   6245720108744660480 &  242.168354 & -19.696900 &  -33.3 & -237.8 &  10.96 &   6245720104449034880 &  242.167800 & -19.695982 &    3.57 &   -6.05 &        0.29 &   3.83 &  234.70 &  124 &       no \\
 \hline
  205111664 &   6245607576304205952 &  243.680091 & -19.346884 &   -3.95 &  -67.2 &  5.36 &   6245607580597734912 &  243.679158 & -19.346593 &         $\cdots$ &         $\cdots$ &        $\cdots$ &   3.43 &         $\cdots$ &           $\cdots$ &      unknown \\
\enddata
\tablenotetext{a}{All designations should be preceded by ``Gaia DR2''.} 
\end{deluxetable*}
\end{longrotatetable}

\clearpage
\begin{deluxetable*}{ccccccc}
\tablecolumns{7}
\tabletypesize{\small}
\tablecaption{Observing Conditions\label{tab:observing}}
\tablehead{
\colhead{} &
\colhead{} &
\colhead{} &
\colhead{Date} &
\colhead{Seeing} &
\colhead{} &
\colhead{} \\
\colhead{Semester} &
\colhead{Instrument} &
\colhead{Program} & 
\colhead{(UT)} &
\colhead{(")} &
\colhead{Weather Conditions} &
\colhead{K2OIs}
}
\startdata
2016A & TSPEC & P08 & 2016 Mar 27 & $0\farcs7 - 1\farcs2$ &  clear & 3 \\
 & TSPEC & P08 & 2016 Mar 28 & $0\farcs8 - 2\farcs1$&  mostly clear  & 25 \\
 \hline
2016B & SpeX & 057 & 2016 Aug 7 & $0\farcs5 - 0\farcs8$ &  closed part of night because of high humidity & 6 \\
   & SpeX & 057 & 2016  Aug 09 & $0\farcs7 - 1\farcs3$ &  clear & 9 \\
 & SpeX & 057 & 2016 Aug 10 & $0\farcs4 - 1\farcs2$ &  clear & 3 \\
 & SpeX & 057 & 2016 Sep 05 & $0\farcs3 - 0\farcs5$  & cirrus & 4 \\
 & SpeX & 057 & 2016 Sep 27 & $0\farcs3 - 0\farcs6$ &  clear & 1 \\
 & SpeX & 057 & 2016 Sep 28 & $0\farcs5 - 0\farcs7$ &  humid & 3 \\
\enddata                  
\end{deluxetable*}

\begin{longrotatetable}
\begin{deluxetable*}{cccccccccccccccccc}
\tablecolumns{18}
\tabletypesize{\tiny}
\tablecaption{Stellar Classifications  \label{tab:classifications}}
\tablehead{
\colhead{} & 
\colhead{} & 
\colhead{} & 
\colhead{} &
\multicolumn{2}{c}{EPIC} &
\multicolumn{4}{c}{Follow-Up Imaging} &
\multicolumn{8}{c}{Gaia} \\
\cline{5-6}
\cline{11-18}
\colhead{} &
\colhead{K2} & 
\colhead{Spectral} &
\colhead{} &
\colhead{} &
\colhead{} &
\colhead{} &
\colhead{} &
\colhead{Eclipsing} &
\colhead{Nearby} &
\colhead{} & 
\colhead{} &
\colhead{Parallax} &
\colhead{Distance} & 
\multicolumn{3}{c}{Flags}\\
\cline{15-17}
\colhead{EPIC} &
\colhead{Name} & 
\colhead{Type} &
\colhead{Campaign} &
\colhead{$Kp$} &
\colhead{$Ks$} &
\colhead{Speckle} &
\colhead{AO} &
\colhead{Binary} &
\colhead{Star} &
\colhead{Match} & 
\colhead{Designation\tablenotemark{a}} &
\colhead{(mas)} &
\colhead{(pc)} &
\colhead{AF\tablenotemark{b}} &
 \colhead{EF\tablenotemark{c}} & 
 \colhead{VF\tablenotemark{d}} &
 \colhead{VP\tablenotemark{e}} 
}
\startdata
201110617 &  K2-156 &       K5V &       10 & 12.947 & 10.391 &         y &      y &     $\cdots$ &         $\cdots$ &  y &   3596250888028092160 &     6.63 &   150 &               $\cdots$ &           $\cdots$ &         $\cdots$ &                      10 \\
201111557 &       $\cdots$ &       K4V &       10 & 11.363 &  9.220 &         y &  $\cdots$ &     $\cdots$ &         $\cdots$ &  y &   3596276829630866432 &    10.26 &    97 &               $\cdots$ &           $\cdots$ &         $\cdots$ &                      11 \\
201119435 &       $\cdots$ &       K5V &       10 & 15.082 & 12.714 &     $\cdots$ &      y &     $\cdots$ &             y &  y &   3595791498326534144 &     1.79 &   551 &               $\cdots$ &           $\cdots$ &         $\cdots$ &                      11 \\
201119435 &       $\cdots$ &       K5V &       10 & 15.082 & 12.714 &     $\cdots$ &      y &     $\cdots$ &             y &     close &   3595791498324496384 &    -0.08 &  1473 &               $\cdots$ &           $\cdots$ &         $\cdots$ &                      10 \\
201127519 &       $\cdots$ &       K4V &       10 & 11.558 &  9.430 &         y &      y &         y &             y &  y &   3597255188821238016 &     8.47 &   118 &               $\cdots$ &           $\cdots$ &         $\cdots$ &                      11 \\
201155177\tablenotemark{f} &   K2-42 &       K5V &        1 & 14.632 & 12.284 &     $\cdots$ &      y &     $\cdots$ &         $\cdots$ &  y &   3599651986730350464 &     2.49 &   397 &               $\cdots$ &           $\cdots$ &         $\cdots$ &                      10 \\
201231064\tablenotemark{g} &  K2-161 &     K1III &       10 & 12.358 & 10.261 &         y &  $\cdots$ &     $\cdots$ &         $\cdots$ &  y &   3598019894861833856 &     1.06 &   916 &               $\cdots$ &           $\cdots$ &         $\cdots$ &                      11 \\
\enddata         
\tablenotetext{a}{All designations should be preceded by ``Gaia DR2''.} 
\tablenotetext{b}{Astrometric Flag. Stars marked as "y" have \tt{astrometric\_sigma5d\_max} $ > 2$~mas in Gaia~DR2.} 
\tablenotetext{c}{Excess Flag. Stars marked as "y" have {\tt  astrometric\_excess\_noise} $ > 2$~mas in Gaia~DR2.} 
\tablenotetext{d}{Visibility Flag. Stars marked as "y" have {\tt  visibility\_periods\_used }$ < 10$ in Gaia~DR2.} 
\tablenotetext{e}{Number of visibility periods used in Gaia~DR2.} 
\tablenotetext{f}{Classified as a dwarf star with a spectral type of K5 but excluded from further analyses because the estimated properties were outside the validity range for the spectroscopic relations established by \citet{newton_et_al2015}.}
\tablenotetext{g}{Validated by \citet{mayo_et_al2018} assuming $T_{\rm eff} = 4972 \pm 50$K and $R_\star = 2.57^{+0.31}_{-0.25}\rsun$.} %K2-161
\tablenotetext{h}{Announced by \citet{vanderburg_et_al2016} and validated by \citet{crossfield_et_al2016}. \citet{vanderburg_et_al2016} classified the star as an M~dwarf with $T_{\rm eff} = 3260$K and $R_\star = 0.23\rsun$, but \citet{crossfield_et_al2016} revised the stellar properties to $T_{\rm eff} = 6133$K and $R_\star = 1.49\pm0.52\rsun$.} %K2-52
\end{deluxetable*}
\end{longrotatetable}

\begin{longrotatetable}
\begin{deluxetable*}{ccccccccccccccccccc}
\tablecolumns{19}
\tabletypesize{\small}
\tablecaption{Spectroscopic Parameters for Cool Dwarfs\label{tab:spectroscopic}}
\tablehead{
\colhead{} & 
\colhead{K2} & 
\colhead{} &
\multicolumn{3}{c}{$T_{\rm eff}$ (K)} &
\multicolumn{3}{c}{$R_\star$ ($\rsun$)} &
\multicolumn{3}{c}{$M_\star$ ($\msun$)} &
\multicolumn{3}{c}{$L_\star$ ($\lsun$)} &
\multicolumn{2}{c}{[Fe/H]} &
\multicolumn{2}{c}{[M/H]} \\
\cline{4-6}
\cline{10-12}
\cline{16-17}
\colhead{EPIC} &
\colhead{Name} &
\colhead{Type} &
\colhead{Val} &
\colhead{-Err} &
\colhead{+Err} &
\colhead{Val} &
\colhead{-Err} &
\colhead{+Err} &
\colhead{Val} &
\colhead{-Err} &
\colhead{+Err} &
\colhead{Val} &
\colhead{-Err} &
\colhead{+Err} &
\colhead{Val} &
\colhead{Err} &
\colhead{Val} &
\colhead{Err} 
}
\startdata
201110617 &  K2-156 &       K5V & 4307 &     135 &     152 & 0.646 & 0.037 & 0.044 & 0.689 &   0.079 &   0.079 & 0.092 &  0.026 &  0.033 &  -0.167 &    0.092 &  -0.060 &    0.090 \\
201119435 &       $\cdots$ &       K5V & 4472 &     121 &     151 & 0.743 & 0.047 & 0.054 & 0.712 &   0.080 &   0.082 & 0.278 &  0.054 &  0.064 &  -0.009 &    0.087 &  -0.011 &    0.087 \\
201264302 &       $\cdots$ &       M2V & 3586 &      92 &      95 & 0.437 & 0.031 & 0.031 & 0.453 &   0.082 &   0.075 & 0.032 &  0.005 &  0.006 &  -0.114 &    0.086 &  -0.157 &    0.085 \\
201367065 &    K2-3 &       M1V & 4191 &     118 &     131 & 0.571 & 0.030 & 0.030 & 0.671 &   0.077 &   0.076 & 0.083 &  0.018 &  0.021 &  -0.321 &    0.090 &  -0.276 &    0.087 \\
201390048 &  K2-162 &       K5V & 4499 &     126 &     130 & 0.646 & 0.038 & 0.041 & 0.715 &   0.081 &   0.081 & 0.199 &  0.026 &  0.029 &  -0.362 &    0.081 &  -0.263 &    0.081 \\
201465501 &    K2-9 &       M3V & 3477 &      98 &      99 & 0.371 & 0.036 & 0.035 & 0.374 &   0.095 &   0.083 & 0.018 &  0.003 &  0.004 &  -0.287 &    0.084 &  -0.245 &    0.083 \\
201596733 &       $\cdots$ &       M1V & 3568 &      90 &      93 & 0.505 & 0.032 & 0.032 & 0.441 &   0.082 &   0.075 & 0.048 &  0.009 &  0.011 &   0.106 &    0.090 &   0.052 &    0.089 \\
201650711 &       $\cdots$ &       K7V & 3782 &     126 &     141 & 0.631 & 0.069 & 0.104 & 0.558 &   0.087 &   0.080 & 0.063 &  0.021 &  0.027 &  -0.108 &    0.102 &  -0.079 &    0.098 \\
\enddata 
\end{deluxetable*}
\end{longrotatetable}

\begin{longrotatetable}
\begin{deluxetable*}{cccccccccccccc}
\tablecolumns{14}
%\tabletypesize{\small}
\tablecaption{Photometric Parameters for Cool Dwarfs\label{tab:photometric}}
\tablehead{
\colhead{} & 
\colhead{K2} & 
\multicolumn{3}{c}{$M_{Ks}$} &
\multicolumn{2}{c}{$T_{\rm eff}$ (K)} &
\multicolumn{2}{c}{$R_\star$ ($\rsun$)} &
\multicolumn{2}{c}{$M_\star$ ($\msun$)} &
\multicolumn{2}{c}{$L_\star$ ($\lsun$)} & 
\colhead{Radius}\\
\cline{3-5}
\cline{8-9}
\cline{12-13}
\colhead{EPIC} &
\colhead{Name} &
\colhead{Val} &
\colhead{-Err} &
\colhead{+Err} &
\colhead{Val} &
\colhead{Err} &
\colhead{Val} &
\colhead{Err} &
\colhead{Val} &
\colhead{Err} &
\colhead{Val} &
\colhead{Err} &
\colhead{Flag\tablenotemark{a}}  \\
}
\startdata
201110617 &  K2-156 &    4.508 &       -0.016 &       -0.016 &   4265 &      78 &    0.706 &     0.021 &    0.673 &     0.017 & 0.148 & 0.006 &           y \\
201119435 &       $\cdots$ &    4.010 &       -0.056 &       -0.057 &   $\cdots$ &     $\cdots$ & $\cdots$ &    $\cdots$ &    0.752 &     0.030 & 0.273 & 0.017 &           $\cdots$ \\
201264302 &       $\cdots$ &    6.074 &       -0.007 &       -0.008 &   3536 &      61 &    0.431 &     0.012 &    0.432 &     0.010 & 0.026 & 0.001 &           $\cdots$ \\
201367065 &    K2-3 &    5.340 &       -0.005 &       -0.005 &   3811 &      66 &    0.541 &     0.015 &    0.551 &     0.013 & 0.055 & 0.002 &           $\cdots$ \\
201390048 &  K2-162 &    4.423 &       -0.011 &       -0.011 &   4532 &      81 &    0.723 &     0.022 &    0.685 &     0.017 & 0.198 & 0.008 &           y \\
201465501 &    K2-9 &    6.901 &       -0.010 &       -0.010 &   3306 &      58 &    0.319 &     0.009 &    0.303 &     0.008 & 0.011 & 0.000 &           $\cdots$ \\
201596733 &       $\cdots$ &    5.497 &       -0.015 &       -0.015 &   3598 &      64 &    0.525 &     0.015 &    0.526 &     0.013 & 0.042 & 0.002 &           $\cdots$ \\
201650711 &       $\cdots$ &    4.504 &       -0.013 &       -0.013 &   4078 &      72&    0.700 &     0.020 &    0.673 &     0.017 & 0.122 & 0.005 &           y \\
\enddata
\tablenotetext{a}{Photometric radius estimated by extrapolating the relations in \citet{mann_et_al2015}.}
\tablenotetext{b}{We do not report a photometric luminosity or a photometric temperature for EPIC~220448185 because we did not find a match for this star in the Carlsberg Meridian Catalogue \citep{muinos+evans2014} and were therefore unable to use the $r-J$ color to compute a bolometric correction.} 
\end{deluxetable*}
\end{longrotatetable}

\begin{longrotatetable}
\begin{deluxetable*}{ccccccccccccccccc}
\tablecolumns{17}
%\tabletypesize{\small}
\tablecaption{Adopted Parameters for Cool Dwarfs\label{tab:adopted}}
\tablehead{
\colhead{} & 
\multicolumn{4}{c}{$T_{\rm eff}$ (K)} &
\multicolumn{4}{c}{$R_\star$ ($\rsun$)} &
\multicolumn{4}{c}{$M_\star$ ($\msun$)} &
\multicolumn{4}{c}{$L_\star$ ($\lsun$)} \\
\cline{2-5}
\cline{10-13}
\colhead{EPIC} &
\colhead{Val} &
\colhead{-Err} &
\colhead{+Err} &
\colhead{Prov} &
\colhead{Val} &
\colhead{-Err} &
\colhead{+Err} &
\colhead{Prov} &
\colhead{Val} &
\colhead{-Err} &
\colhead{+Err} &
\colhead{Prov} &
\colhead{Val} &
\colhead{-Err} &
\colhead{+Err} &
\colhead{Prov}
}
\startdata
201110617 &       4265 &            78 &            78 &    phot &    0.706 &       0.021 &       0.021 &  phot &    0.673 &       0.017 &       0.017 &  phot &     0.145 &        0.002 &        0.002 &   phot \\
201119435 &       4331 &           121 &           151 &    spec &    0.743 &       0.047 &       0.054 &  spec &    0.752 &       0.030 &       0.030 &  phot &     0.258 &        0.015 &        0.015 &   phot \\
201264302 &       3536 &            61 &            61 &    phot &    0.431 &       0.012 &       0.012 &  phot &    0.432 &       0.010 &       0.010 &  phot &     0.032 &        0.005 &        0.006 &   spec \\
201367065 &       3811 &            66 &            66 &    phot &    0.541 &       0.015 &       0.015 &  phot &    0.551 &       0.013 &       0.013 &  phot &     0.083 &        0.018 &        0.021 &   spec \\
201390048 &       4532 &            81 &            81 &    phot &    0.723 &       0.022 &       0.022 &  phot &    0.685 &       0.017 &       0.017 &  phot &     0.220 &        0.002 &        0.002 &   phot \\
201465501 &       3308 &            58 &            58 &    phot &    0.319 &       0.009 &       0.009 &  phot &    0.303 &       0.008 &       0.008 &  phot &     0.018 &        0.003 &        0.004 &   spec \\
201596733 &       3598 &            64 &            64 &    phot &    0.525 &       0.015 &       0.015 &  phot &    0.526 &       0.013 &       0.013 &  phot &     0.048 &        0.009 &        0.011 &   spec \\
201650711 &       4078 &            72 &            72 &    phot &    0.700 &       0.020 &       0.020 &  phot &    0.673 &       0.017 &       0.017 &  phot &     0.099 &        0.001 &        0.001 &   phot \\
\enddata      
\end{deluxetable*}
\end{longrotatetable}

\bibliography{mdwarf_biblio_oct2018}

\clearpage
\enddocument